\documentclass{aa}  
\usepackage{graphicx}
\usepackage{epsfig}
\usepackage{natbib}
\bibpunct{(}{)}{;}{a}{}{,}
\usepackage{graphicx, amsmath}

\usepackage{txfonts}
\newcommand{\cnt}{C^\mathrm{nt}}

\newcommand{\parcD}[2]{\frac{\partial #1}{\partial #2}}

\newcommand{\dd}{\,\mathrm{d}}
\newcommand{\LparcD}[2]{\partial #1 / \partial #2}

\begin{document}
   \title{Response of optical hydrogen lines to beam heating: \\ 
                  I. Electron beams}
   \titlerunning{Response of optical hydrogen lines to beam heating: \\ 
                  I. Electron beams} 
   \authorrunning{Ka\v{s}parov\'{a}, J. et al.}

   \author{J. Ka\v{s}parov\'{a}
          \inst{1}
          \and
          M. Varady
          \inst{2,1}    
          \and
          {P. Heinzel \inst{1}
          \and
          M. Karlick\'{y}\inst{1}
	  \and
          Z. Moravec \inst{2}}}

   \offprints{Ka\v{s}parov\'{a}}
    
   \institute{Astronomick\'{y} \'{u}stav Akademie v\v{e}d \v{C}esk\'{e} republiky, v.v.i., Fri\v{c}ova 298, 
    251 65 Ond\v{r}ejov, Czech Republic\\
             \email{kasparov@asu.cas.cz}
         \and
    Katedra fyziky, Universita J. E. Purkyn\v e, \v Cesk\'e ml\'ade\v ze 8,
    400 24 \'Ust\'\i~nad Labem, Czech Republic\\
              \email{mvarady@physics.ujep.cz}
             }
  \abstract
  {Observations of hydrogen Balmer lines in solar flares remain an important 
  source of information on flare processes in the 
  chromosphere during the impulsive phase of flares. The intensity profiles of optically thick hydrogen 
  lines are determined by the temperature, density, and ionisation structure of the flaring atmosphere, 
  by the plasma velocities and by the velocity distribution of particles 
  in the line formation regions.}
  {We investigate the role of non-thermal electrons in the formation regions
  of H$\alpha$, H$\beta$, and H$\gamma$ lines in order to unfold their influence on the formation of these lines. 
  We concentrate on pulse-beam heating varying on a subsecond timescale.
  Furthermore, we theoretically explore possibility that a new diagnostic tool exists indicating the 
  presence of non-thermal electrons in the 
  flaring chromosphere based on observations of optical hydrogen lines.
  }
  {To model the evolution of the flaring atmosphere and the time-dependent
  hydrogen excitation and ionisation, we used a 1-D radiative hydrodynamic code combined 
  with a test-particle code that simulates the propagation, scattering, and 
  thermalisation of a power-law electron beam in order to obtain the flare 
  heating and the non-thermal collisional rates due to the interaction of the
  beam with the hydrogen atoms. To not bias the results by other 
  effects, we calculate 
  only short time evolutions of the flaring atmosphere and neglect the plasma 
  velocities in the radiative transfer.} 
  {All calculated models have shown a time-correlated response of
 the modelled Balmer line intensities on a subsecond timescale,
 with a subsecond timelag behind the beam flux. 
 Depending on the beam parameters, both line centres and wings can show
 pronounced intensity variations. The non-thermal 
 collisional rates generally result in an increased emission from a
 secondary region formed in the chromosphere.
}
  {Despite the clear influence of the non-thermal electron beams on the 
 Balmer line intensity profiles, we were not able on the basis of our simulations  
 to produce  any unambiguous diagnostic of non-thermal electrons in the line-emitting region, 
 which would be based on comparison of individual Balmer line
 intensity profiles. 
However, fast line intensity variations, well-correlated with the beam flux variations, 
represent an indirect indication of pulsating beams.

}


   \maketitle
%

\section{Introduction}
In the context of interpreting flare loop hard X-ray footpoint sources 
\citep{hoy81,hufa02}, all contemporary flare models \citep{stu68,kopne76, 
shi96,turk05,fle08}, regardless of their nature,
assign a fundamental role during the flare energy release, 
transport and deposition to the high-energy non-thermal particle beams. In the
impulsive phase of flares, the beams formed by charged particles are also guided 
from the acceleration site (wherever it is located) downwards along the 
magnetic field lines into the transition region, chromosphere and possibly 
photosphere. At lower atmospheric layers 
due to the high density of local plasma, their kinetic 
energy is efficiently dissipated by Coulomb collisions, the corresponding regions 
are rapidly heated, and dramatic changes of temperature and ionisation occur.
This results in explosive evaporation \citep{dos96}. 
The manifestations of the early flare processes can be observed in the 
microwaves, soft and hard-X rays, and  optical lines \citep{ta88}. Later on in the thermal  
(gradual) phase, the heating leads to the evaporation of chromospheric 
and transition region plasma into the corona which is gradually filled with 
relatively dense (up to $\sim10^{10}$~cm$^{-3}$) and hot ($\sim 10^7$~K) flare 
plasma \citep{czay99}. The radiation from the flare region is
now dominated by soft-X rays, EUV, and again by emission in the optical
spectral lines.    

We concentrate on modelling the formation of optically thick hydrogen spectral 
lines H$\alpha$, H$\beta$, and H$\gamma$ in the early phases of solar flares by 
the means of 
numerical radiative hydrodynamics combined with a test particle approach to 
simulate the propagation, scattering and energy loss of an electron beam with
the power-law spectrum and prescribed time-dependent energy flux
propagating through the solar atmosphere and depositing its energy into the
solar plasma. In this context we address three main questions: 
\begin{enumerate}
\item {Does rapidly varying electron beam flux manifest itself in the Balmer line intensities?}
\item {How do the non-thermal particles in the Balmer lines formation regions influence the line profiles and intensities?}
\item {Can an unambiguous diagnostic method be developed  that is applicable to
  observations of Balmer lines recognising the presence of the
  non-thermal particles in the line formation regions?}
\end{enumerate}

Due to the complexity of simultaneously treating non-LTE radiative 
transfer in deep layers of the solar atmosphere and the hydrodynamics 
(radiative hydrodynamics), only a few attempts have been made 
to model the optical emission of flares. 
First models of pulse-beam heating were developed by \citet{can84} and 
\citet{ficacly85}. Recently, \citet{abb99} and  \citet{all05} studied 
emission in several lines and continua using complex radiative
hydrodynamic simulations of electron beam heating on a time scale up to several tens of seconds.

In this paper we concentrate on fast time variations on a subsecond time scale.
In previous works on this topic, plasma dynamics was neglected. Simplified 
time-dependent non-LTE 
simulations of H$\alpha$ line were then performed e.g. by 
using a prescribed time evolution of a flare atmosphere from independent hydrodynamic simulations 
of pulse-beam heating \citep{hein91} or by solving approximate energy equation \citep{ding01}.
Both results showed significant H$\alpha$ line response to pulse-beam heating on subsecond time scales.
Here, we solve 1-D radiative hydrodynamics of a solar atmosphere subjected to a subsecond electron beam heating
and study emissions in the H$\alpha$, H$\beta$, and H$\gamma$ lines.

The paper is organised as follows:
Section~\ref{sec:model} describes the numerical code and models of beam heating. Results of simulations
concerning flare atmosphere dynamics and Balmer line emission are presented in Section~\ref{sec:results}.
There, we also analyse several proposed diagnostic methods for recognising the presence of electron beams in line
formation regions. Section~\ref{sec:discussion} summarises our results.

\section{Model}\label{sec:model}
 
The model covers three important classes of processes whose importance was 
identified in flares:
\begin{enumerate}
\item{Propagation of charged, high-energy particle beams with power-law
  spectra and time-dependent energy flux downwards through the solar
  atmosphere and their gradual thermalisation due to the Coulomb collisions 
  with the ambient plasma in the solar atmosphere 
   \citep{bai82,em78}.} 
\item{The hydrodynamic response of low-$\beta$ solar plasma corresponding 
  to the energy deposited by the beam.}
\item{Time evolution of the ionisation structure and formation of optical emission in the
chromosphere and photosphere where non-LTE conditions apply.}
\end{enumerate} 
The individual classes of flare processes are modelled using three 
computer codes, each modelling one class of the processes 
identified above. The codes have been integrated into one 
radiative hydrodynamic code.
   
\subsection{Flare heating}
\begin{figure}
\begin{center}
\includegraphics[width=6.5cm]{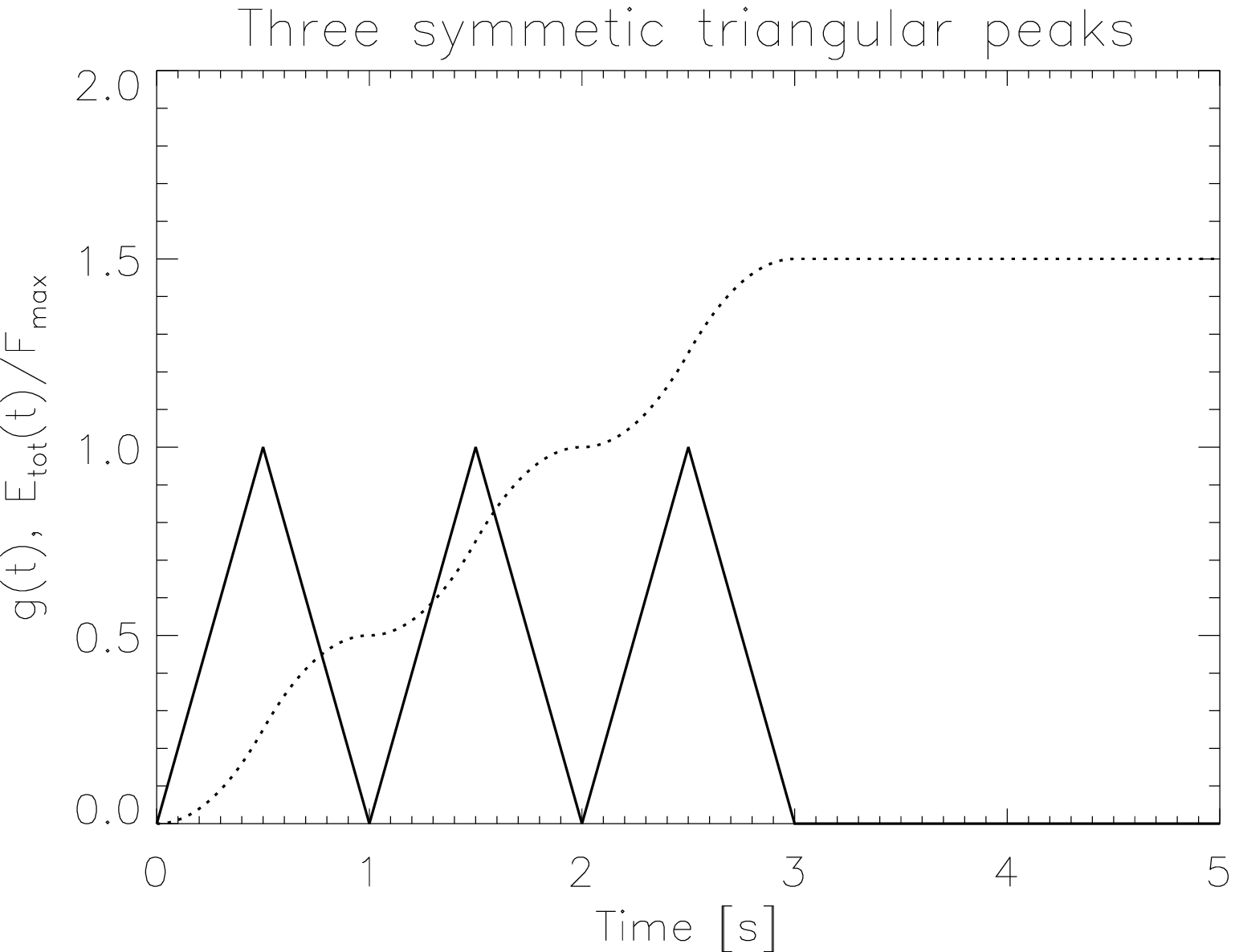}

\includegraphics[width=6.5cm]{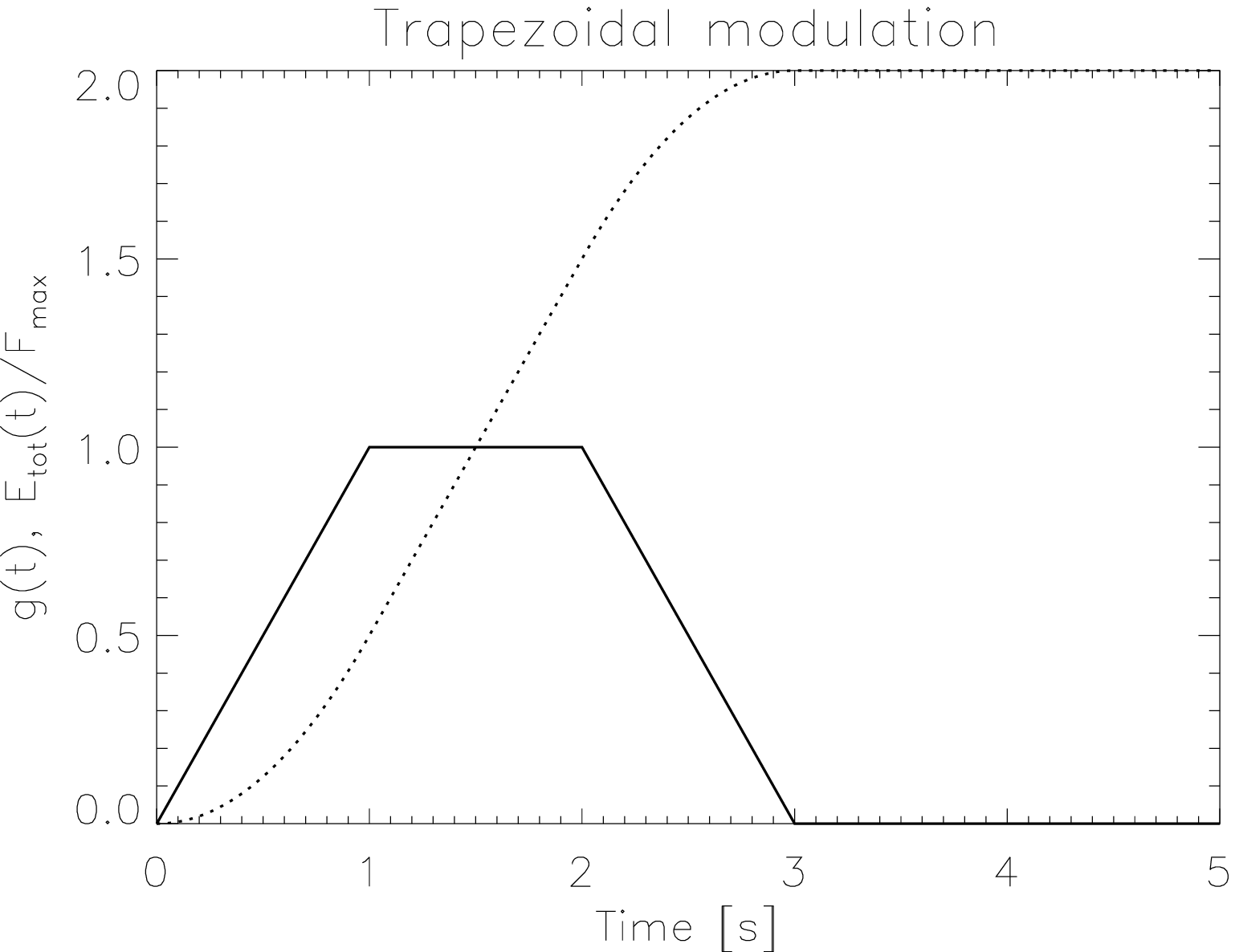}
\end{center}
\caption{Time modulations of the beam flux. Solid lines show $g(t)$, 
dotted lines denote $\int_0^t g(t')\dd t'= E_\mathrm{tot}(t)/ F_\mathrm{max}$,
see Eqs.~(\ref{eq:powerlaw}) and (\ref{eq:etot}).}
\label{fig:timemod}
\end{figure}

The flare heating caused by an electron beam propagating from 
the top of the loop located in the corona ($s=9.5\times10^{3}$~km
corresponding to $T=1$~MK)
downwards is calculated using a test-particle code (TPC) based on
\citet{ka90,ka92}. We assume an electron beam with a power-law 
electron flux spectrum [electrons~cm$^{-2}$~s$^{-1}$ per unit energy] 
\citep{naem84}
\begin{equation}\label{eq:powerlaw}
F(E,t) = g(t)F(E) =g(t)\ (\delta-2) 
\frac{F_\mathrm{max}}{E_0^2}\left(\frac{E}{E_0}\right)^{-\delta} \ ,
\end{equation}
where $\delta$ is the power-law index,
$g(t)\in \langle 0,1\rangle$ is a function describing the time modulation of the 
beam flux,
$F_\mathrm{max}$ is the maximum energy flux, i.e. energy flux of electrons with $E\ge E_0$ at $g(t)=1$.
In order to model the
electron spectrum by the test particles the beam electron energy
is limited by a low and a high-energy cutoff, $E_0=20$~keV and $E_1 = 150$~keV, respectively.
We present results for two types of time modulation $g(t)$: three symmetric triangular peaks 
and a trapezoidal modulation (see Fig.~\ref{fig:timemod})
and two total deposited energies $E_\mathrm{tot}$ (see Table~\ref{modelparam})
\begin{equation}
\label{eq:etot}
E_\mathrm{tot} = F_\mathrm{max}\int_0^{t_1}g(t) \dd t\ ,
\end{equation}
where $t_1$ is the duration of the energy deposit.
The  energy fluxes $F_\mathrm{max}$ have been chosen in such a way 
that for both time modulations the total deposited energy $E_\mathrm{tot}$ is the same. 
The model parameters are specified in Table~\ref{modelparam}, 
their values  are consistent with common beam characteristics derived from hard X-ray observations.

\begin{table}
\caption{Model parameters}
\begin{tabular}{lcccc}
\hline\hline
Model & $E_\mathrm{tot}$ [erg~cm$^{-2}$]& $g(t)$ & $F_\mathrm{max}$ [erg~cm$^{-2}$~s$^{-1}$]& $\delta$\\
\hline
H\_TP\_D3 & $9\times10^{10}$& trapezoid& $4.5\times10^{10}$& 3\\
H\_TP\_D5 & $9\times10^{10}$& trapezoid& $4.5\times10^{10}$& 5\\
H\_3T\_D3 & $9\times10^{10}$& 3 triangles& $6\times10^{10}$& 3\\
H\_3T\_D5 & $9\times10^{10}$& 3 triangles& $6\times10^{10}$& 5\\
\hline
L\_TP\_D3 & $1.5\times10^{10}$& trapezoid& $0.75\times10^{10}$& 3\\
L\_TP\_D5 & $1.5\times10^{10}$& trapezoid& $0.75\times10^{10}$& 5\\
L\_3T\_D3 & $1.5\times10^{10}$& 3 triangles& $1\times10^{10}$& 3\\
L\_3T\_D5 & $1.5\times10^{10}$& 3 triangles& $1\times10^{10}$& 5\\
\hline
\end{tabular}

\label{modelparam}
\end{table}
The TPC simulates the propagation, scattering and energy loss 
of an electron beam with a specified energy flux and
power-law index as it propagates through partly ionised hydrogen
plasma in the solar atmosphere. 
The losses of the beam electron kinetic energy
caused by Coulomb collisions due
to electron and neutral (hydrogen) components of solar plasma
are given by \citet{em78}
\begin{eqnarray}
\Delta E_\mathrm{ee} &=&-\frac{2\pi e^4}{E}\Lambda (x+\varepsilon)
n_\mathrm{H} \varv_\mathrm{e}\Delta t_\mathrm{B}\ , \\
\Delta E_\mathrm{en}&=&-\frac{2\pi e^4}{E}\Lambda^{'}(1-x) n_\mathrm{H} \varv_\mathrm{e}\Delta t_\mathrm{B}\ ,
\end{eqnarray}
where $E$ is the kinetic energy of the non-thermal electron, $\varv_\mathrm{e}$ is the non-thermal velocity,
$n_\mathrm{H}=n_\mathrm{p} + n_\mathrm{n}$ is the number density of equivalent hydrogen atoms,
$n_\mathrm{p}$ and $n_\mathrm{n}$ are the proton and neutral hydrogen number densities,  
$x\equiv n_\mathrm{p}/n_\mathrm{H}$ is the hydrogen ionisation degree and 
$\varepsilon = 1.4\times10^{-4}$ accounts for the contribution from the 
metals to the plasma electron density \citep{heinkar92}.
The metal contribution to electron densities is critical around the temperature minimum where
the hydrogen is almost neutral. We account for it in this approximate way. On the other hand, we neglect
the helium contribution which can reach maximum 20\% of total electron density in higher altitudes.
The Coulomb logarithms 
$\Lambda$ and $\Lambda^{'}$ are given by \citet{em78} and
$\Delta t_\mathrm{B}$ is a constant TPC timestep which has to be chosen to satisfy
the condition that the total beam electron energy loss $\Delta E = \Delta E_\mathrm{ee} + \Delta E_\mathrm{en}$
per a timestep is negligible relative to its kinetic energy, i.e. $\Delta E/E \ll 1$.

The scattering of the beam due to Coulomb collisions is taken into account
using a Monte-Carlo method combined with the analytical expressions for the
cumulative effects described by \citet{bai82}. The relation between the mean square of the beam
electron deflection angle $\langle\theta^2\rangle$ and the corresponding energy loss $\Delta E$  
(which holds if $\langle\theta^2\rangle \ll 1$ or equivalently if $\Delta E/E\ll 1$)
is given by formula
\begin{equation}
\langle\theta^2\rangle = \left(\frac{\Delta E }{E}\right)\left(\frac{4}
{\gamma_\mathrm{L}+1}\right
)\ ,
\label{eq:bai_1}
\end{equation}
where $\gamma_\mathrm{L}=1/\sqrt{1-\varv_e^2/c^2}$ is the Lorentz factor.
The new electron pitch angle $\theta_0 + \Delta\theta$ is given by 
\begin{equation}
\cos(\theta_0 + \Delta\theta) = \cos{\theta_0}\cos{\theta_s} + 
\sin{\theta_0}\sin{\theta_s}\cos{\phi}\ ,
\end{equation}
where $\theta_0$ is the original pitch angle at the beginning of the time step,
$\theta_s$ is given by equation~(\ref{eq:bai_1}) and using a 2-D 
Gaussian distribution. The distribution of the azimuthal angle $\phi$ is uniform,
$\phi\in\langle 0, 2\pi)$.

The TPC in principle follows the motion of statistically 
important number of test particles representing clusters of electrons
in the time varying atmosphere which responses through the 1-D HD code and
the non-LTE radiative transfer code to the flare heating by TPC. 
The test particles with a time-dependent power-law spectra are
generated in the corona at the loop-top and at
each timestep the positions, energies and pitch angles 
of the particle clusters are calculated. 
The macroscopic energy deposits into the electron ${\cal E}_\mathrm{ee}$
and neutral hydrogen ${\cal E}_\mathrm{en}$ component of solar plasma are 
obtained by summing the energy losses ($\Delta E_\mathrm{ee}$ and $\Delta E_\mathrm{en}$)
of a huge number of particle clusters for each position in the atmosphere using a fine
equidistant grid.
This approach allows not only to
calculate the total flare heating 
$$
{\cal H} = {\cal E}_\mathrm{ee} + {\cal E}_\mathrm{en}
$$
but also to distinguish between the beam energy
deposited into the electron and hydrogen component of solar plasma and
therefore to calculate the non-thermal contribution to the transition rates
in hydrogen atoms which is the crucial point for the present study. 
The test-particle approach used here naturally takes into account 
propagation effects of the beam and time evolution
of ionisation structure of the atmosphere. 
This leads us to a more realistic description of beam energy losses as compared to the approach
of \citet{abb99} or \citet{all05} who used an analytic heating function corresponding to a stationary solution
of beam propagation through the atmosphere \citep{1994ApJ...426..387H}.
\subsection{1-D plasma dynamics}\label{hydro}

The state and time evolution of originally hydrostatic low-$\beta$ plasma 
along magnetic field lines is calculated using a 1-D hydrodynamic code. 
The temperature, density and ionisation profiles of the initial atmosphere 
correspond to the VAL~3C atmosphere \citep{val81} with a hydrostatic 
extension into the corona. 
The half-length of the loop is 10~Mm.
The time evolution of the atmosphere is initiated by the energy deposited by 
the beam. The main processes that determine plasma evolution in flare loops 
are: convection and conduction (both in 1-D due to the magnetic field),
radiative losses and indeed the dominant factor is the flare heating here
calculated by the TPC. The evolution of plasma in the flare loop can be 
described by a system of hydrodynamic conservation laws  
\begin{equation}
\parcD{\rho}{t} + \parcD{}{s}(\rho u) = 0\ ,
\end{equation}
\begin{equation}
\parcD{\rho u}{t} + \parcD{}{s}(\rho u^2) = - \parcD{P}{s} + F_\mathrm{g}\ ,
\end{equation}
\begin{equation}
\parcD{E}{t} + \parcD{}{s}(u E) = - \parcD{}{s}(u P) -
\parcD{}{s}{\mathcal{F}}_\mathrm{c}  +
\mathcal{S}\ ,
\end{equation}
where $s$ is the position and $u$ the macroscopic plasma velocity along the
magnetic field line and $\rho$ is the plasma density. The gas pressure and the \
total plasma energy are
\begin{equation}
P = n_\mathrm{H} (1 + x +\varepsilon)k_\mathrm{B} T\,, \qquad E=
\frac{P}{\gamma-1} +\frac{1}{2}\rho u^2\ ,
\end{equation}
where $\gamma\equiv c_\mathrm{p}/c_\mathrm{v} = 5/3$ is the specific heats ratio, $k_\mathrm{B}$ the Boltzmann constant.
The hydrogen ionisation degrees $x$ in the photosphere and chromosphere is calculated at each timestep 
by the time-dependent non-LTE radiative transfer code.  In the transition region and corona we assume $x\equiv1$.
The source terms on the right hand sides of the system of 
equations are: $F_\mathrm{g}$ the parallel component of the gravity force in respect to the semicircular
field line, ${\mathcal F}_\mathrm{c}$ the heat flux, calculated using
the Spitzer's classical formula, and
$$
{\cal S} = {\cal H} - {\cal R} + {\cal Q}
$$
includes all other considered energy sources and sinks, i.e. the dominant flare heating
${\cal H}$ which drives the time 
evolution of the atmosphere, 
the quiet heating ${\cal Q}$ assuring the stability of the initial quiescent
 unperturbed
(hydrostatic) atmosphere and the radiative losses ${\cal R}$. The radiative losses 
are calculated according to \citet{rtv78} for optically thin regions and according 
to \citet{pes82} for optically thick regions. 

The 1-D gas dynamics is treated using the explicit LCPFCT solver 
\citep{orbo87}, the Crank-Nicolson algorithm for the heat transfer and the 
time step splitting
technique to couple the individual source terms of the energy equation with
hydrodynamics \citep{orbo87book}.

\subsection{Time-dependent non-LTE radiative transfer}

Using the instant values of $T$, $n_\mathrm{H}$, and ${\mathcal E}_\mathrm{en}$ obtained by the hydrodynamic
and test-particle codes, a time-dependent non-LTE radiative transfer 
for hydrogen is solved in lower part of the loop in a 1-D plan-parallel approximation.
The hydrogen atom is approximated by a five 
level plus continuum  atomic model.

The level populations  $n_i$ are determined by the solution of a time-dependent 
system of equations of statistical equilibrium (ESE) 
\begin{equation}
\label{ese}
\frac{\partial n_i}{\partial t} = \sum\limits_{j\ne
  i}n_jP_{ji}-n_i\sum\limits_{j\ne i}P_{ij}\ ,
\end{equation}
where $P_{ij}$
contain sums of thermal collisional rates $C_{ij}$ and radiative rates $R_{ij}$, and
$R_{ij}$ are preconditioned in the frame of MALI method \citep{ryhu91}.
The excitation and ionisation of hydrogen by the non-thermal electrons from the
beam are also included into $P_{ij}$ using the non-thermal collisional rates 
$C^{\mathrm{nt}}_{ij}$  following the approach of \citet{fang93}
\begin{alignat}{2}
\label{eqcnt}
\cnt_{1c} & =1.73\times 10^{10}\frac{{\cal{E}}_{\mathrm{en}}}{n_1}\ , & \qquad
\cnt_{12} & =2.94\times 10^{10}\frac{{\cal{E}}_{\mathrm{en}}}{n_1} \ , \notag \\
\cnt_{13} & =5.35\times 10^{9}\frac{{\cal{E}}_{\mathrm{en}}}{n_1}  \ , & \qquad 
\cnt_{14} & =1.91\times 10^{9}\frac{{\cal{E}}_{\mathrm{en}}}{n_1} \ .
\end{alignat}
For transitions from the ground level we thus get
\begin{equation}
P_{1j} = R_{1j} + C_{1j} + C^{\mathrm{nt}}_{1j}\ .
\end{equation}
Non-thermal collisional rates from excited levels as well as three-body recombination rates
are not considered here since \citet{ka04a} and \citet{st07} found them to be negligible compared to thermal ones.

In order not to bias the effects of the non-thermal collisional processes by effects caused 
by macroscopic plasma velocities, we excluded the advection term $\LparcD{(n_i u)}{s}$ from 
Eq.~(\ref{ese}). The omission of the advection term can be justified by small velocities 
($\sim 10$~km~s$^{-1}$) in the Balmer line formation regions \citep{tomnej98} 
attained 
during the first few seconds of the flare atmosphere evolution. On the other hand, the benefit is 
a significant simplification of the radiative transfer code.
The system of ESE (\ref{ese}) is closed by 
charge and particle conservation equations
\begin{equation}\label{conser}
n_{\mathrm e} = n_{\mathrm p} + \varepsilon n_{\mathrm{H}}\,, \qquad \sum_{j=1}^5 n_j + n_{\mathrm p} = n_{\mathrm{H}} \ ,
\end{equation}
where $n_{\mathrm e}$ is the electron density. The contribution of helium to ionisation is neglected. 
Because the electron density is not known in advance, the system of preconditioned ESE is nonlinear due 
to products of atomic level populations (including protons) with $n_\mathrm{e}$ or $n_\mathrm{e}^2$.
Therefore, the ESE and conservation conditions
(\ref{conser}) are linearised with respect to the level populations and electron 
density. The complete system of equations is then solved using the 
Crank-Nicolson algorithm and Newton-Raphson iterative method 
\citep{hein95,kas03}. 

The non-LTE transfer is solved on the shortest time step given by the time step 
splitting technique in the hydrodynamic part, see Section~\ref{hydro}. Resulting electron density 
(ionisation) is then fed
back to the hydrodynamic equations and the TPC.

\section{Results of flare simulations}\label{sec:results}
We computed the atmosphere dynamics and time evolution of the H$\alpha$, H$\beta$, and H$\gamma$ 
line profiles resulting from a time-dependent electron beam heating of an initially hydrostatic VAL C atmosphere. 
\subsection{Flare dynamics}\label{sec:dynamics}
\begin{figure*}
\centerline{\includegraphics[width=16cm]{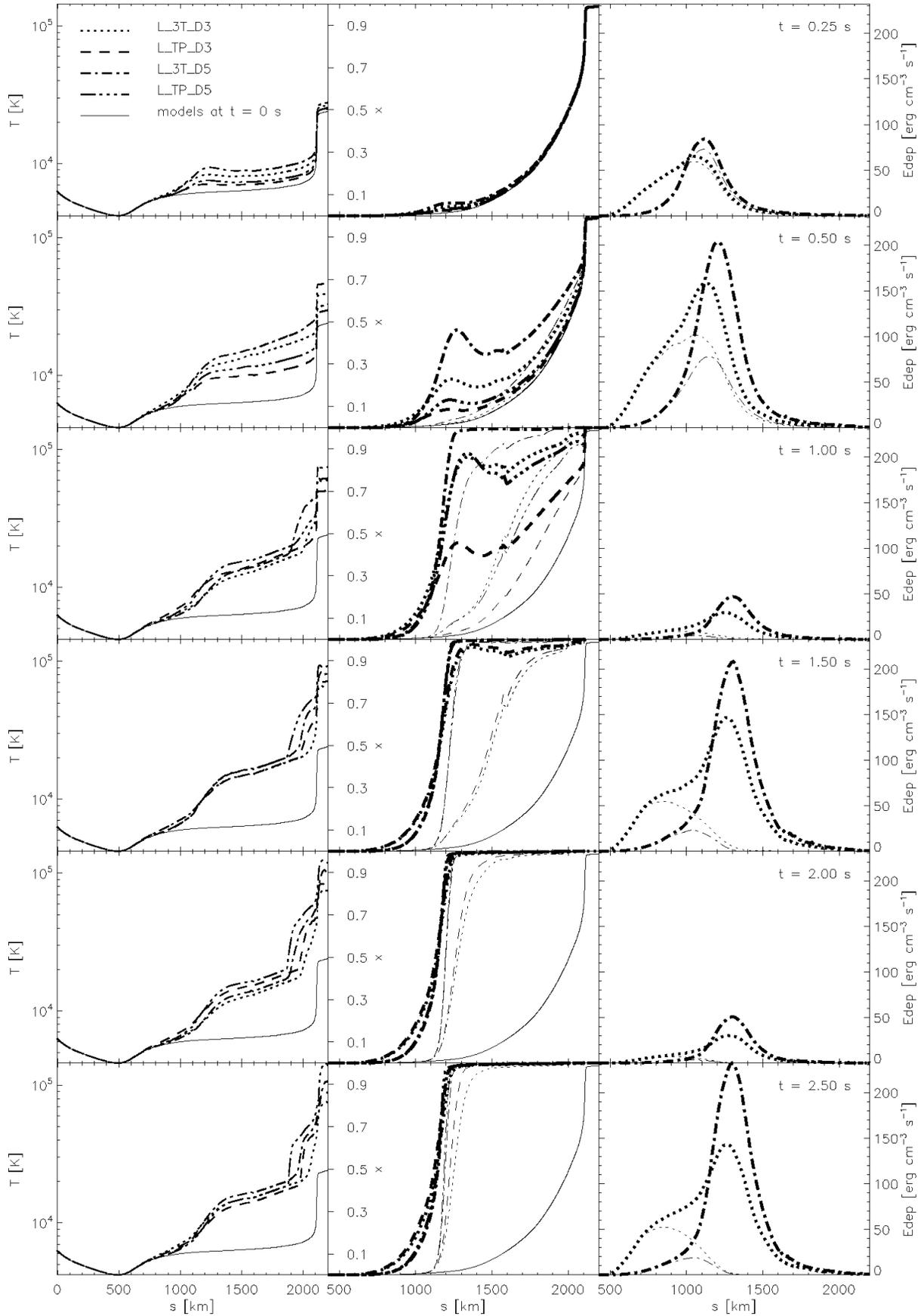}}
\caption{
Temperature, ionisation, and energy deposit corresponding to the low-flux models (L)
and their evolution in time (from top to bottom). Different line styles denote four low-flux models.
Thin solid line shows initial VAL C temperature and ionisation structure.
{\em Left:} Temperature. {\em Middle:} Ionisation. Thick lines denote the models with $\cnt$, thin lines without $\cnt$.
{\em Right:} Total energy deposit ${\cal H}$ (thick lines) and energy deposit to 
hydrogen ${\cal E}_\mathrm{en}$ (thin lines). Only deposits corresponding to the triangular modulation models (3T) are
displayed.
}
\label{fig:L_atms}
\end{figure*}

\begin{figure*}
\centerline{\includegraphics[width=16cm]{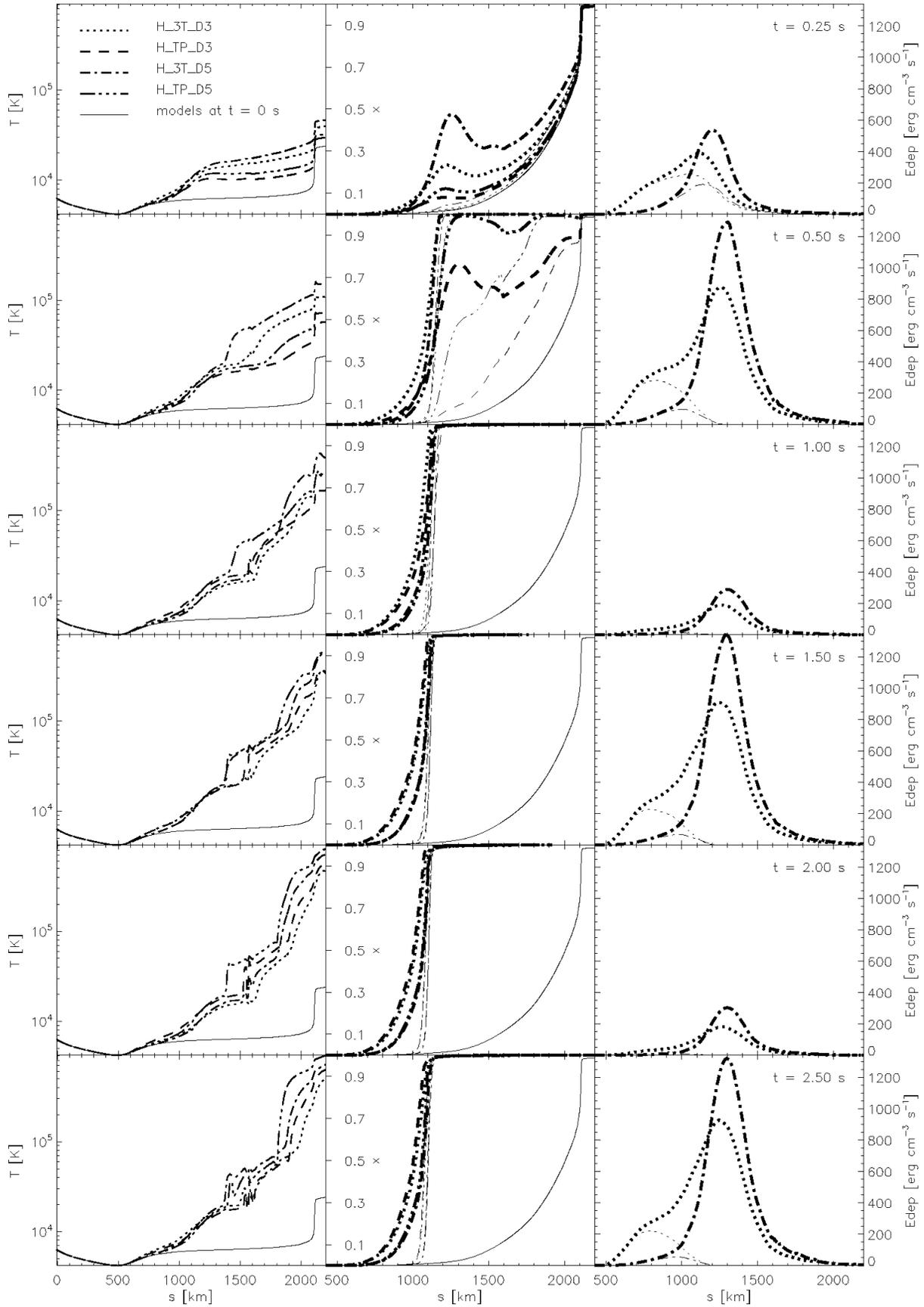}}
\caption{
Temperature, ionisation, and energy deposit corresponding to the high-flux models (H)
and their evolution in time (from top to bottom). The notation is the same as in Fig.~\ref{fig:L_atms}.
}
\label{fig:H_atms}
\end{figure*}

Figures~\ref{fig:L_atms} and \ref{fig:H_atms} show the time evolution of temperature, hydrogen ionisation,
energy deposit $\cal{H}$, and energy deposit to  neutral hydrogen ${\cal E}_\mathrm{en}$
of models specified in Table~\ref{modelparam}.
Shortly after the beam injection at the loop top at $t = 0$~s, the chromosphere is at $t\sim 0.25$~s heated mainly 
at heights between $\sim$ 1000~--~2000~km. The temperature rise is higher for models 
with triangular peak modulation (3T)
than for the trapezoidal one (TP) due to the higher energy flux injected into the atmosphere (compare 
the time evolution of total injected energy in Fig.~\ref{fig:timemod}). 
Since the low-energy electrons are stopped higher 
in the atmosphere, the total energy deposit for steeper beams (higher $\delta$) is larger than for flatter beams
at these heights (compare ${\cal H}$ at $t=0.25$~s in 
Figs.~\ref{fig:L_atms} and \ref{fig:H_atms}) and the temperature rise is most 
significant for  
models L\_3T\_D5 and H\_3T\_D5, see the first panel in 
Figs.~\ref{fig:L_atms} and ~\ref{fig:H_atms}, respectively.
This is generally true for all times during the heating;  the temperature above $s\sim 1000$~km is larger
for larger $\delta$ when comparing models with the same time modulation $g(t)$ and $F_\mathrm{max}$.
On the contrary, at lower heights $s < 1000$~km due to  the larger heating for flatter beams, see 
Figs.~\ref{fig:L_atms} and \ref{fig:H_atms}, the temperature at those atmospheric layers rises more for models with $\delta=3$ 
than with $\delta=5$.

\begin{figure*}
\centerline{
\includegraphics[width=6cm]{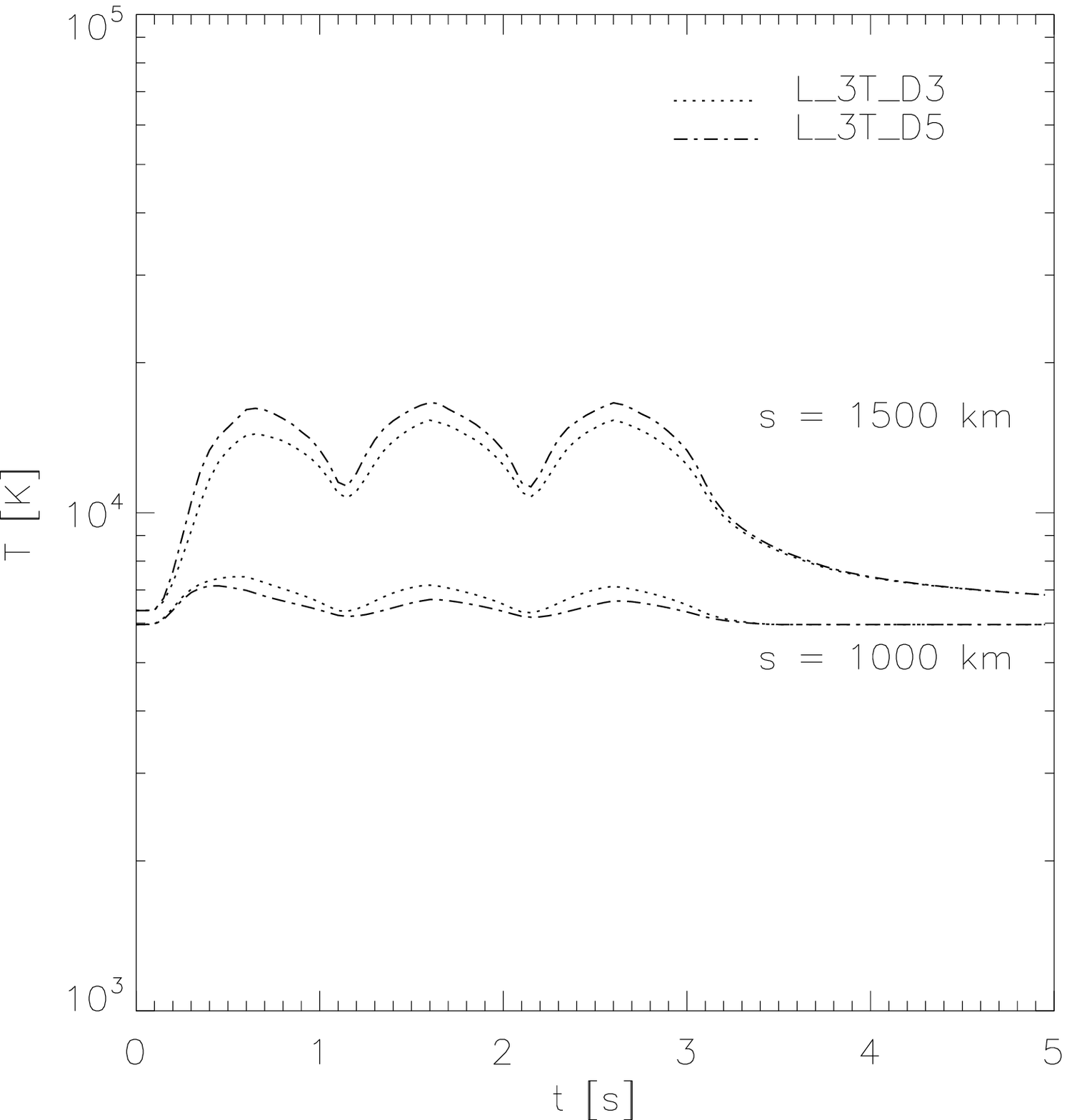}
\hspace{2cm}
\includegraphics[width=6cm]{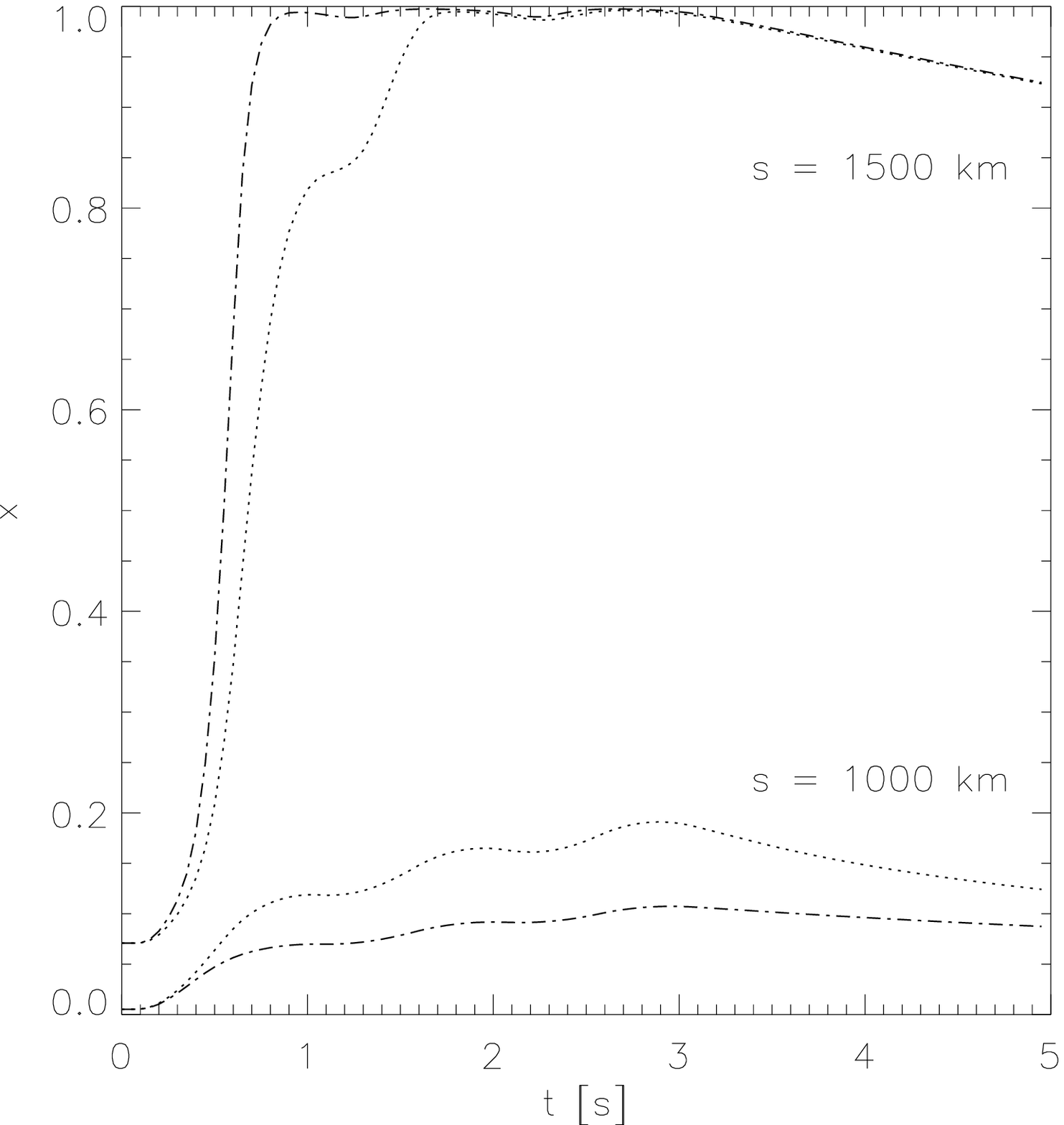}
}
\centerline{
\includegraphics[width=6cm]{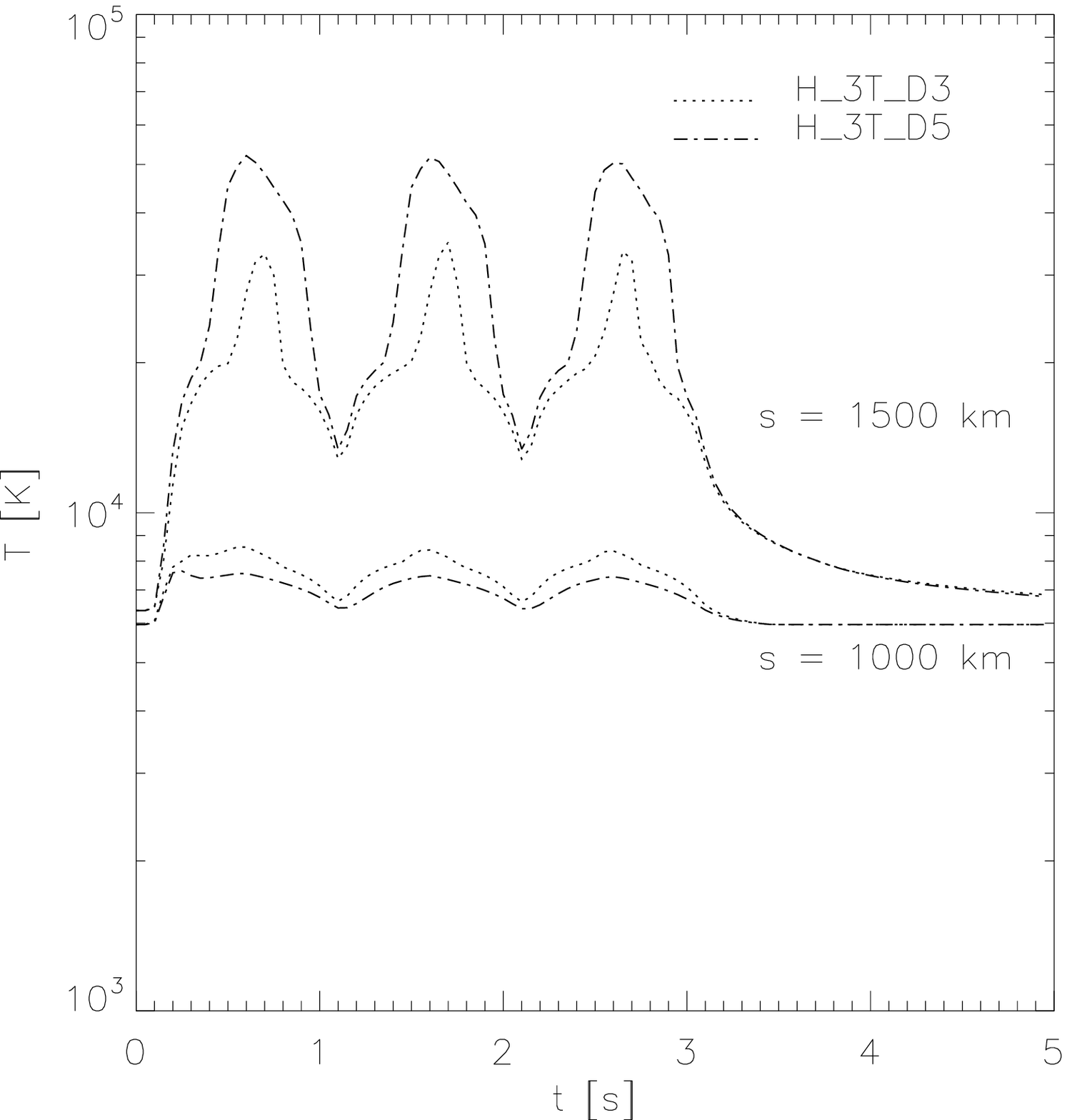}
\hspace{2cm}
\includegraphics[width=6cm]{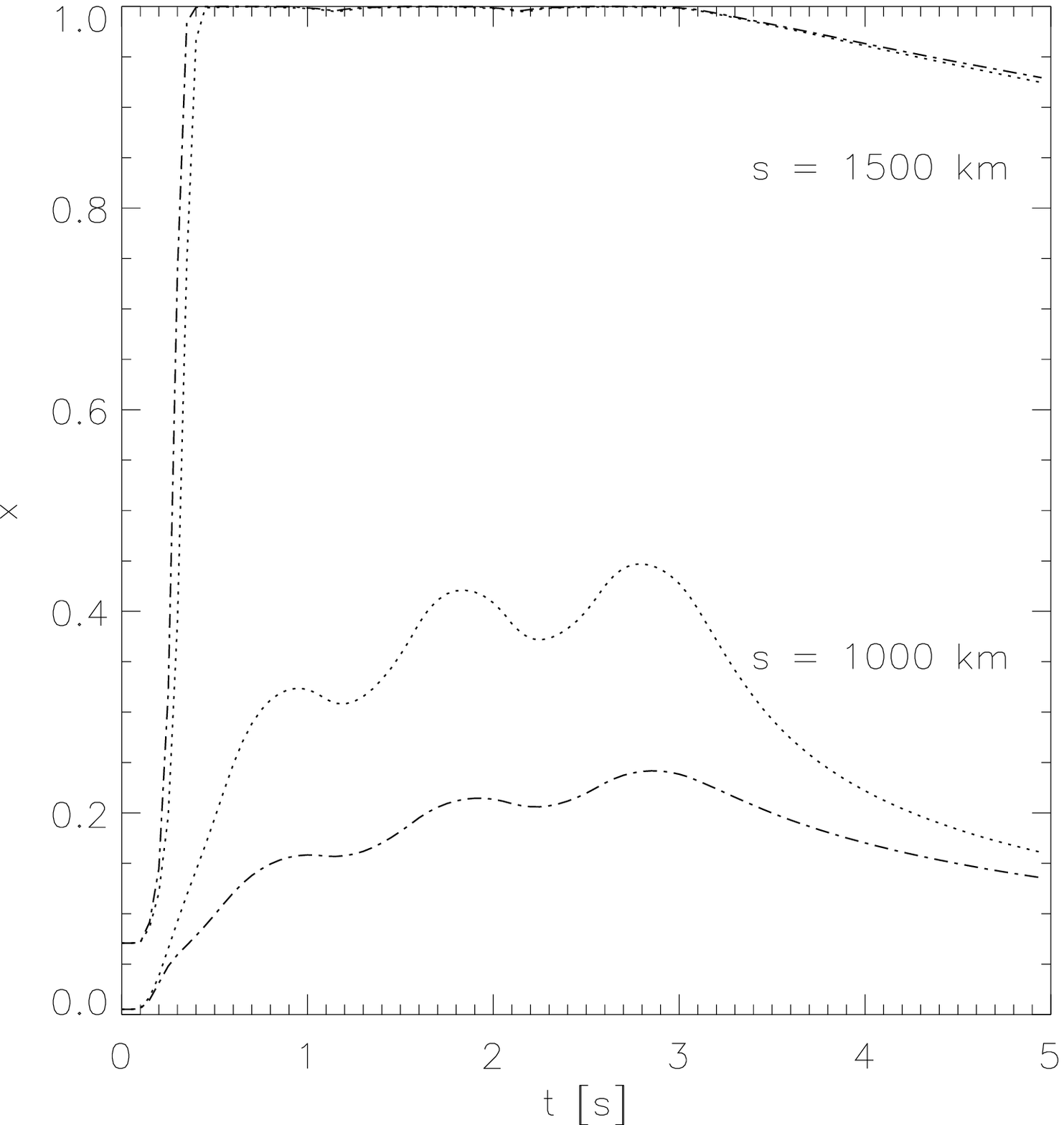}
}
\caption{Time evolution of temperature (left) and ionisation (right) at two heights $s=1000$ and 1500~km.
Top: low-flux models L\_3T\_D3 (dotted) and L\_3T\_D5 (dash-dotted). Bottom: high-flux models H\_3T\_D3 
(dotted) and H\_3T\_D5 (dash-dotted). $\cnt$ were included.}
\label{fig:temp_x_time}
\end{figure*}

In the low-flux models the heating leads to a gradual increase of temperature above $\sim 1000$~km. 
The steep rise of temperature from chromospheric to coronal values is shifted by about 200~km from 
the preflare height to $s\sim 1900$~km, see Fig.~\ref{fig:L_atms}. 
Heating by a higher flux results in a secondary region of a steep temperature rise at $s\sim 1400$~km 
which is formed at $t\sim 1$~s, see Fig.~\ref{fig:H_atms}. 
Due to the locally efficient radiative losses, the temperature structure 
in all models up to $s\sim 2000$~km
follows the time modulation of the beam flux; i.e. it rises and drops -- compare the temperature 
structure e.g. at $t=1.5$~s and $t=2$~s for 3T models and TP models in 
 Figs.~\ref{fig:L_atms} and \ref{fig:H_atms} and the time evolution of temperature at two selected heights 
in Fig.~\ref{fig:temp_x_time}.

Similarly to the temperature evolution, ionisation increases above $s\sim 1000$~km shortly after the beam 
injection, see the middle panel in Figs.~\ref{fig:L_atms} and \ref{fig:H_atms}. Again,
due to beam flux modulation and the dependence of the energy deposit on $\delta$, the increase of ionisation at those layers 
is most significant for 3T models with $\delta=5$. As the heating continues, the layers 
above $\sim 1200$~km become completely ionised (at $t\sim 2$~s and
$t\sim 1$~s for low and high flux models, respectively -- see thick lines in 
Figs.~\ref{fig:L_atms} and \ref{fig:H_atms})
and ionisation does not change during further heating -- see also Fig.~\ref{fig:temp_x_time}.

On the contrary, lower heights, below $s\sim 1000$~km, exhibit most significant increase 
of ionisation for models with $\delta=3$.
The ionisation at these layers reacts to the beam flux time modulation 
more for the flatter beams -- see Fig.~\ref{fig:temp_x_time} which also demonstrates that the
relaxation of ionisation to preheating values lags behind the time evolution of temperature.

This effect previously shown by \citet{hein91} and \citet{heinkar92} is due to 
time evolution of the ratio of the number of recombinations to photoionisations.
Detailed behaviour of photoionisations followed by photorecombinations
was discussed by \citet[][Section~4]{abb99}.

\subsection{Influence of non-thermal collisional rates}
\label{sec:cnteffect}
To evaluate the influence of the non-thermal collisional rates, two separate runs with and 
without $\cnt$ (Eq.~\ref{eqcnt}) were made for each model in Table~\ref{modelparam}.

Taking into account $\cnt$ leads only to marginal changes of temperature and density structure of the atmosphere
(up to 17\%). 
On the contrary,  hydrogen ionisation and emission in Balmer lines may significantly differ in models with and without $\cnt$. 
Generally, the effect of $\cnt$ is stronger for models with larger $F_\mathrm{max}$  or lower $\delta$. 
In the low-flux models, $\cnt$ significantly modify the time evolution of ionisation structure. They
lead to a faster complete ionisation of layers above $s\sim 1200$~km and cause an increase of ionisation in the layers below, 
compare thin and thick lines in Fig.~\ref{fig:L_atms}. 
The influence of $\cnt$ in the high-flux models is localised mainly 
in the layers below $s\sim 1000$~km where the flatter beams increase the ionisation. The upper parts of the atmosphere
are affected by $\cnt$ only temporarily, till $t\sim 0.5$~s, when they contribute to the fast ionisation of those layers --
see Fig.~\ref{fig:H_atms}.

\begin{figure*}
\begin{center}
\includegraphics[width=5.5cm]{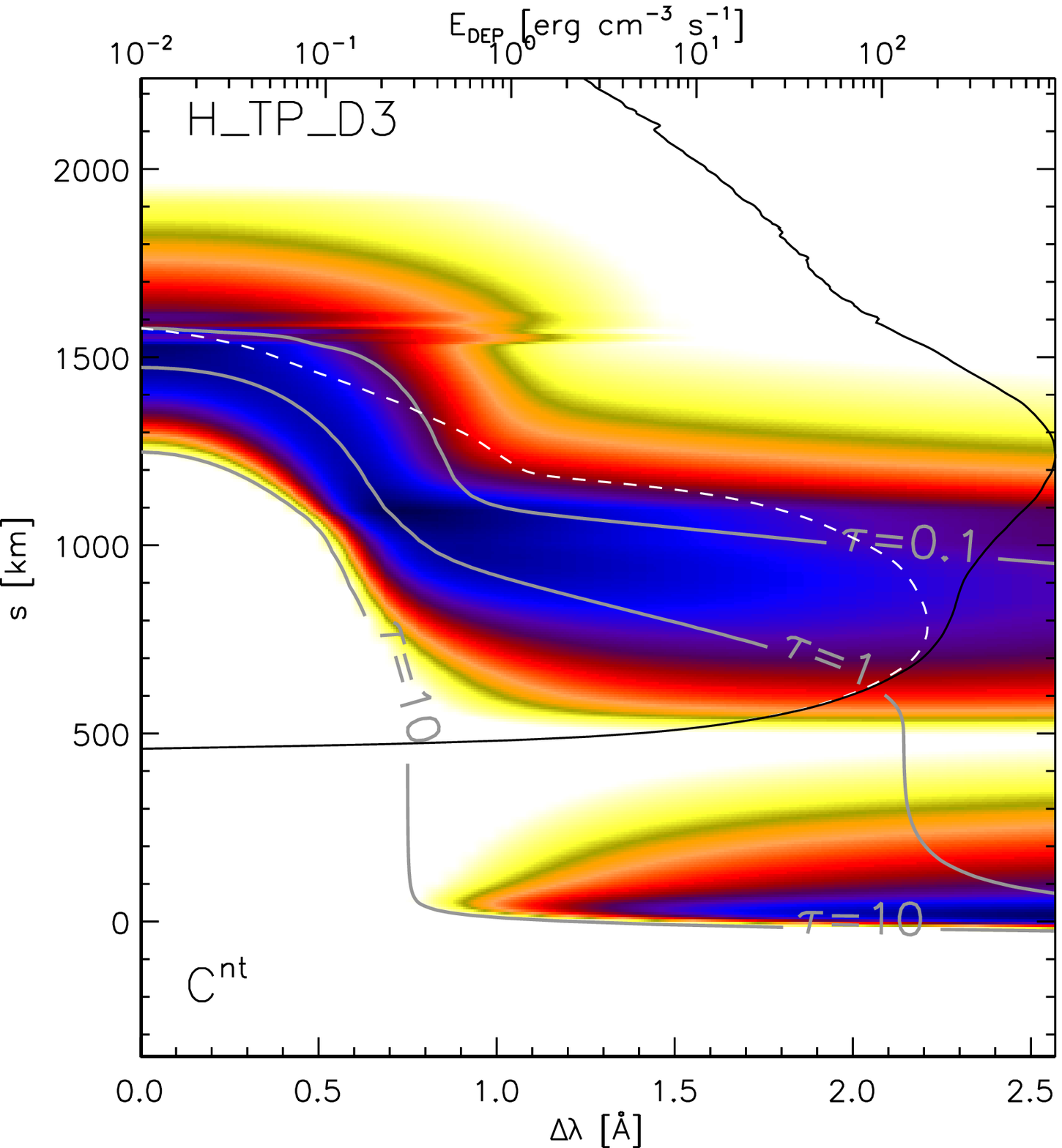}
\includegraphics[width=5.5cm]{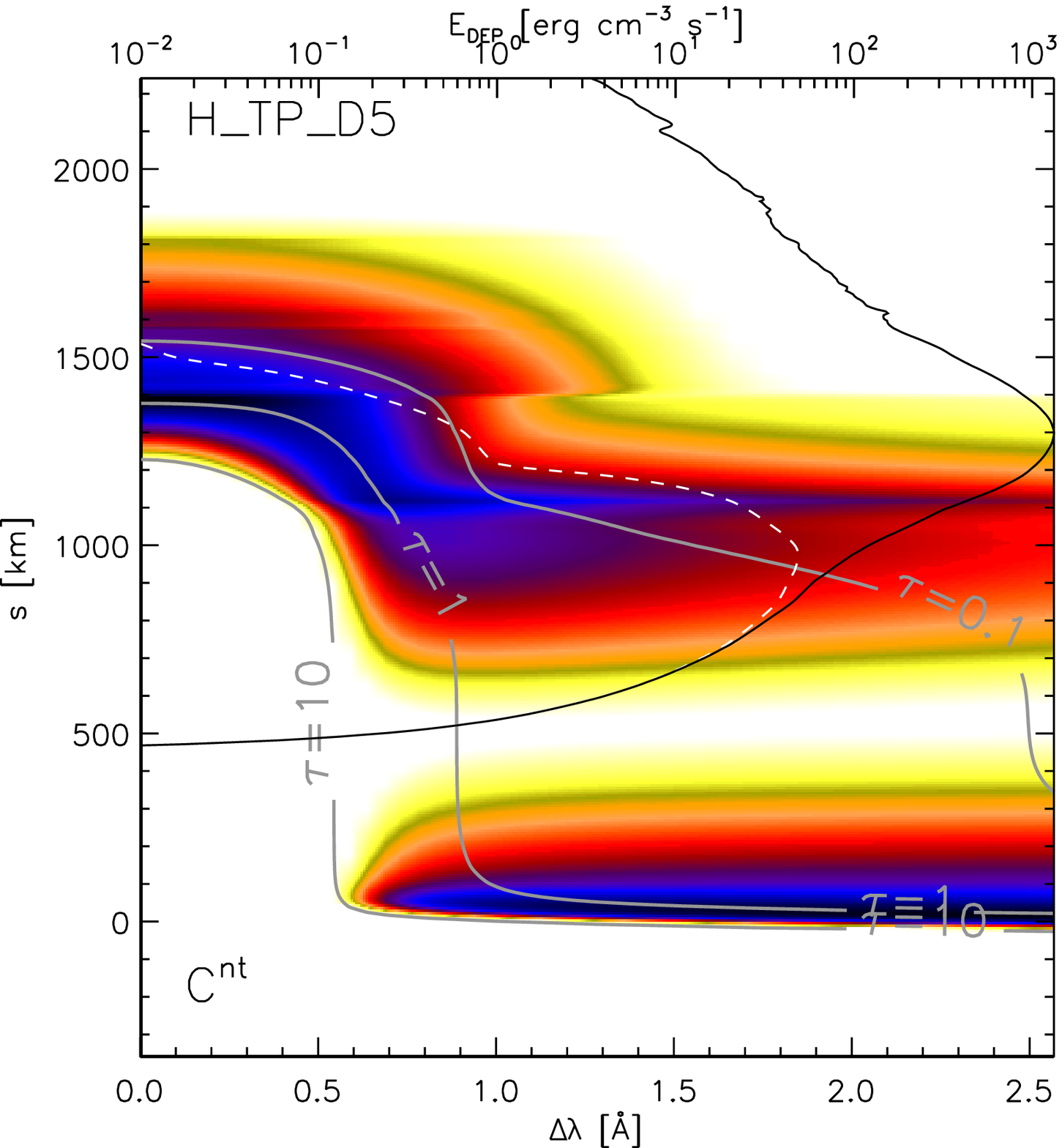}
\includegraphics[width=5.5cm]{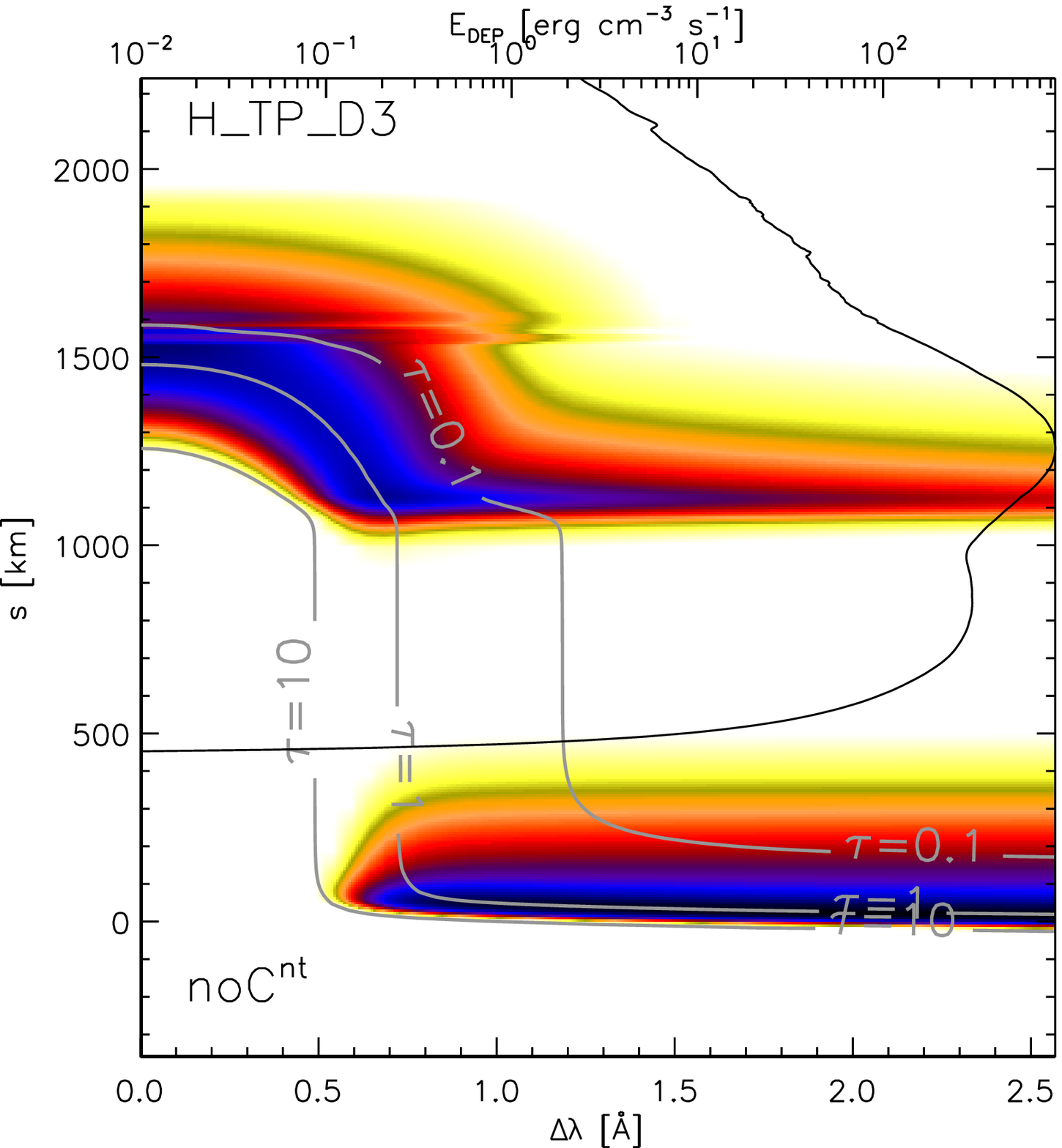}

\includegraphics[width=5.5cm]{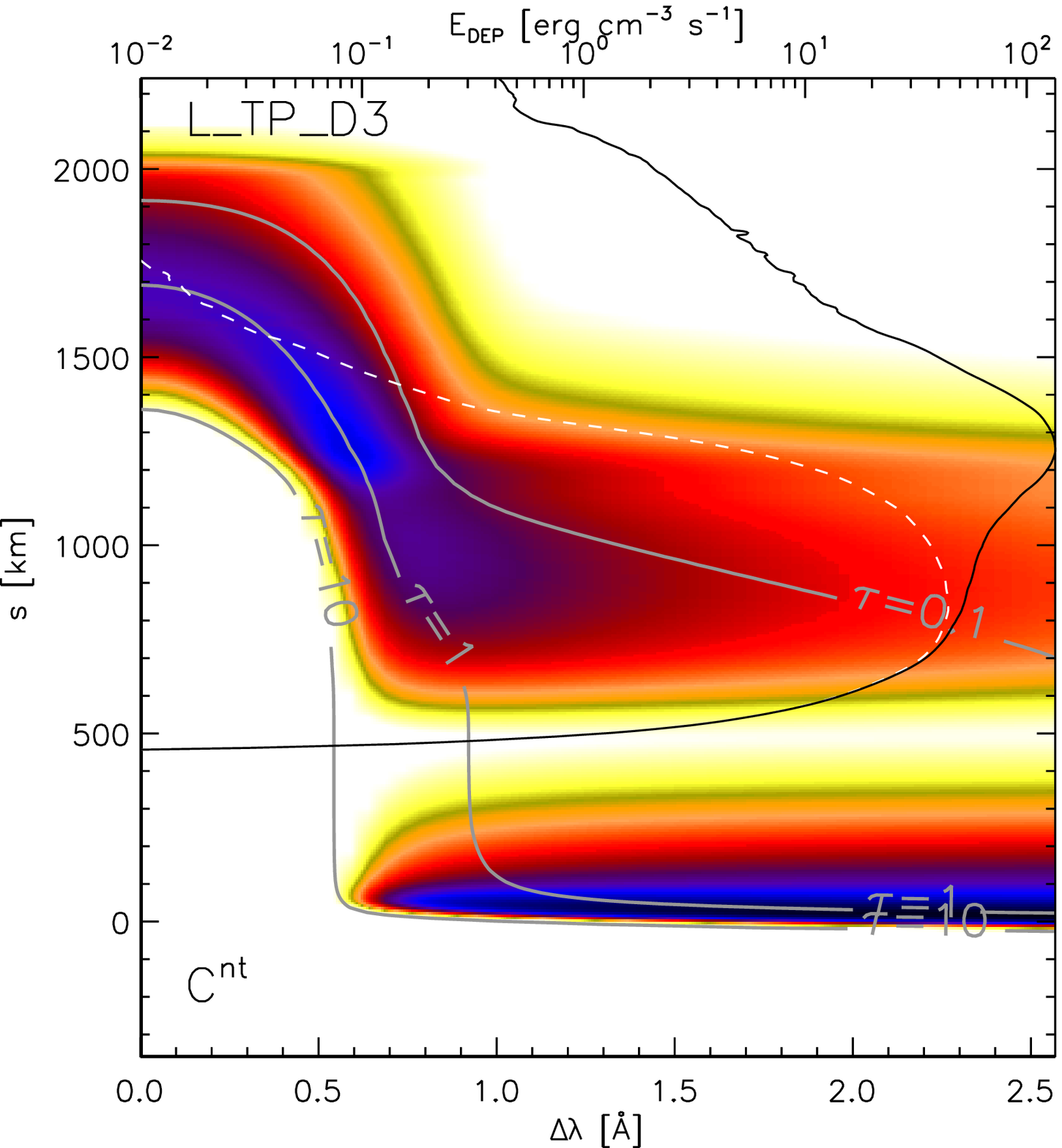}
\includegraphics[width=5.5cm]{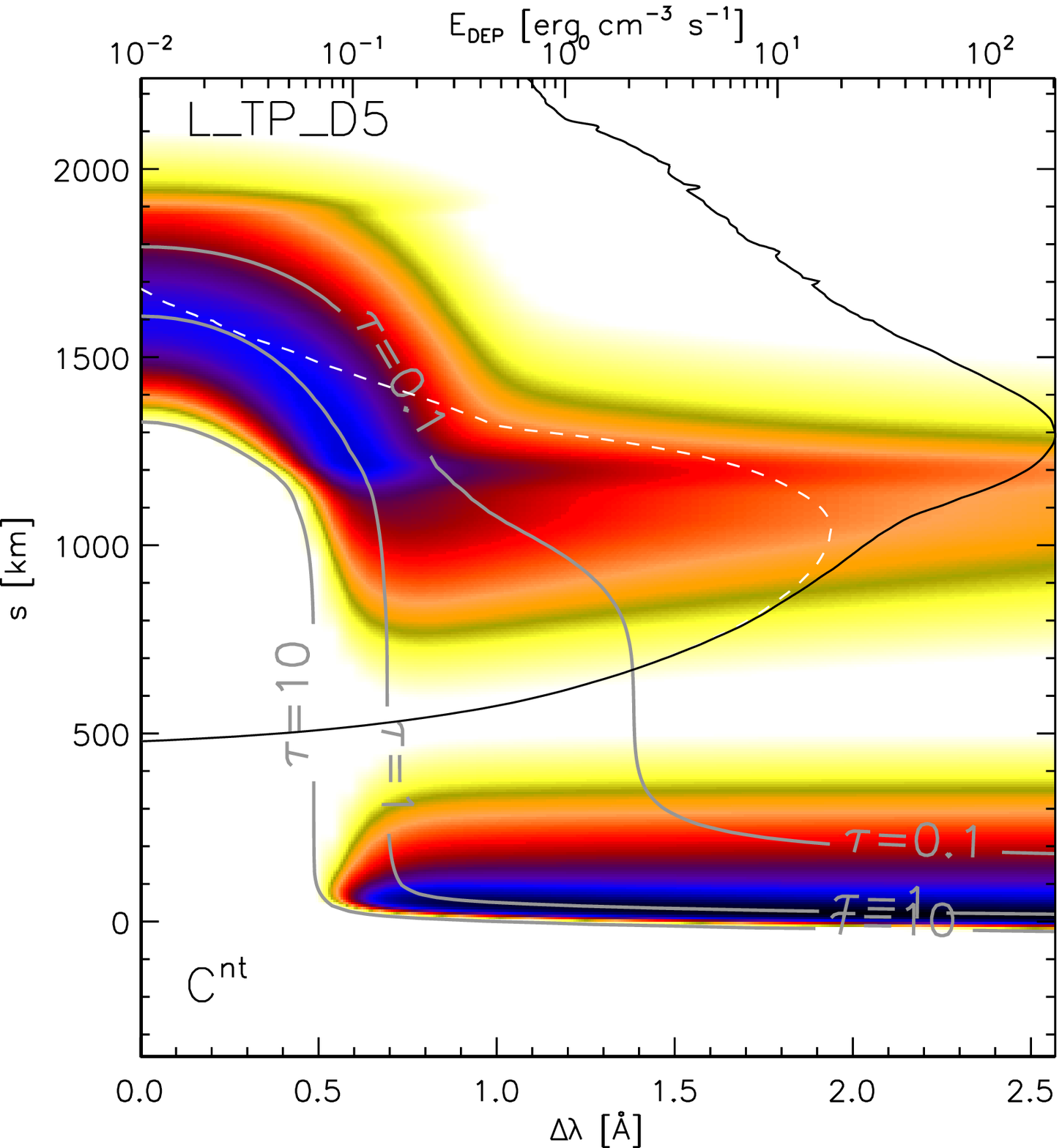}
\includegraphics[width=5.5cm]{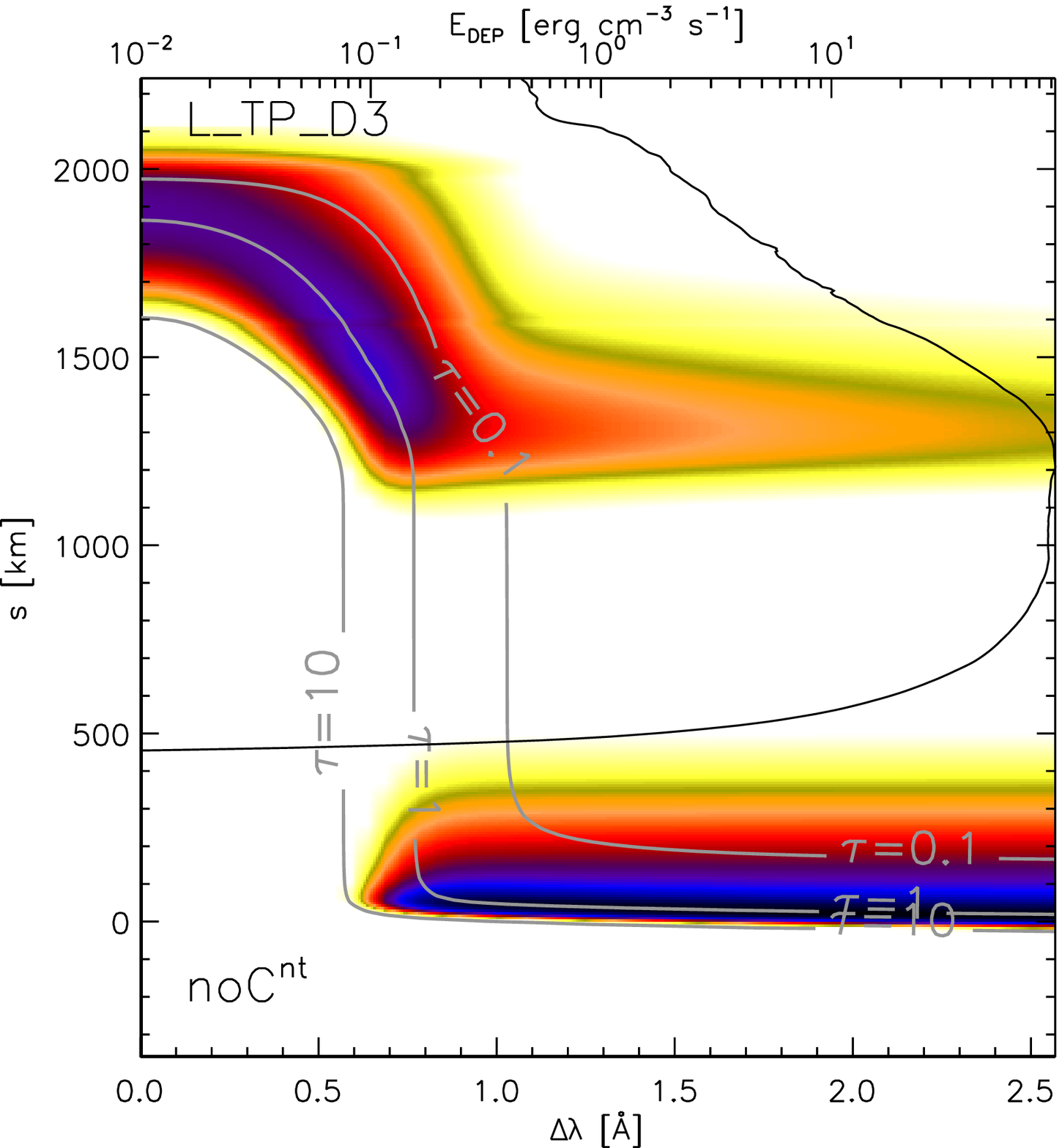}

\medskip
\includegraphics[width=6cm]{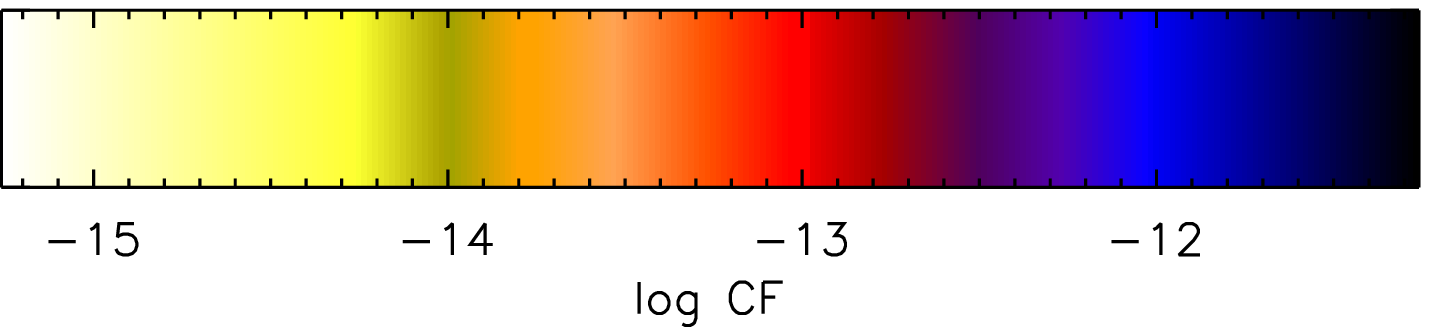}
\end{center}
\caption{H$\alpha$ contribution functions for high (top) and low (bottom) flux models 
for trapezoidal beam flux modulation.
Black solid  curves indicate the total energy deposit, white dashed curves denote the energy deposit on hydrogen.
Gray lines display contours of optical depth $\tau=$~0.1, 1, 10.
{\em Top left:} Model H\_TP\_D3 with $\cnt$. {\em Top centre:} Model H\_TP\_D5 with $\cnt$. {\em Top right:} 
Model H\_TP\_D3 without $\cnt$.
{\em Bottom left:} Model L\_TP\_D3 with $\cnt$. {\em Bottom centre:} Model L\_TP\_D5 with $\cnt$. {\em Bottom right:} 
Model L\_TP\_D3 without $\cnt$. Colour scale denotes $\log C\!F$. 
All plots correspond to $t=1.8$~s -- maxima of line intensities, see Fig.~\ref{fig:lineint_time}.
}
\label{fig:CF}
\end{figure*}

Since $\cnt$ are directly proportional to the energy deposit on hydrogen  $\cal{E}_{\mathrm{en}}$ (Eq.~\ref{eqcnt}),
their influence is strongly linked to the $\cal{E}_{\mathrm{en}}$ as a function of height. Consequently, $\cnt$ 
affect Balmer line intensities according to their formation heights. That can be understood in terms of 
the contribution function $C\!F$ to the outgoing intensity $I_\lambda$ 
\begin{equation}\label{eq:cf}
I_\lambda=\int_{s_\mathrm{min}}^{s_\mathrm{max}} C\!F_\lambda \dd s = 
\int_{s_\mathrm{min}}^{s_\mathrm{max}} \eta_\lambda(s) e^{-\tau_\lambda(s)} \dd s \ ,
\end{equation}
where $\eta_\lambda$ is the emissivity and $\tau_\lambda$ is the optical depth.
Figure~\ref{fig:CF} demonstrates the effect of $\cnt$ on the H$\alpha$ line for a low and high-flux model.
In high-flux models, $\cnt$ affect mainly the line wings. A new wing formation region appears at heights of maximum 
of energy deposit on hydrogen, lower $\delta$ producing a stronger contribution to $C\!F$. 
In the low-flux models, $\cnt$ influence also the line centre due to change of ionisation of upper layers
where e.g. the H$\alpha$ line centre is formed. The optical depth in the line centre at these heights is decreased
and the H$\alpha$ line centre formation region is shifted deeper. Similarly as in the high-flux models, 
a new region of wing formation region again  appears at the height of the ${\cal E}_\mathrm{en}$ maximum, however the dominant
part of the wing emission still comes from the photospheric layers, i.e. from $s\sim 100$~km -- see Fig.~\ref{fig:CF}.
\subsection{Time variation of line intensities}
\begin{figure*}
\begin{center}
\includegraphics[width=4.9cm]{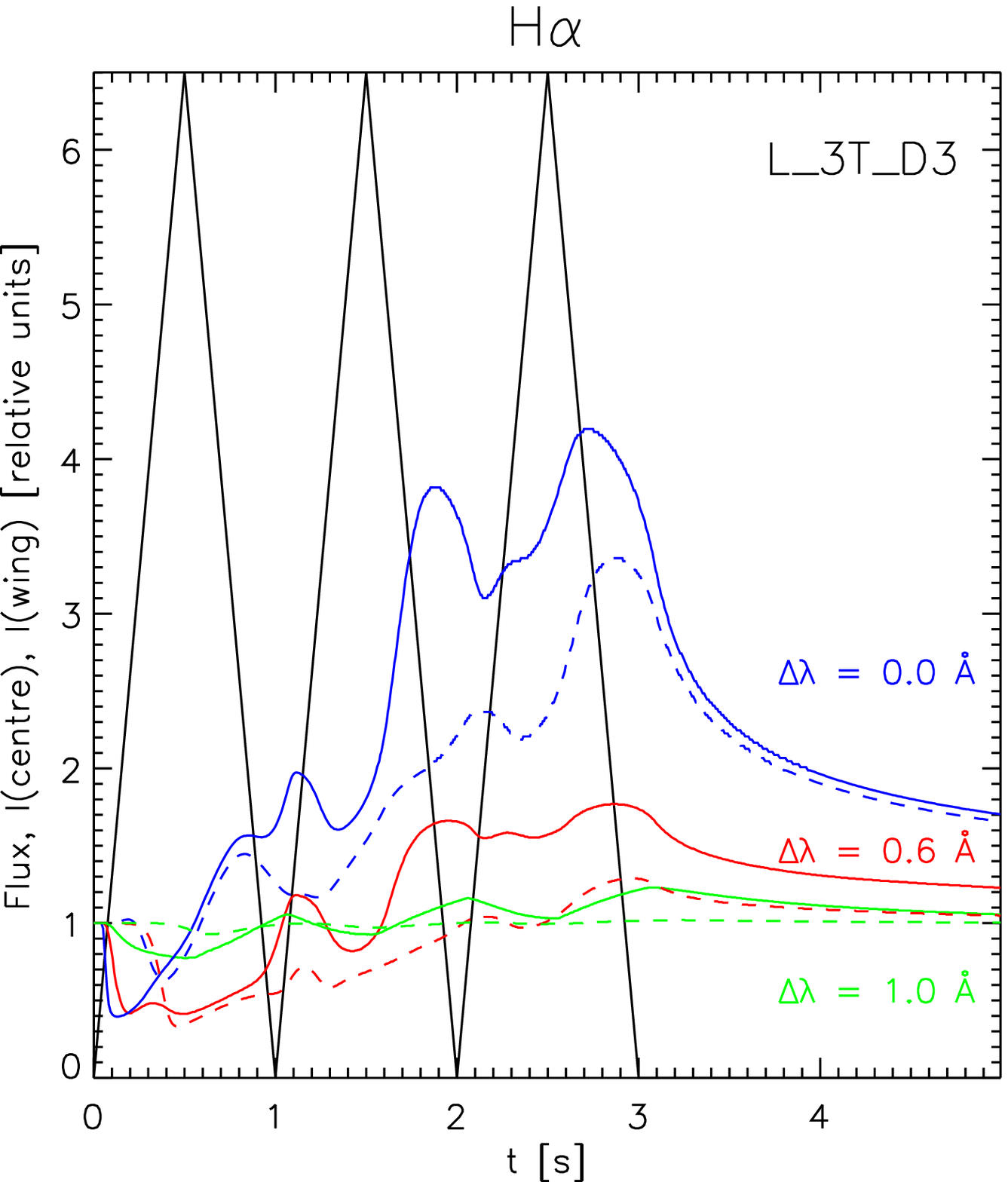}
\includegraphics[width=4.9cm]{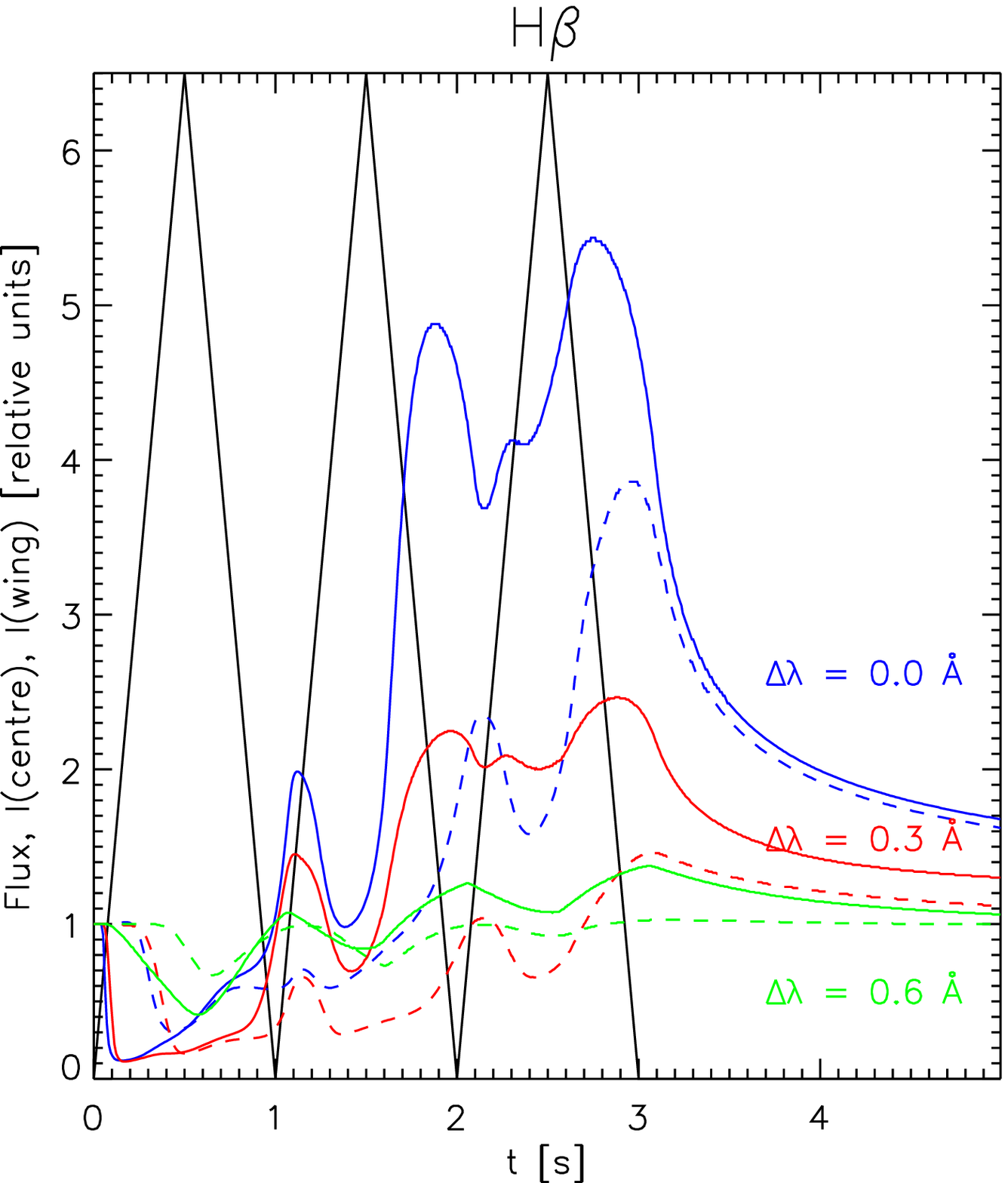}
\includegraphics[width=4.9cm]{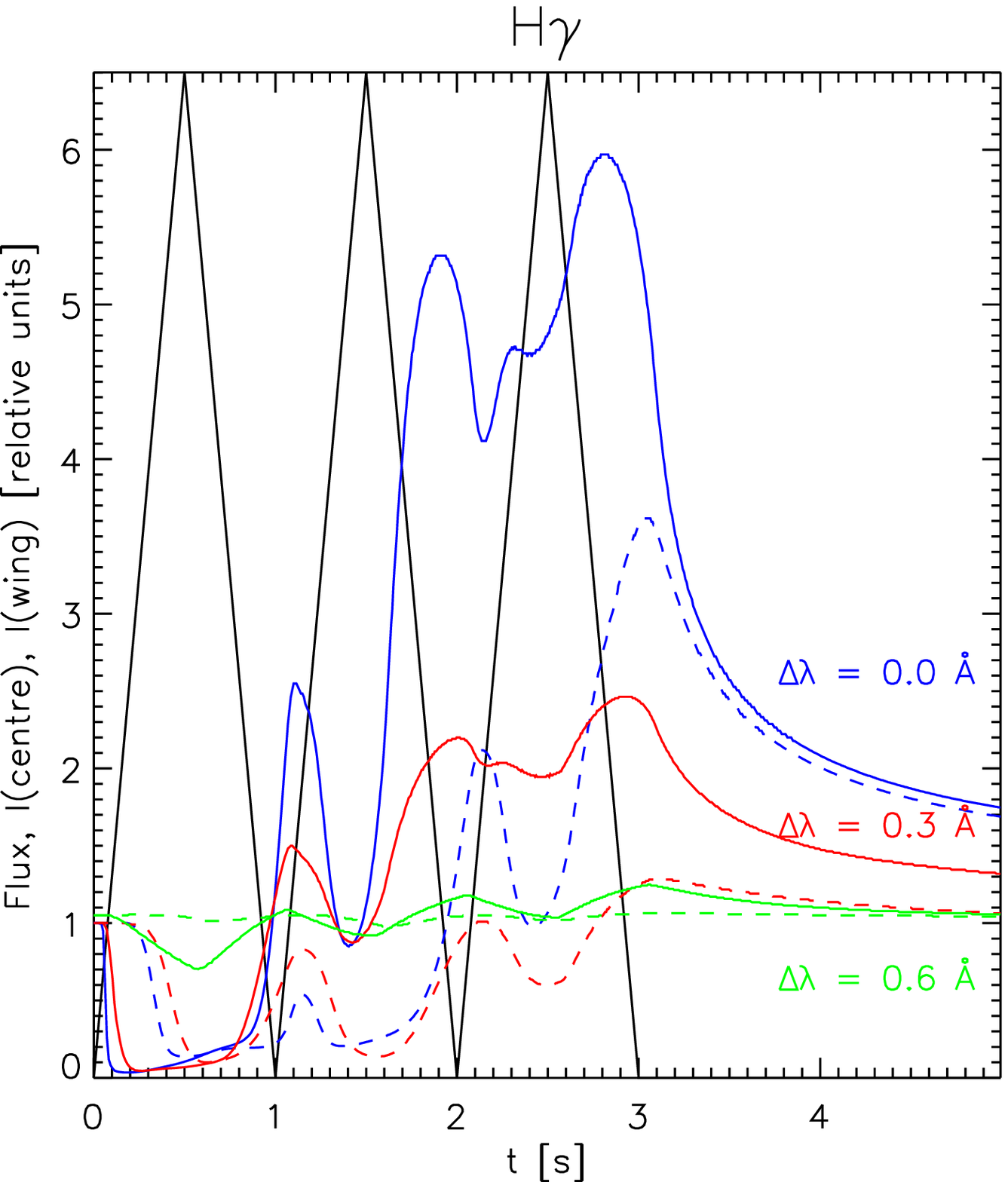}

\includegraphics[width=4.9cm]{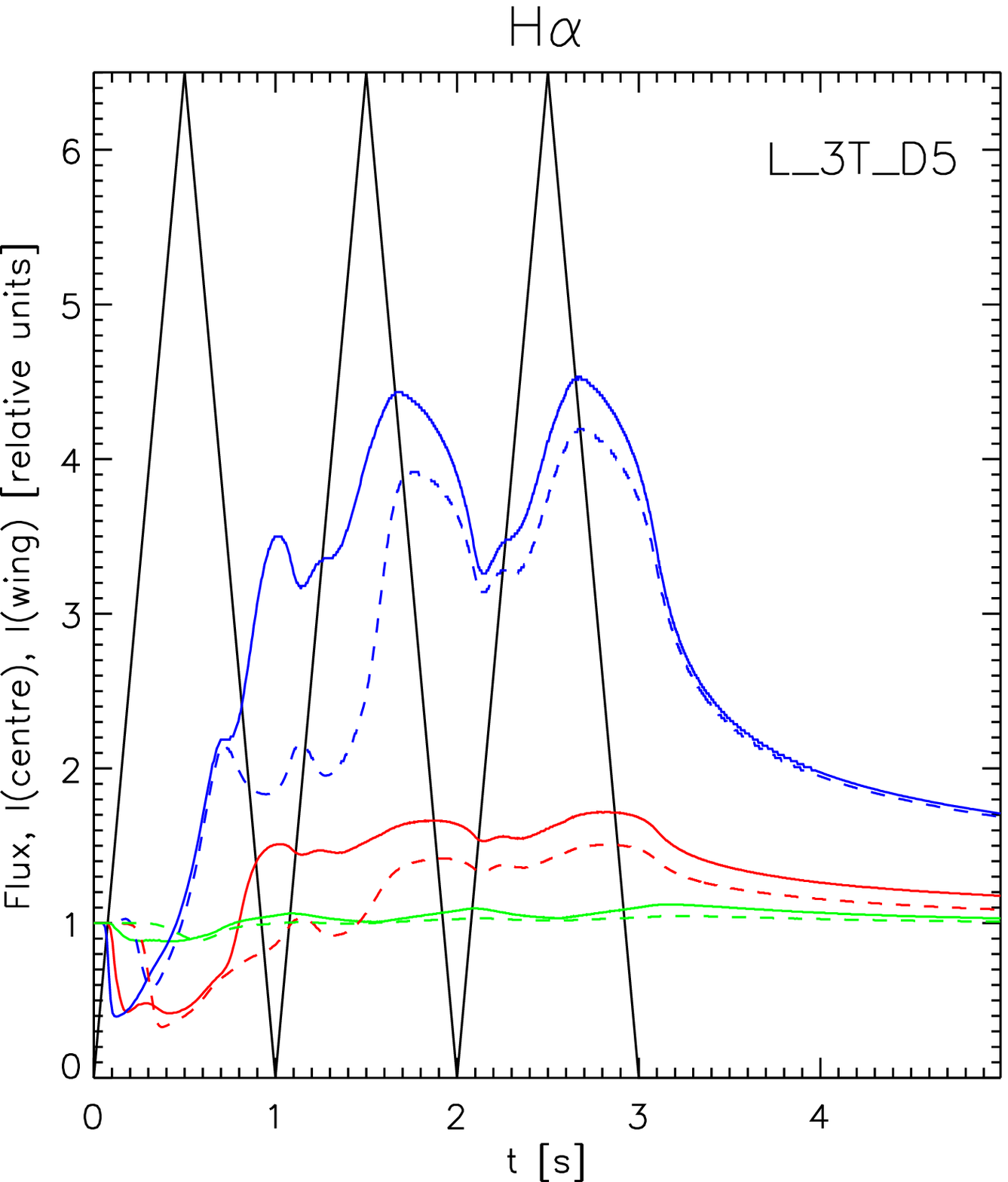}
\includegraphics[width=4.9cm]{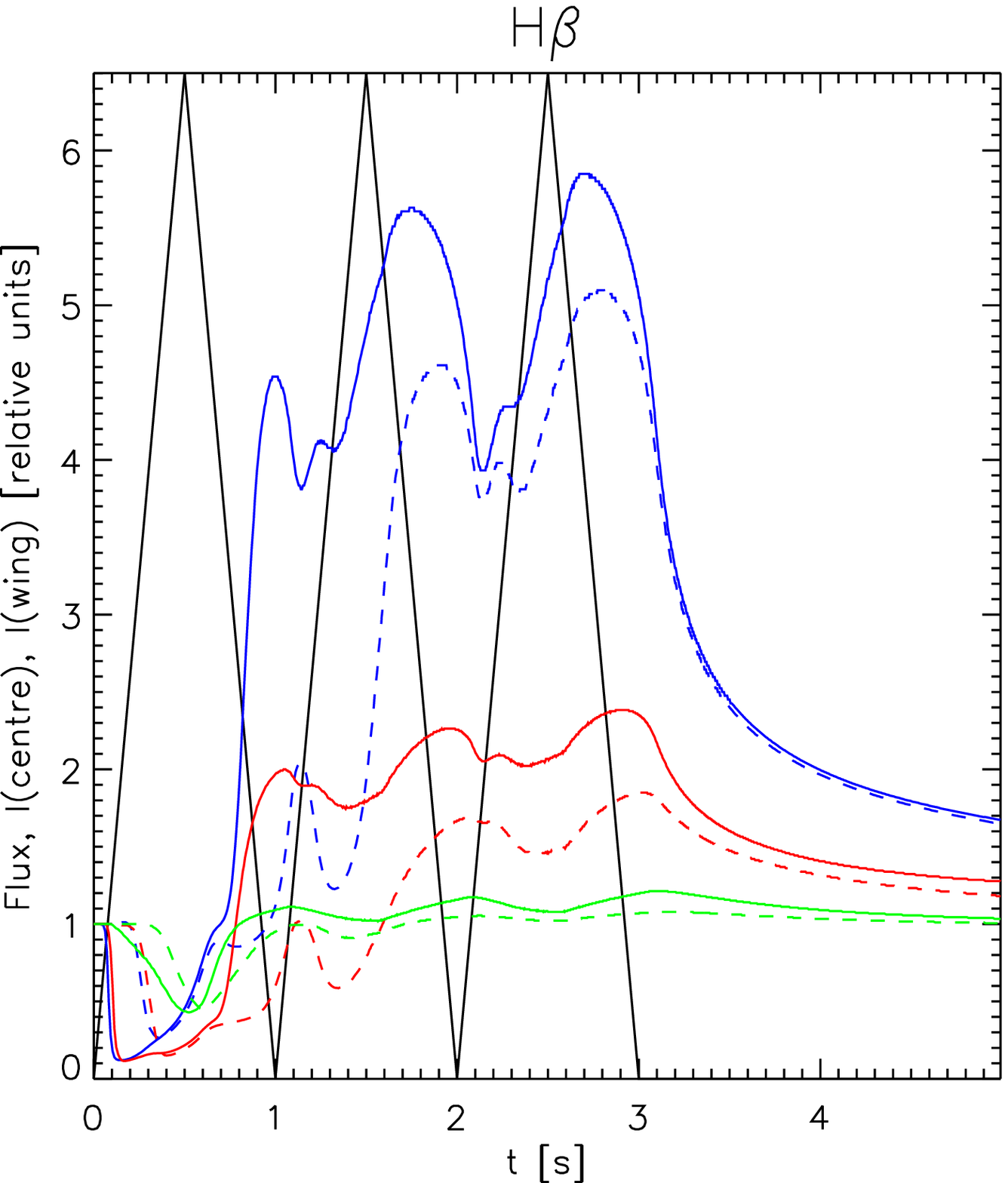}
\includegraphics[width=4.9cm]{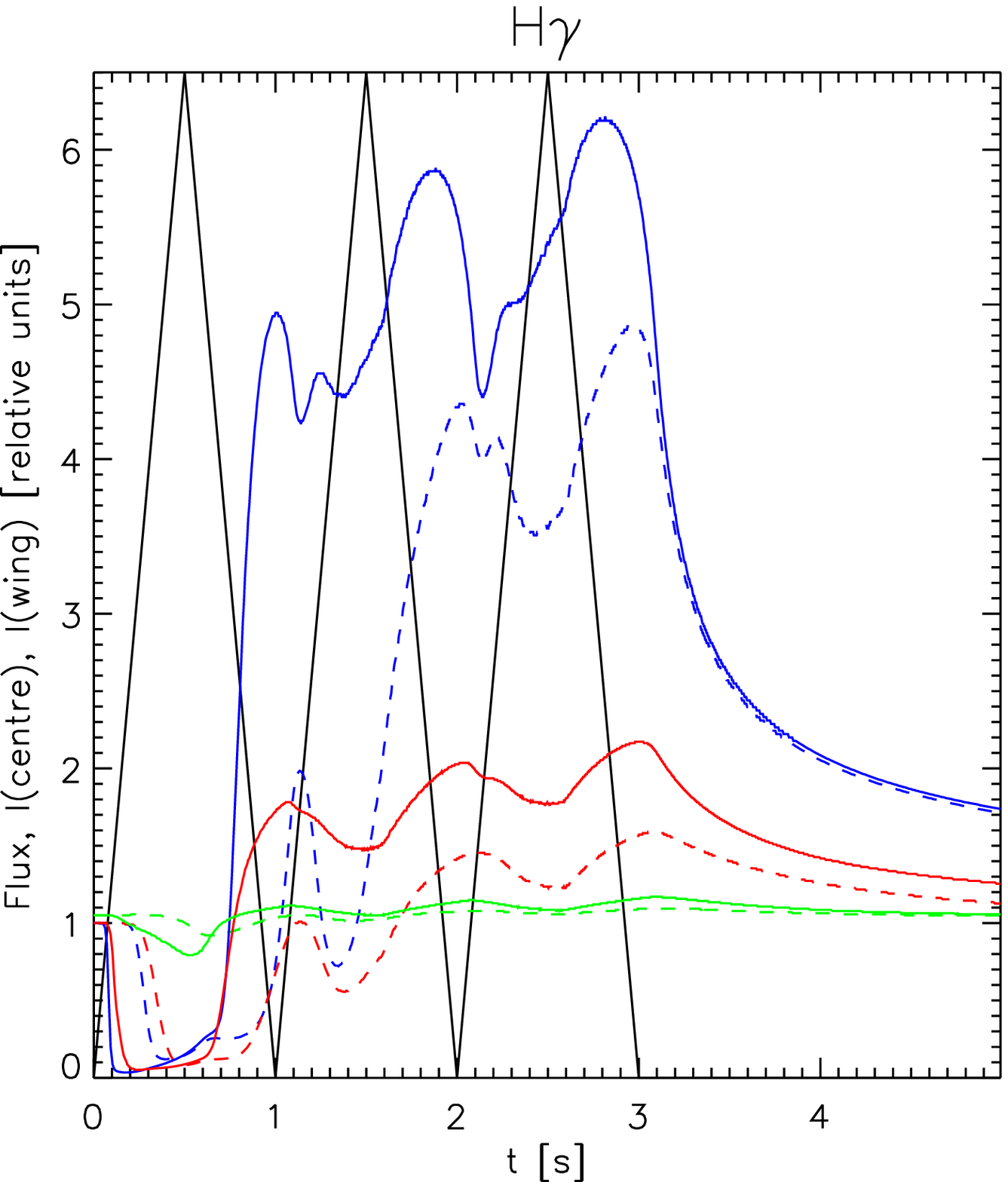}

\includegraphics[width=4.9cm]{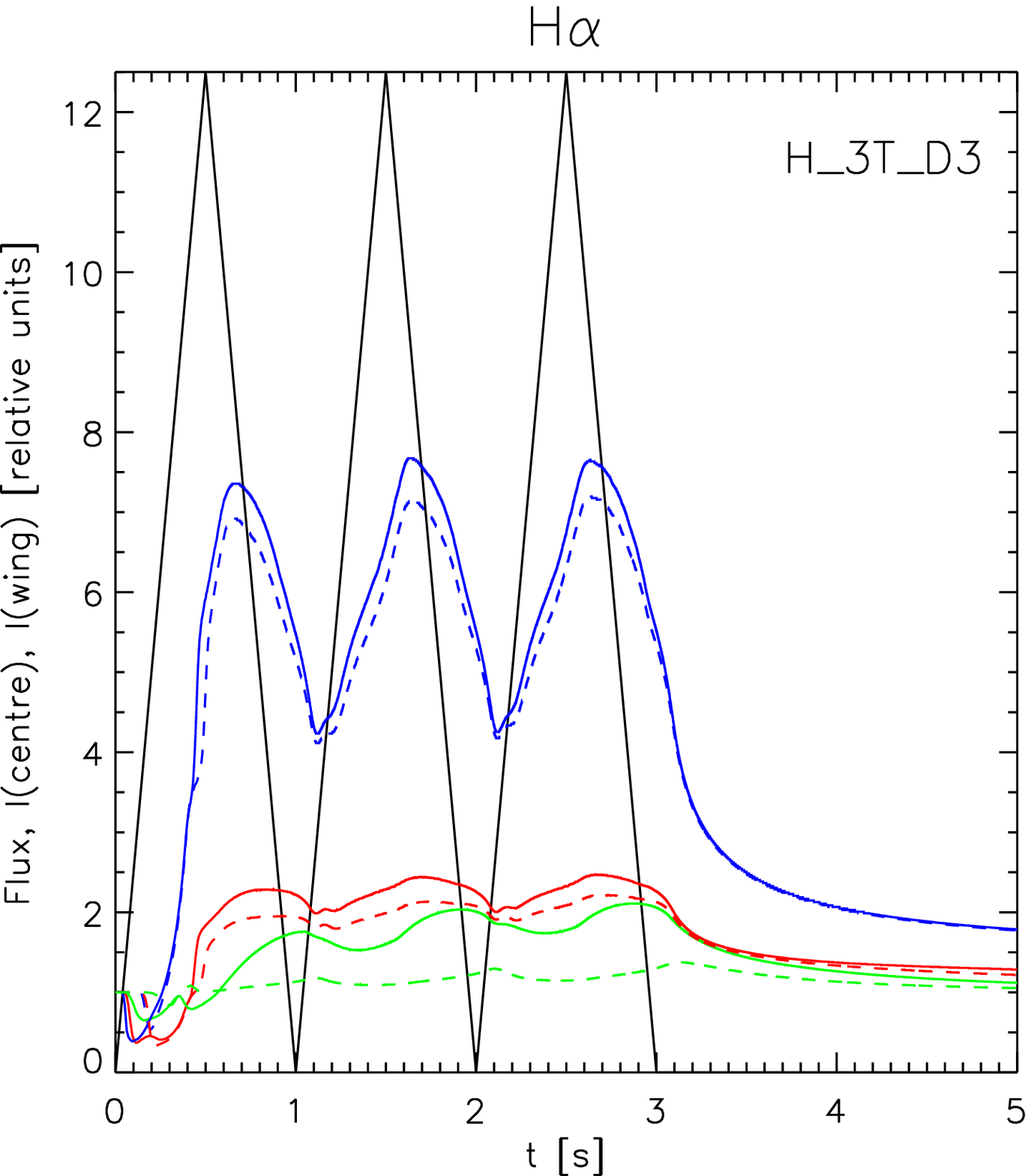}
\includegraphics[width=4.9cm]{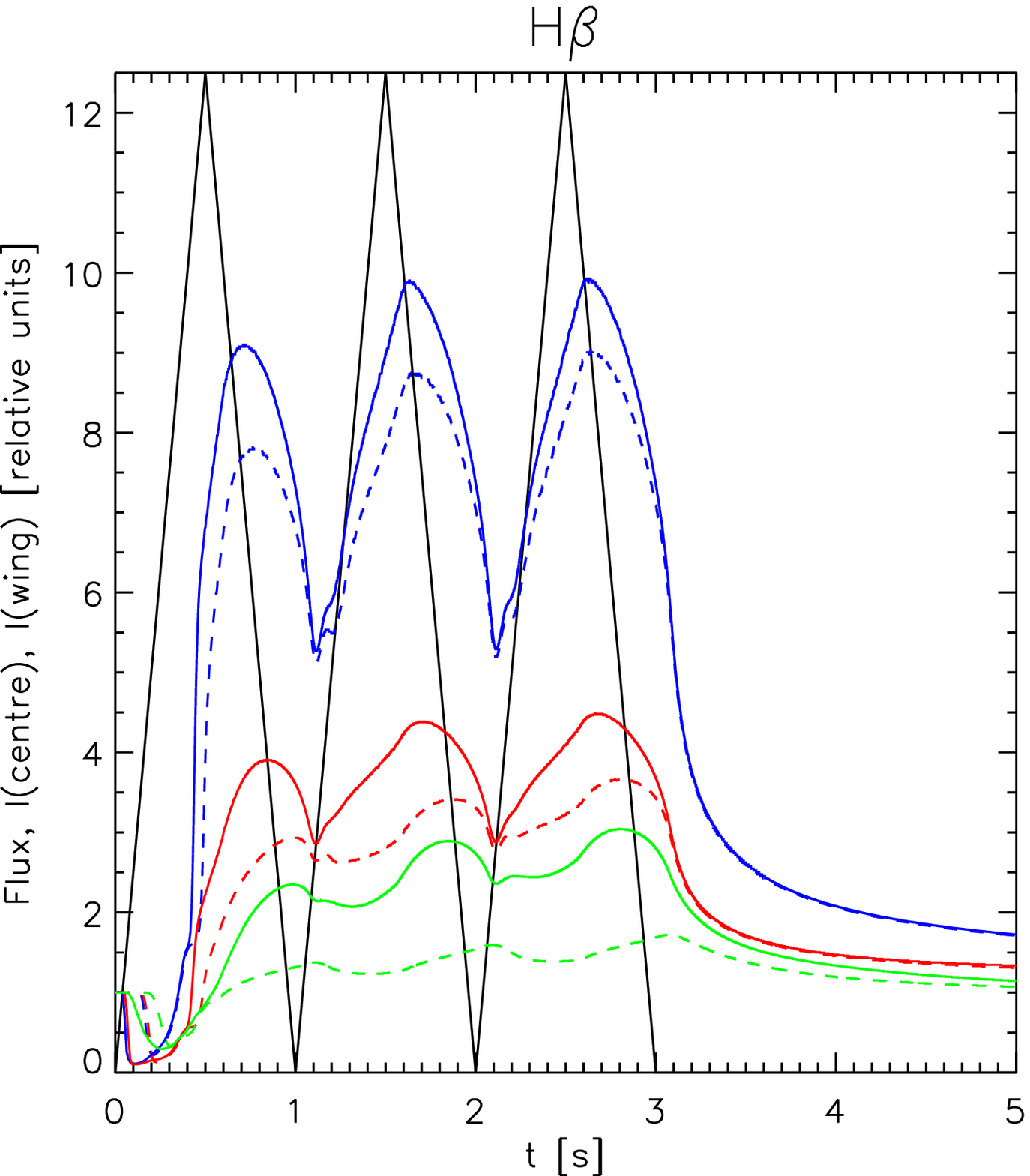}
\includegraphics[width=4.9cm]{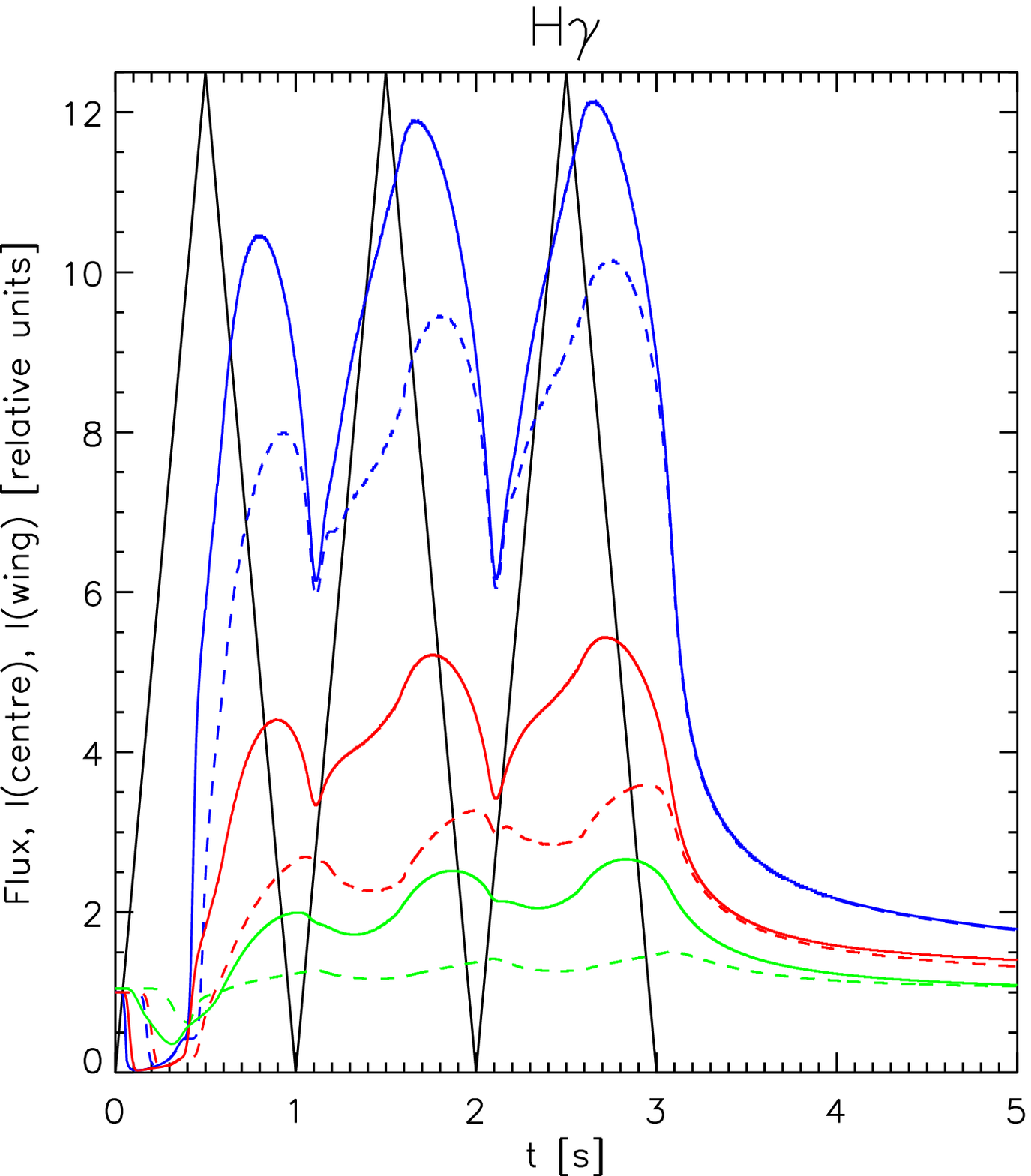}

\includegraphics[width=4.9cm]{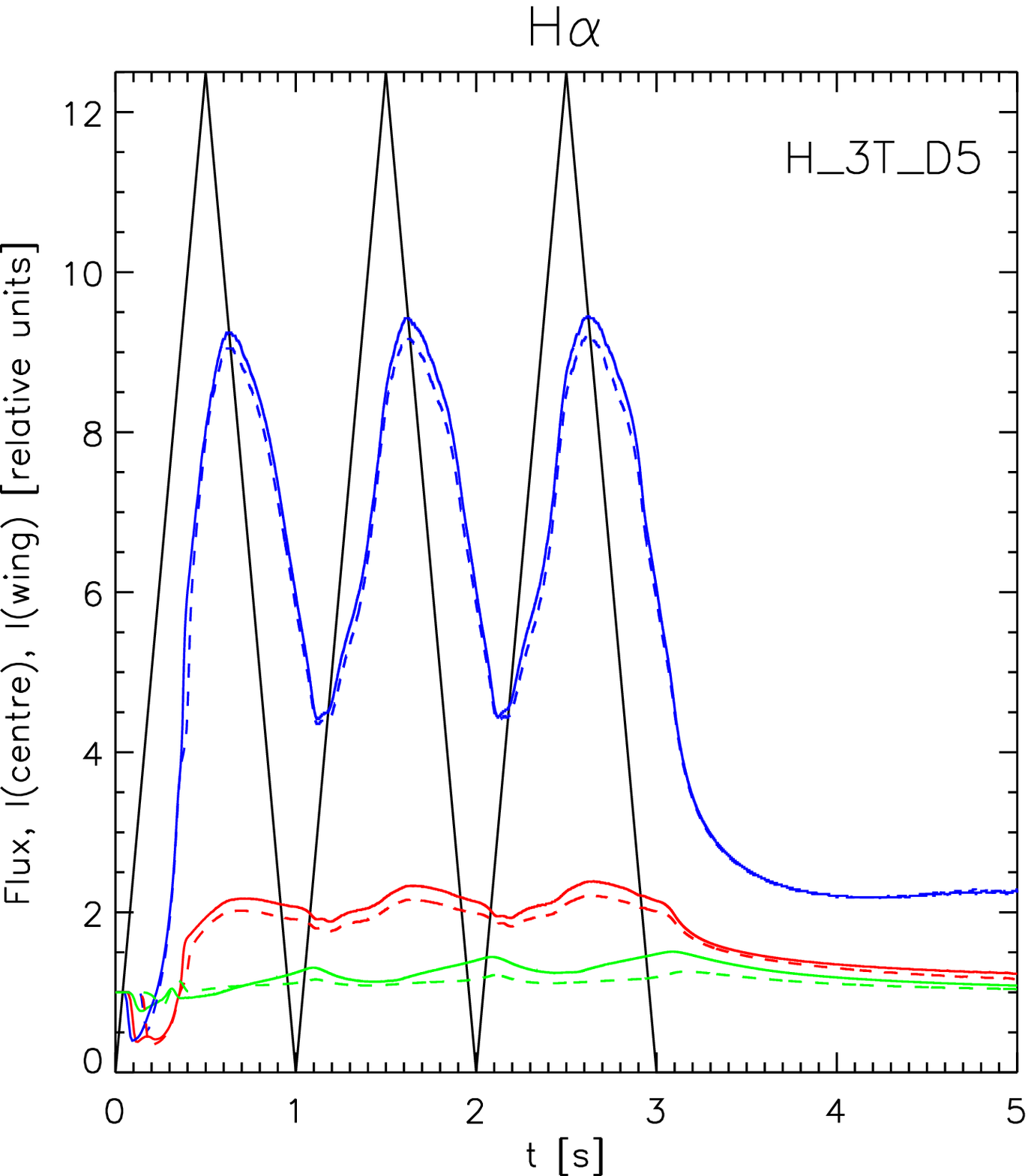}
\includegraphics[width=4.9cm]{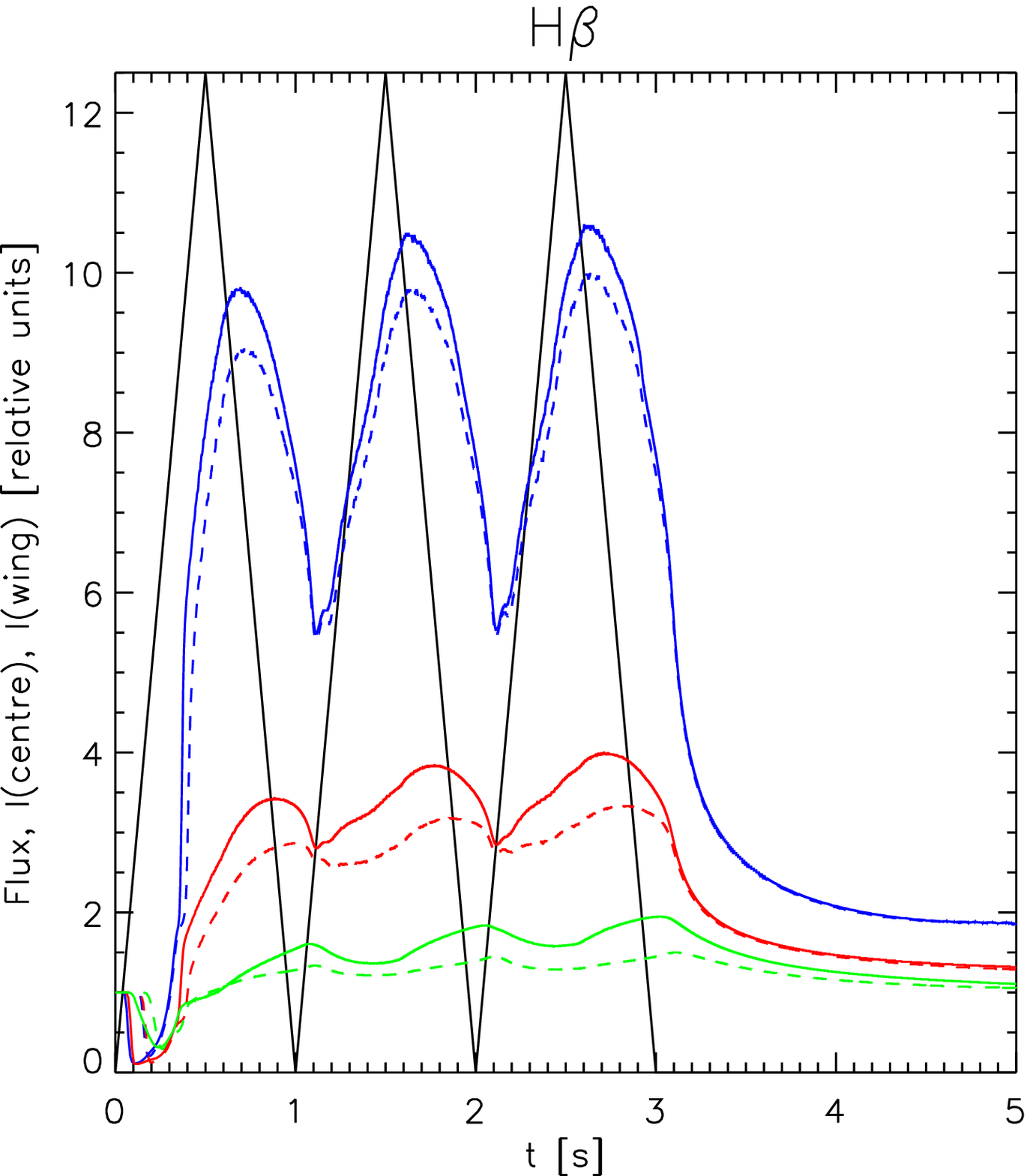}
\includegraphics[width=4.9cm]{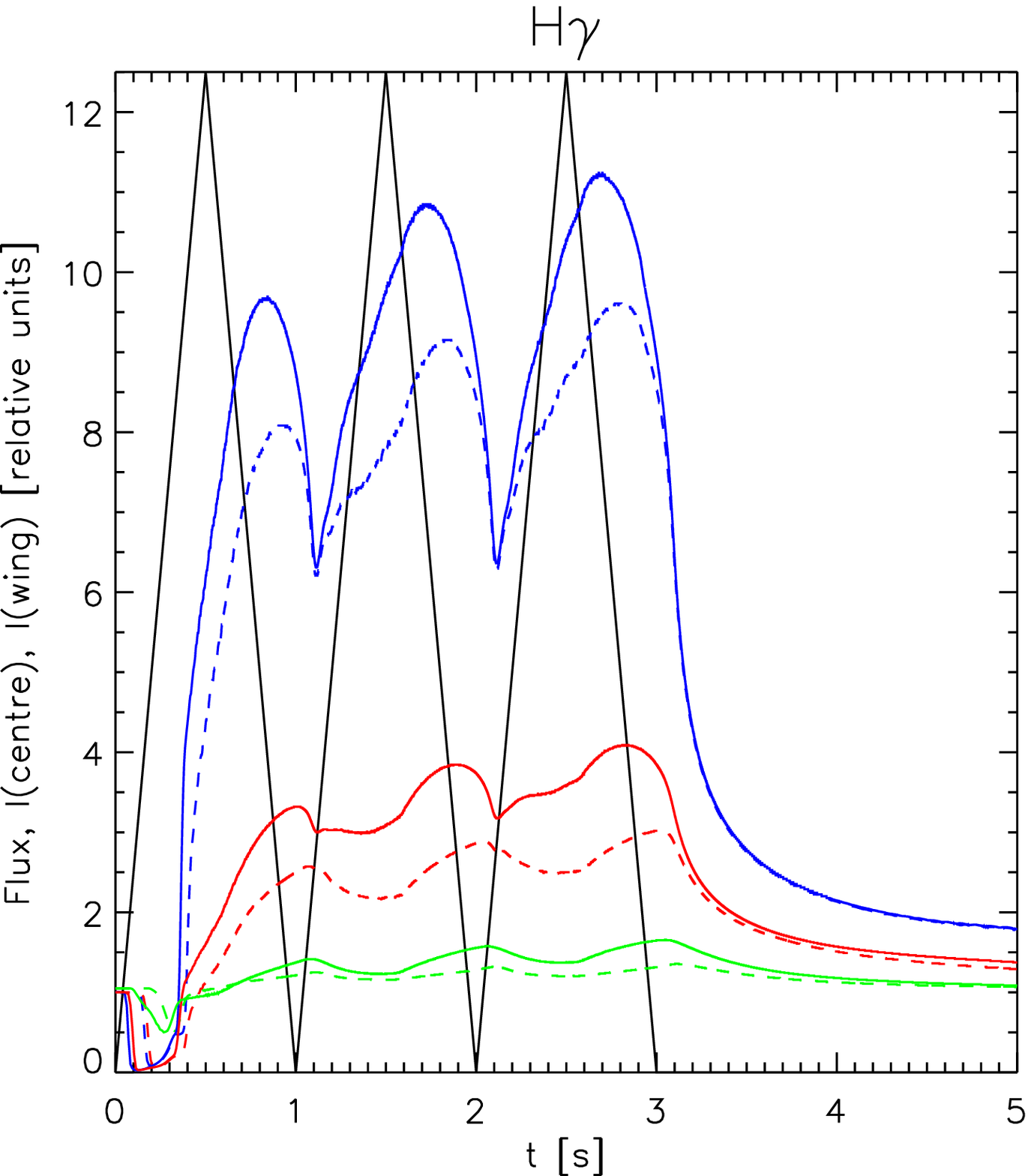}
\end{center}
\caption{
Time evolution of the H$\alpha$,  H$\beta$, and H$\gamma$
line intensities in three selected wavelengths for each line. 
Solid lines denote models with $\cnt$, dashed without $\cnt$. 
The black solid line shows the beam flux time modulation in relative units.
From top to bottom: L\_3T\_D3, L\_3T\_D5, H\_3T\_D5, and H\_3T\_D5 model.
}
\label{fig:lineint_time}
\end{figure*}

The intensity variations depend on the maximum beam flux. The low beam flux results in gradual increase of intensities 
(models L), whereas high beam flux (models H) 
causes rapid heating of the atmosphere and hence fast and larger increase of line intensities
-- see Fig.~\ref{fig:lineint_time} which compares time variations of H$\alpha$, H$\beta$, and H$\gamma$ 
for models L\_3T\_D3,  L\_3T\_D5, H\_3T\_D3, and H\_3T\_D5. 
(Owing to use of the five level plus continuum atomic model, results concerning H$\gamma$ line should be regarded
only qualitatively.)
Since larger $\delta$ results in larger heating of the upper parts of the atmosphere, 
it leads to higher line centre intensities of H$\alpha$ and H$\beta$ which are formed 
in the upper parts of the atmosphere.
On the other hand, the whole H$\gamma$ line and H$\alpha$ and H$\beta$ wings are formed 
in deeper layers, therefore
they show more prominent intensity variations for flatter beams ($\delta=3$) --  
compare D3 and D5 models in Fig.~\ref{fig:lineint_time}.

\begin{figure*}
\begin{center}
\includegraphics[width=4cm]{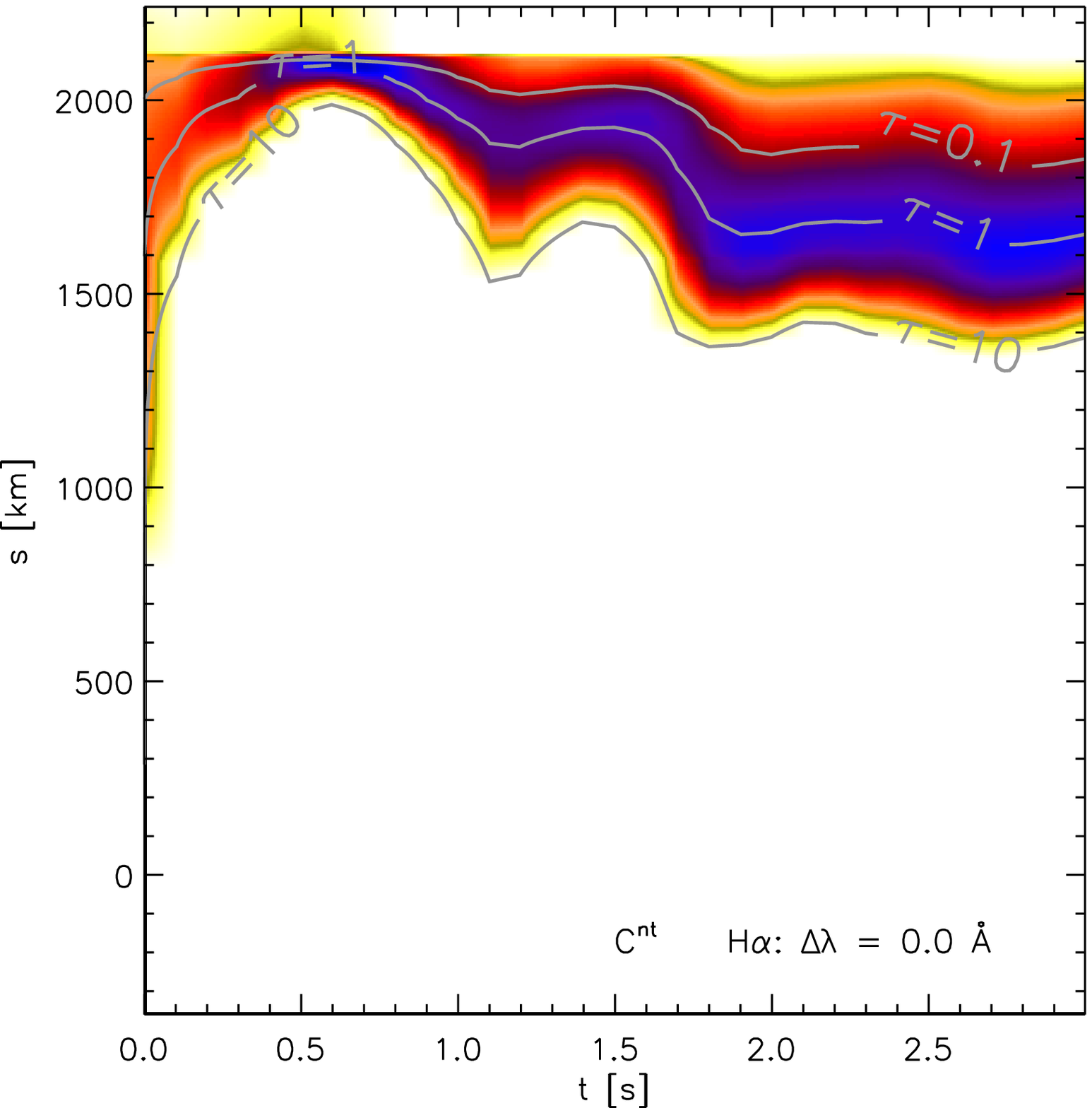}
\includegraphics[width=4cm]{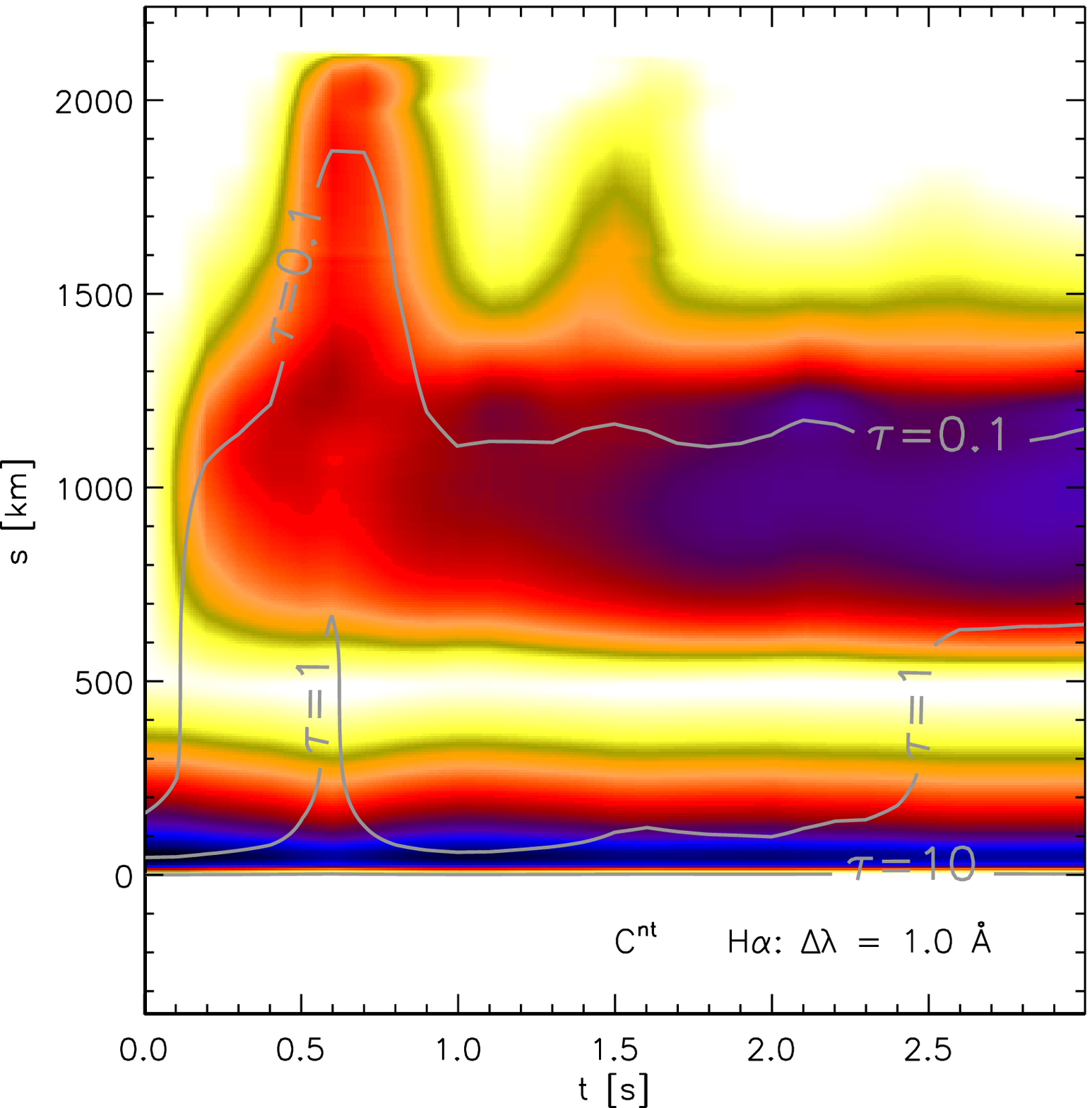}
\includegraphics[width=4cm]{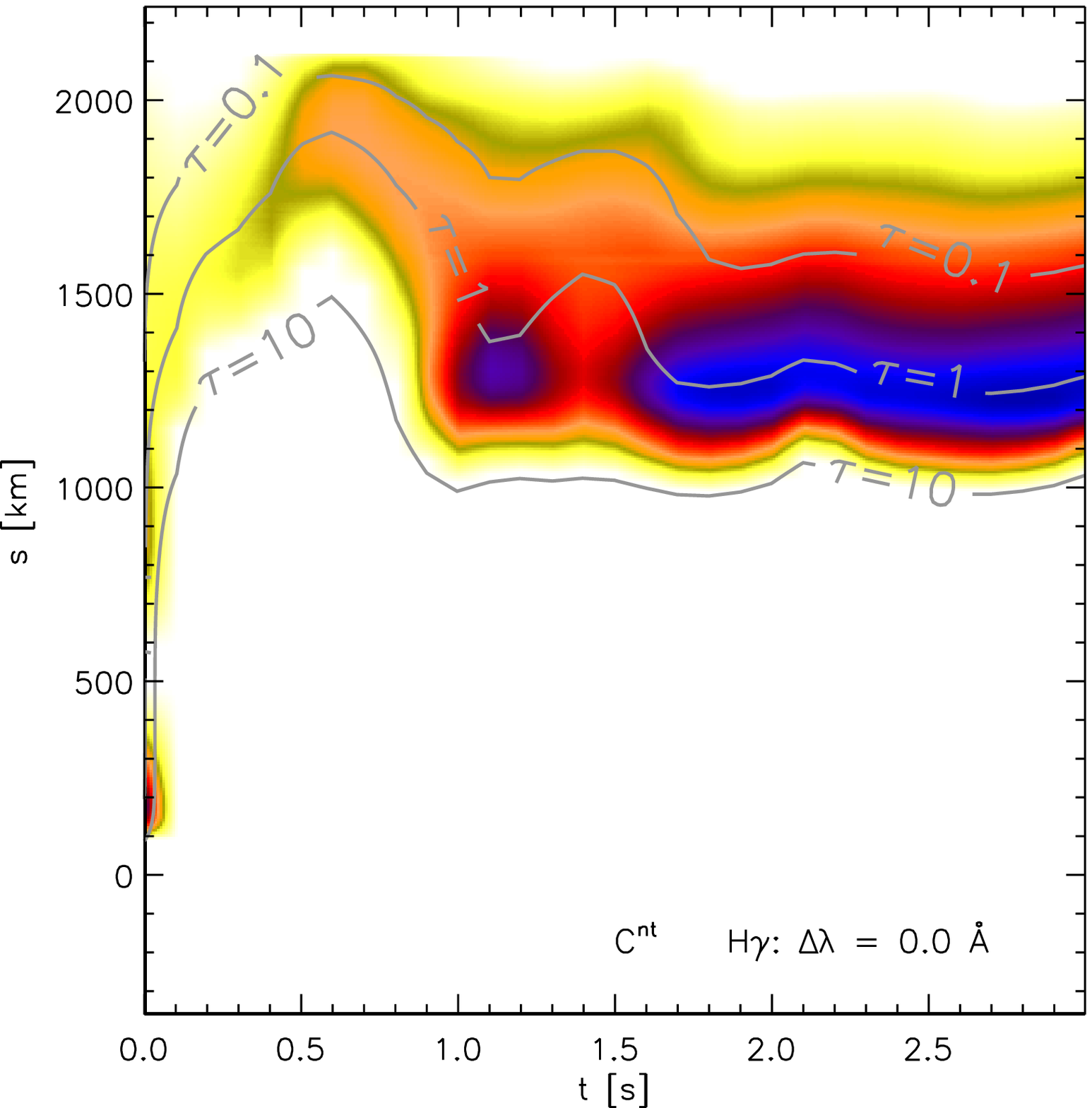}
\includegraphics[width=4cm]{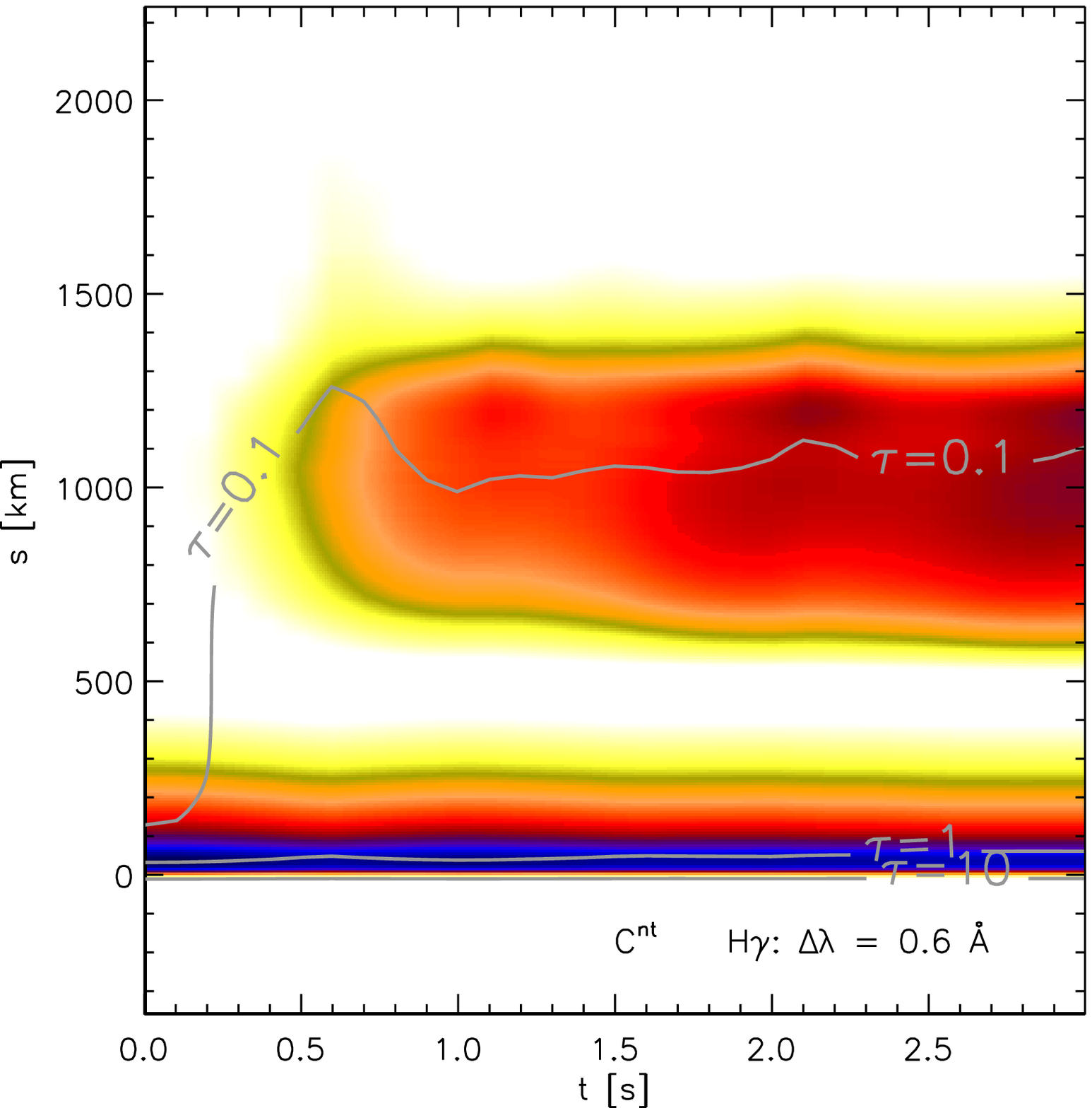}

\includegraphics[width=4cm]{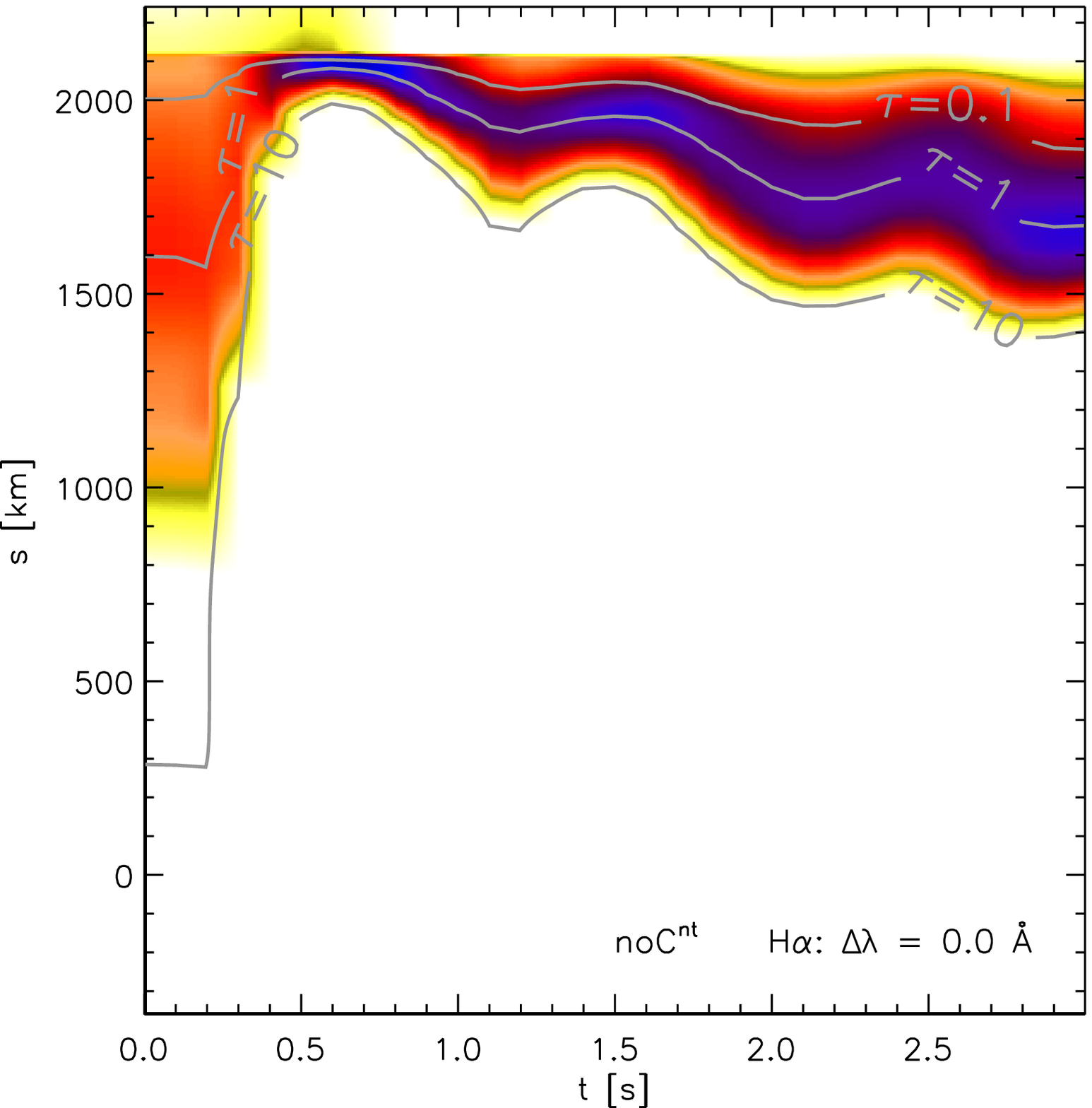}
\includegraphics[width=4cm]{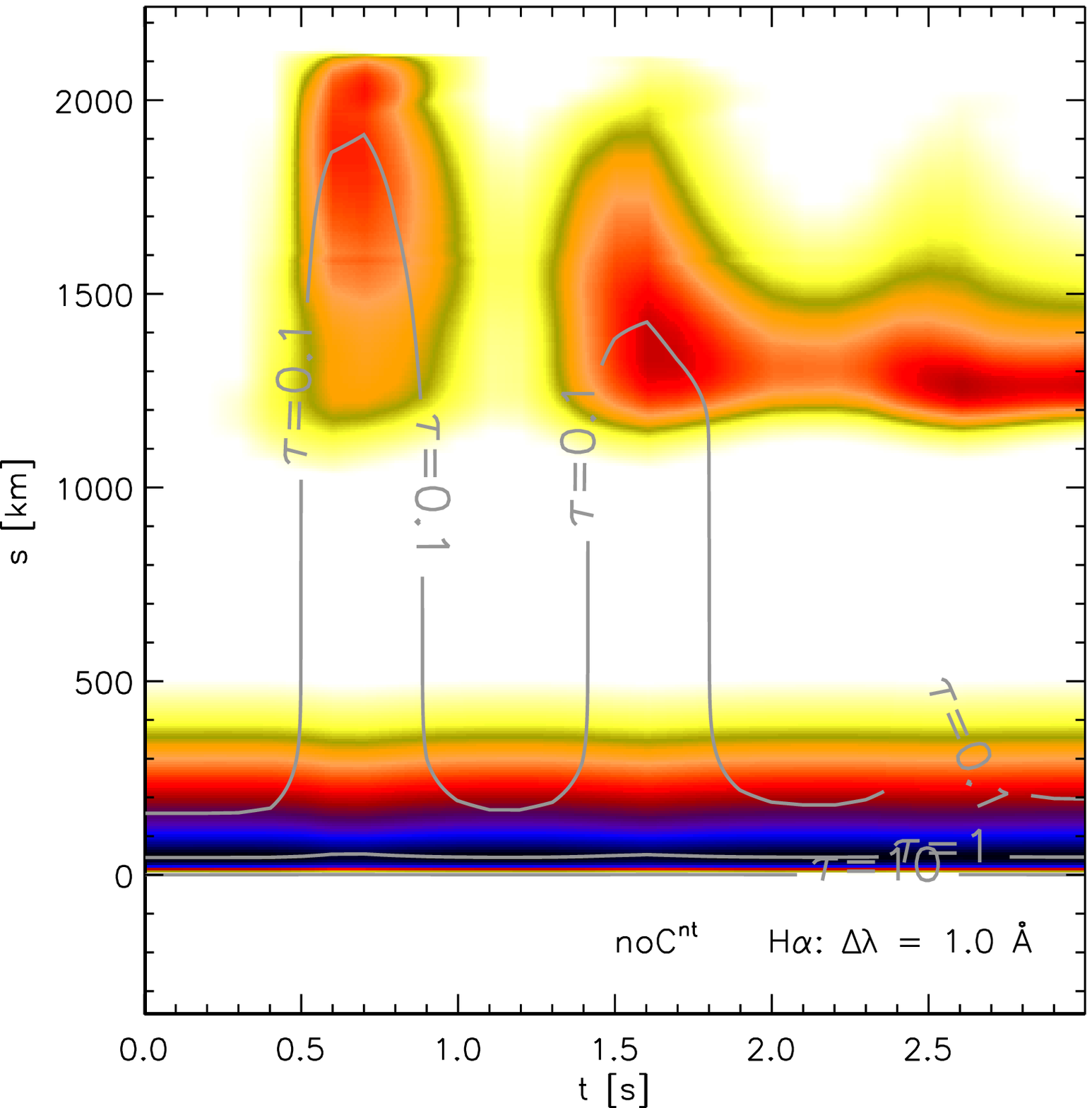}
\includegraphics[width=4cm]{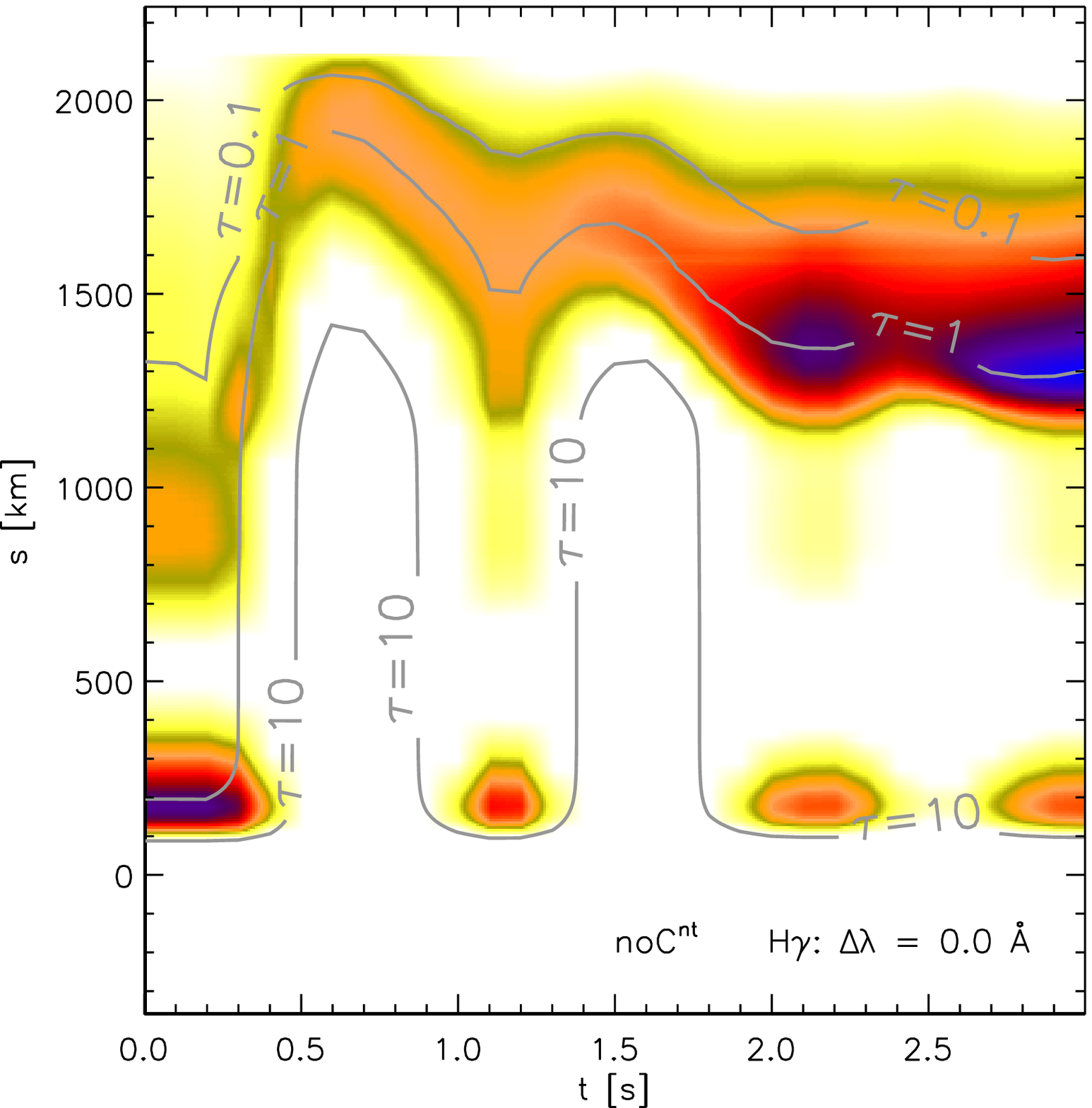}
\includegraphics[width=4cm]{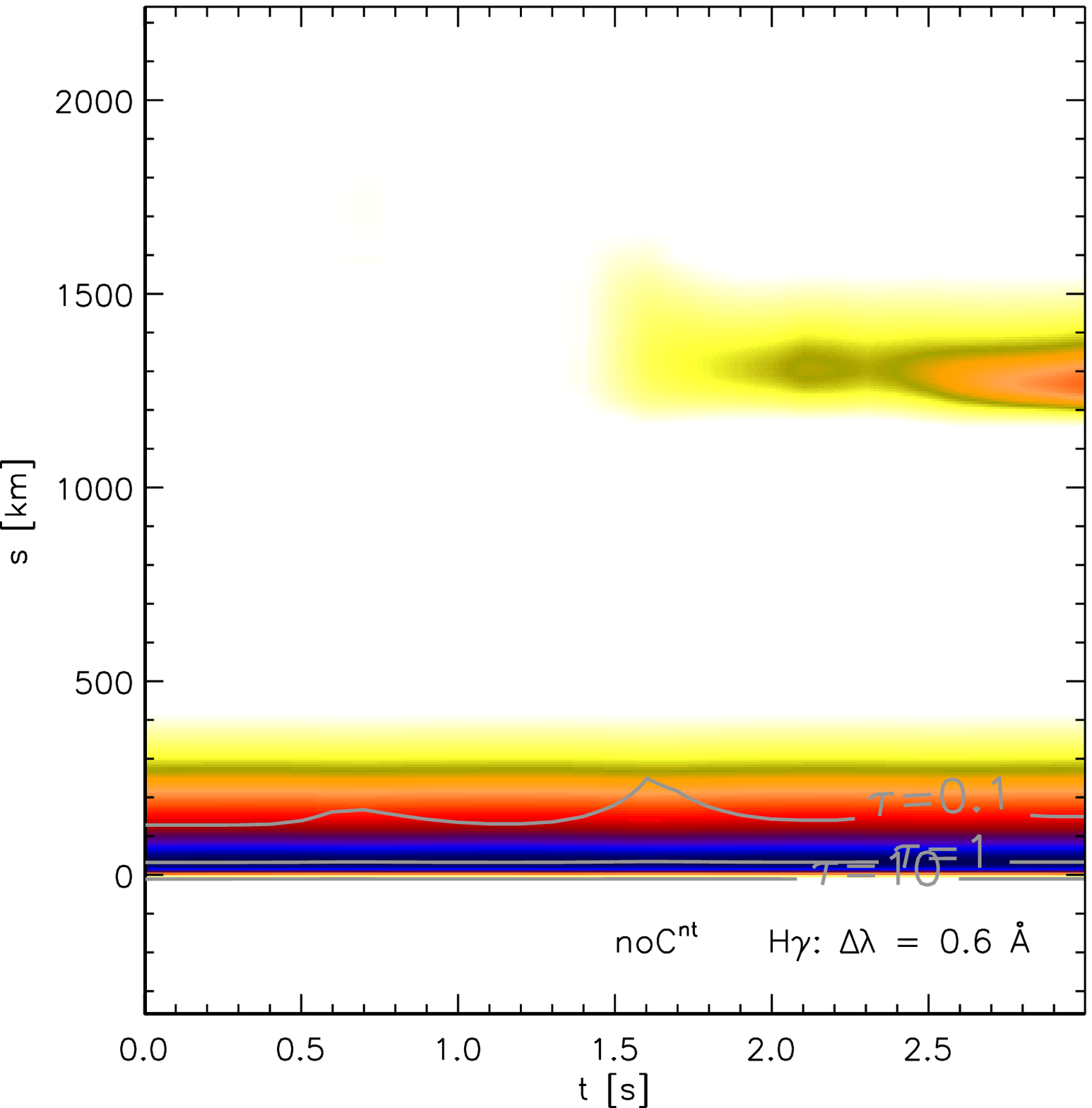}

\medskip
\includegraphics[width=4cm]{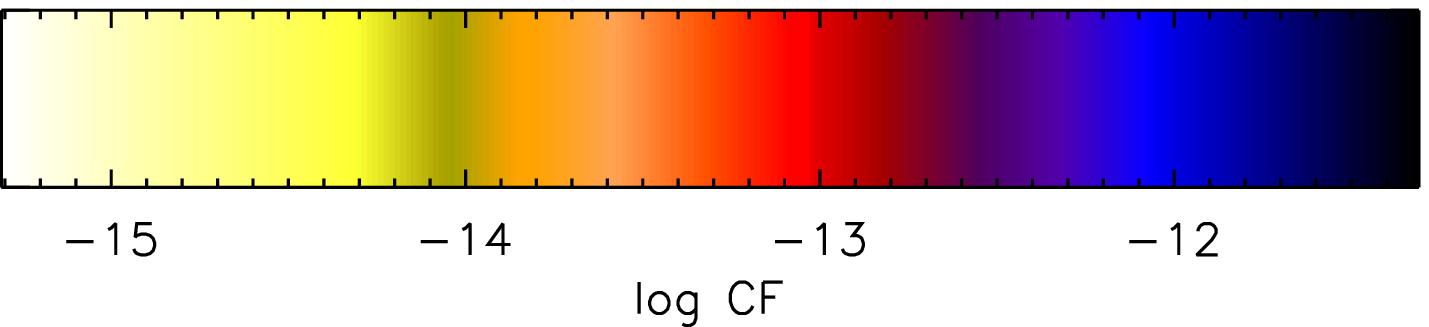}
\end{center}
\caption{Time evolution of $C\!F$ of H$\alpha\ (\Delta\lambda=$~0~\AA), H$\alpha\ (\Delta\lambda=$~1~\AA), 
H$\gamma\ (\Delta\lambda=$~0~\AA), and H$\gamma\ (\Delta\lambda=$~0.6~\AA)
for L\_3T\_D3 model. Gray lines denote contours of $\tau_\lambda=(0.1,1,10)$.
{\em Top:} with $\cnt$, {\em Bottom:} without $\cnt$. Colour scale denotes $\log C\!F$. 
}
\label{fig:cftime}
\end{figure*}

Figure~\ref{fig:cftime}  demonstrates the time variation and changes induced by  $\cnt$ of formation regions of 
the H$\alpha$ line centre ($\Delta\lambda=$~0~\AA), line wing $\Delta\lambda=$~1~\AA, H$\gamma$ line centre ($\Delta\lambda=$~0~\AA),
and H$\gamma$ line 
wing ($\Delta\lambda=$~0.6~\AA) for L\_3T\_D3 model. 
Shortly after the first beam injection into the VAL C atmosphere, at $t\sim 0.1$~s, line intensities decrease by a factor 
of $\sim$~2 and more -- see Fig.~\ref{fig:lineint_time}.  Such a dip appears due to a temporal increase in optical 
depth, $\tau$, which is caused by $\cnt$. 
A similar decrease appearing later, at $t\sim 0.4$~s is present also in the models without $\cnt$ -- compare $C\!F$ with and 
without $\cnt$ in Fig.~\ref{fig:cftime}. 
That dip is a result of relative importance of the thermal collisional rates $C_{1j}$. 
This behaviour was explained by \citet{hein91} for the case of a 3-level model of hydrogen as a consequence of steeper rise of 
second level population, $n_2$, in comparison to $n_3$ with time.
Figure~\ref{fig:cftime} also shows that a region of secondary H$\alpha$ wing emission is formed shortly after the beam injection
and its intensity slowly increases in time. 

In the case of the high-flux model, the emission from the secondary formation 
region dominates and photospheric contribution to the wing intensities diminishes; this behaviour is typical for all studied Balmer lines.  
As regards H$\gamma$ line, formation of the H$\gamma$ line centre is due to $\cnt$ completely moved from the photosphere
to the layers above $s\sim 1200$~km. Contrary to the model without $\cnt$, no emission from the photosphere 
contributes to the outgoing intensity.
Furthermore, $\cnt$ again lead to the prominent secondary wing formation region at $s\sim 1000$~km which occurs 
in the model without $\cnt$ at much later time $t\sim 2$~s -- see last panel in Fig.~\ref{fig:cftime}.

Intensities of all three studied lines show a good correlation with the beam 
flux on a time scale of the beam flux variation, i.e. on a subsecond time scale. 
Depending on the amount of heating, time variations are caused by the time-dependent temperature structure, e.g. H$\alpha$
line centre in the high-flux models, and influence of $\cnt$.
Line intensities which are affected by $\cnt$ (line centres in the low-flux models and 
line wings in the high-flux models) show more significant time variations.
Maxima of line intensities lag behind the beam flux maxima, the 
time lags are generally larger for line wing intensities than for
line centre intensities. Such a time lag is not a beam propagation effect but it is related to different 
time variations of electron density at different heights.
We are aware of the fact that velocities would cause asymmetry of lines and modify the line intensities. 
Here we concentrate on the beam influence on line formation. For detailed comparison with observed Balmer lines,
velocity should be considered in calculating the line intensities.

\subsection{Diagnostic tools}
Having obtained the time evolution of Balmer line profiles for various electron beam parameters, 
we can search for observable signatures that could provide a method for diagnostic of the electron beam presence
in the Balmer line formation regions. In order to distinguish between the flare energy transport by the non-thermal electrons 
and other agents (e. g. Alfv\'{e}n waves -- \citet{fle08}), 
we consider a method suitable for such a diagnostics only if $\cnt$ lead to significant and systematic differences
in measured quantities.
\subsubsection{Intensity ratios}
Recently, \citet{kash08} reported on so-called sidelobes in H$\alpha$/H$\beta$ intensity ratio
(i.e. increased value of H$\alpha$/H$\beta$ at $\Delta\lambda\sim 0.5$~\AA\ with respect to other 
wavelength positions) observed in flare kernels associated with radio and hard X-ray bursts. 
They attributed the appearance of such sidelobes to the effects of non-thermal electrons.

\begin{figure}
\begin{center}
\includegraphics[width=4.4cm]{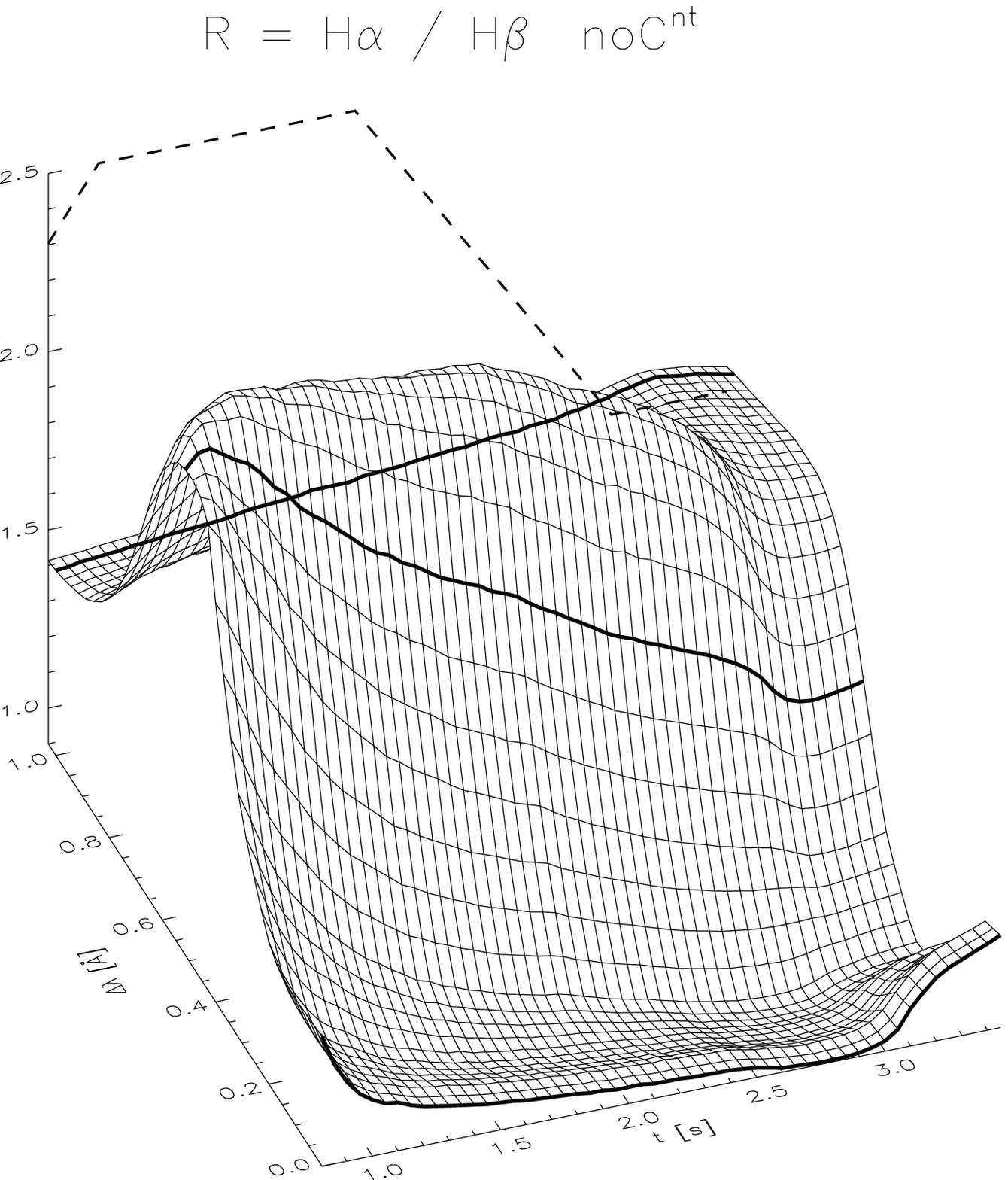}
\includegraphics[width=4.4cm]{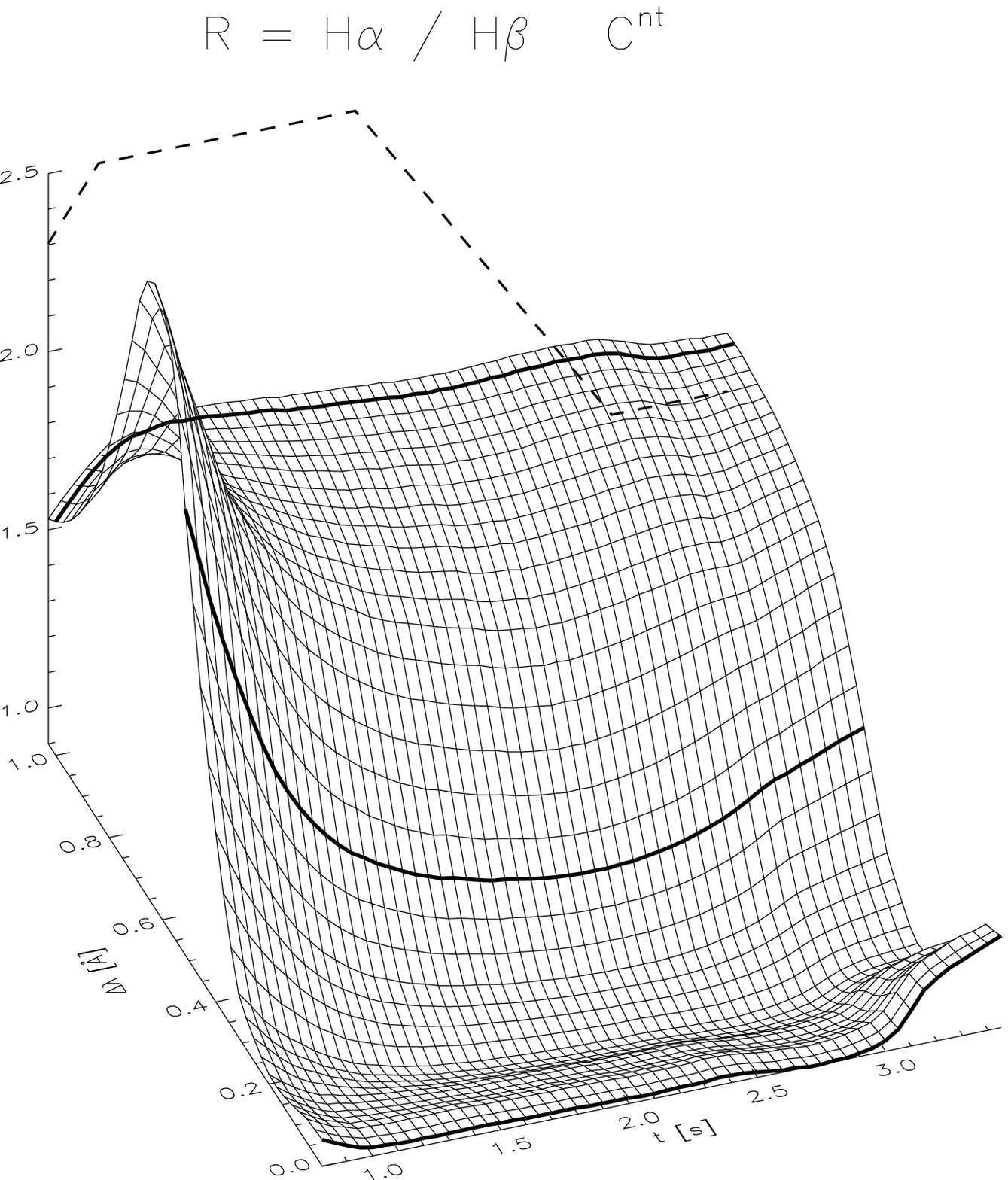}
\end{center}
\caption{
Intensity ratio $R_{\alpha\beta}(\Delta\lambda,t)$ as a 2D function of wavelength and time for the H\_TP\_D3 model.
{\em Left:} without $\cnt$, {\em right:} with $\cnt$. The black dashed line denotes time modulation of the beam flux.
Thick solid lines represent  $R_{\alpha\beta}$ at selected $\Delta\lambda=0,0.5,1$~\AA\ which are shown for all models in 
Fig.~\ref{fig:lineratios}. For display purposes, the time evolution is shown from $t=0.8$~s.}
\label{fig:3Dratio}
\end{figure}

Using our simulations we are able to check whether the observed sidelobes are a feature related to the 
electron beams. 
An example of a sidelobe is shown in Fig.~\ref{fig:3Dratio} (left). It appears as a local maximum
at $\Delta\lambda\approx 0.6$~\AA\ and varies on a timescale similar to the beam flux modulation.
Due to velocities, such sidelobes could be asymmetric with respect to the line centre. On the other hand,
there are observations of almost symmetric Balmer lines and H$\alpha$/H$\beta$ ratios in the flare kernels
associated with hard X-ray emission \citep{2008ESPM...12.2.61K},
thus our models which do not consider velocities in radiative transfer can be applied to such situations.

\begin{figure*}
\begin{center}
\includegraphics[width=4.cm]{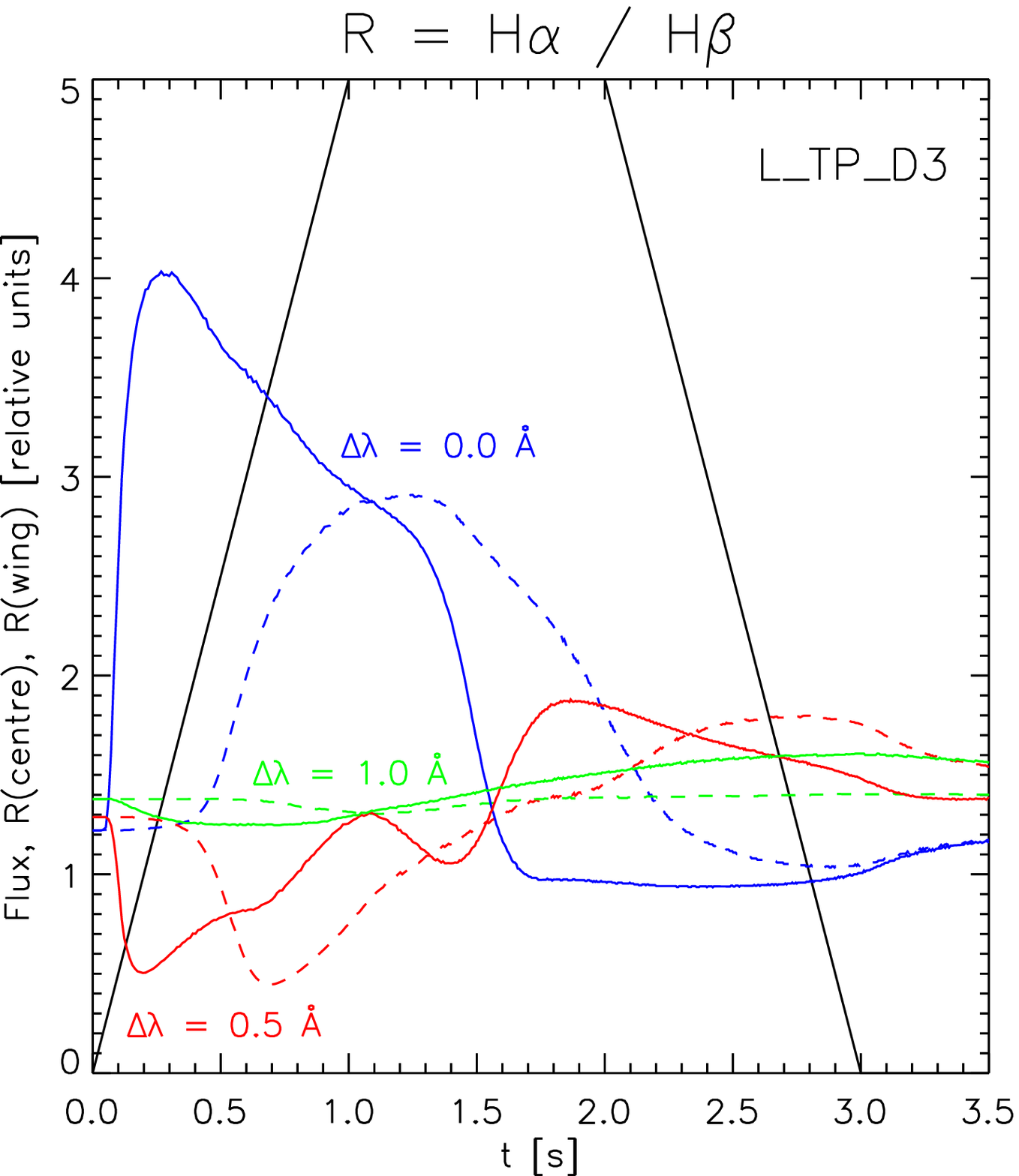}
\includegraphics[width=4.cm]{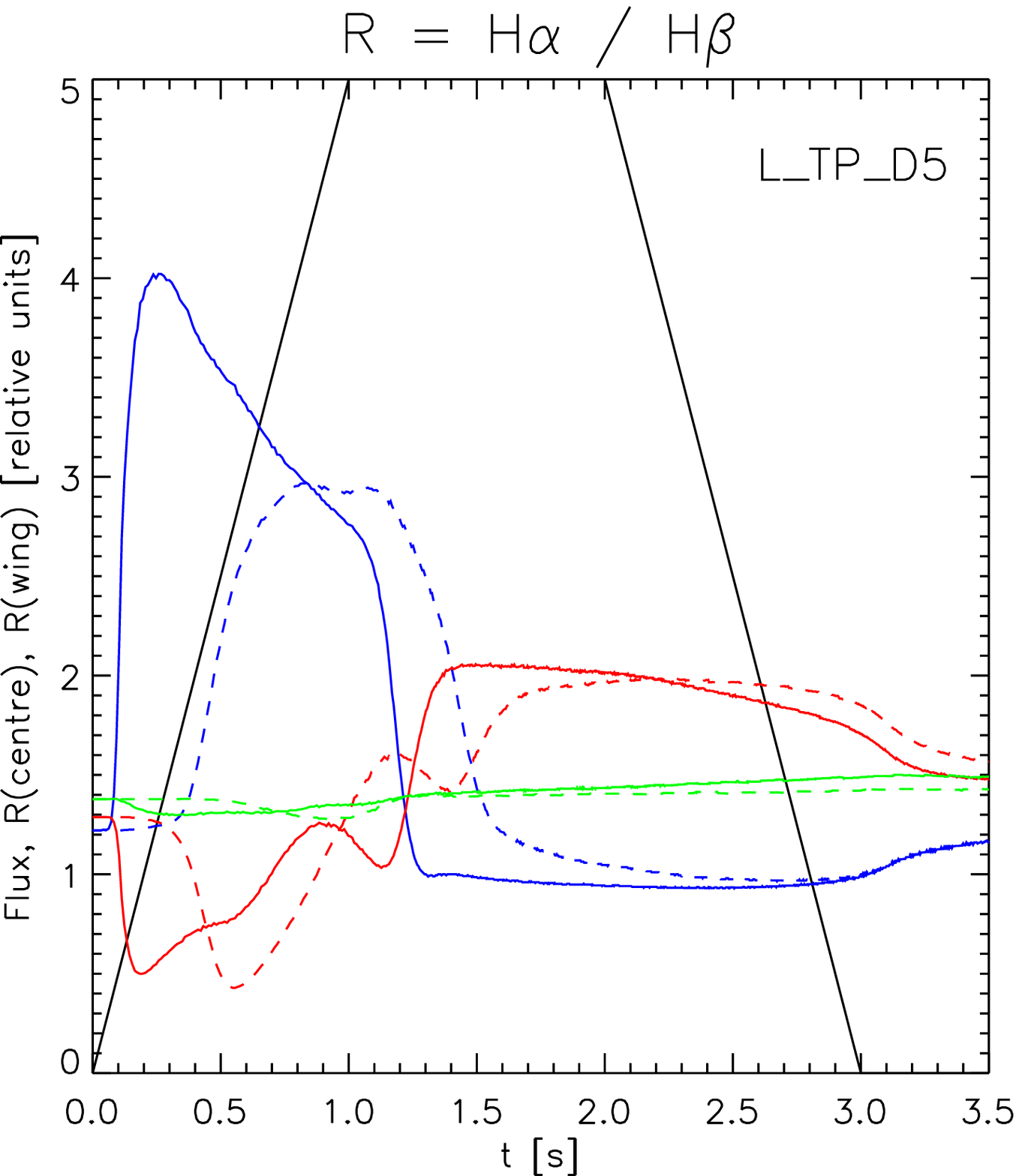}
\includegraphics[width=4.cm]{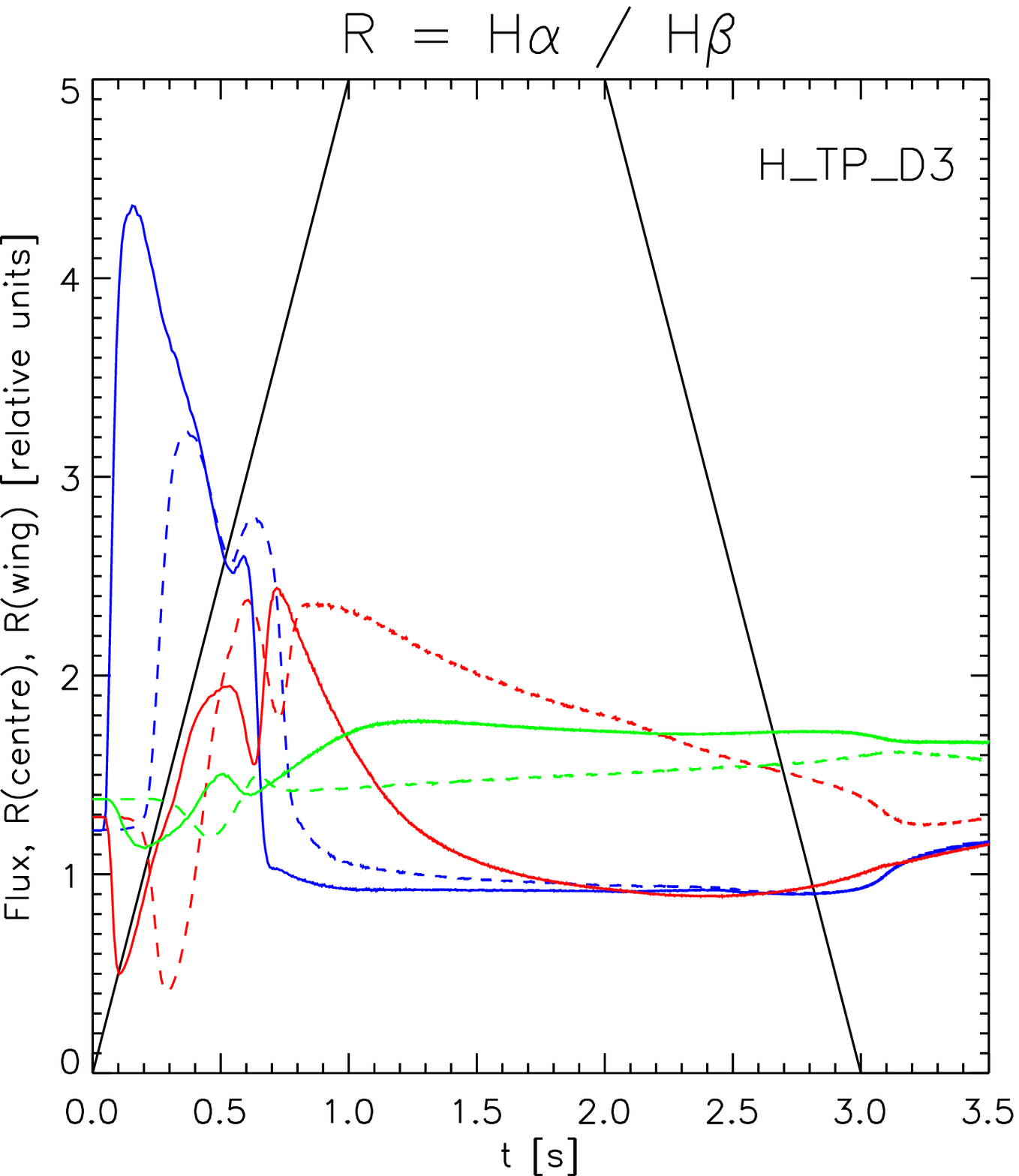}
\includegraphics[width=4.cm]{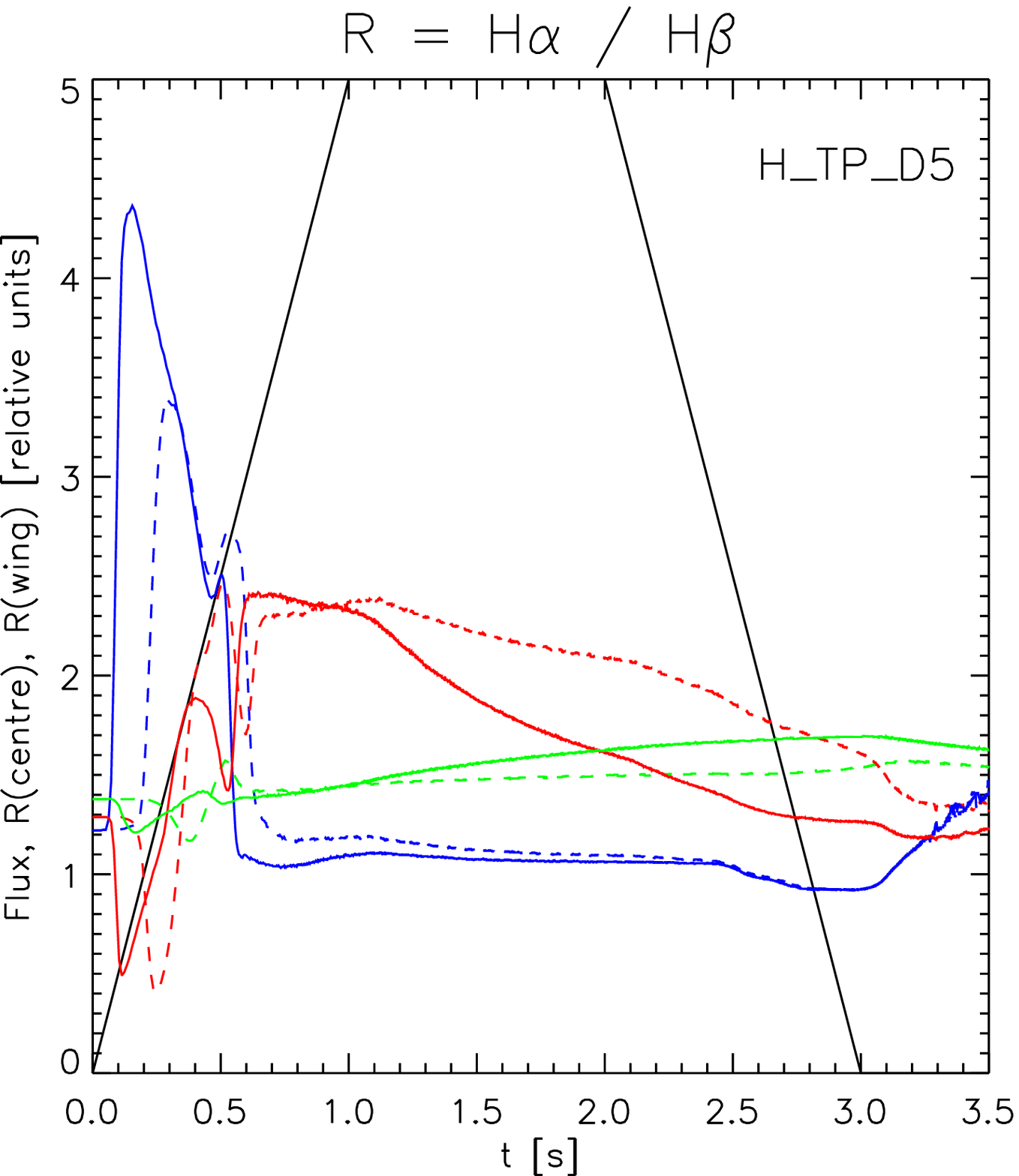}

\includegraphics[width=4.cm]{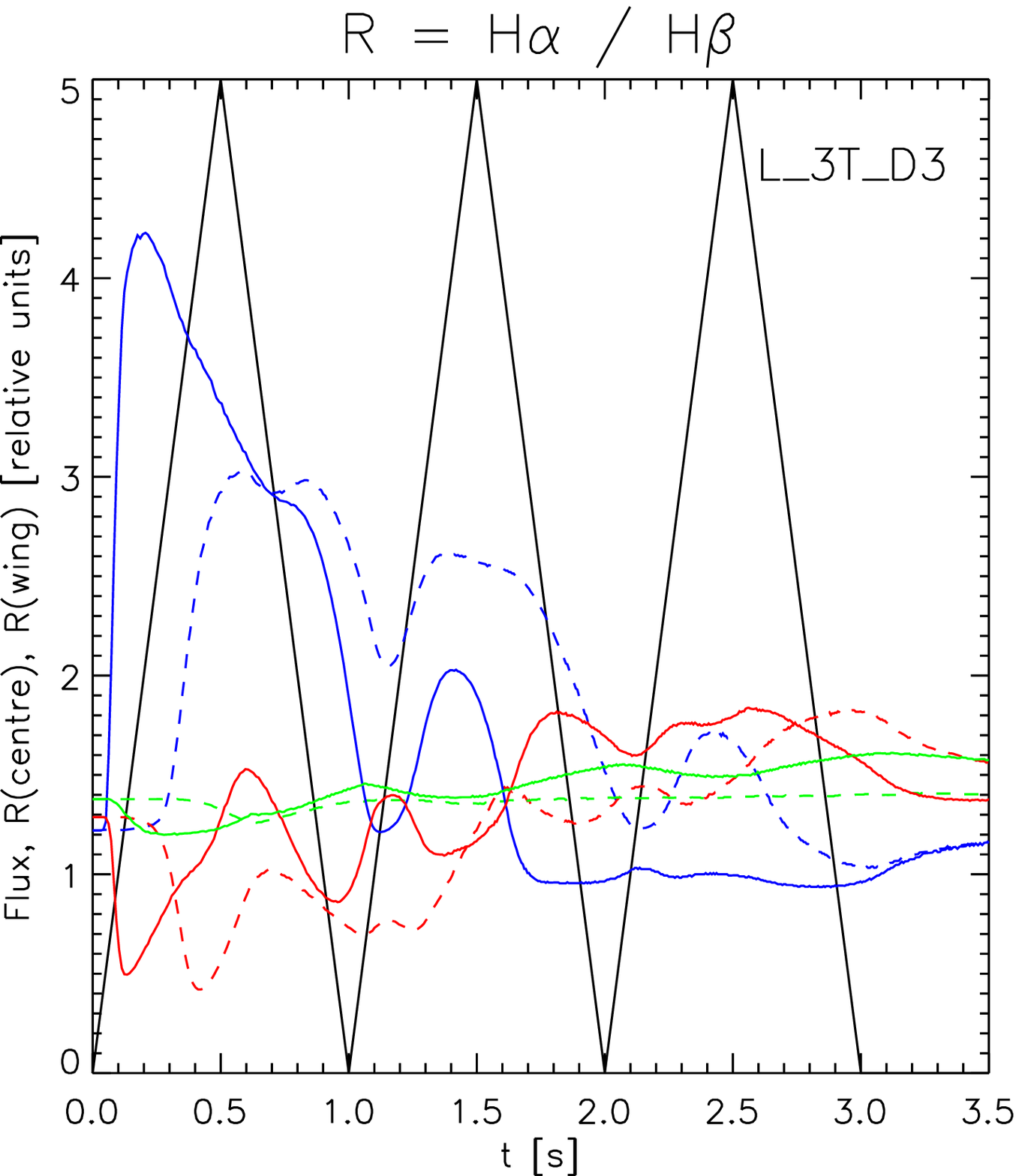}
\includegraphics[width=4.cm]{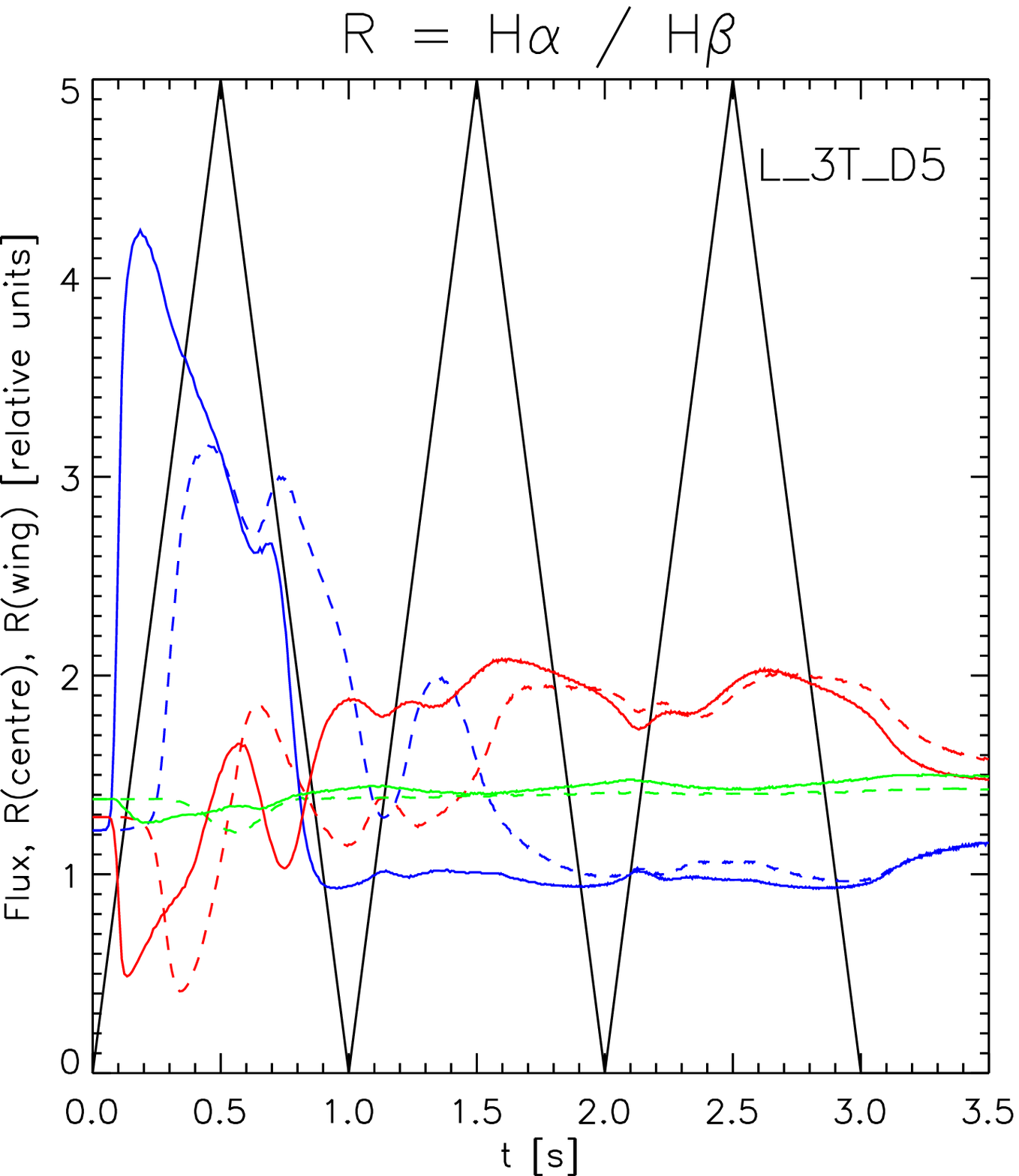}
\includegraphics[width=4.cm]{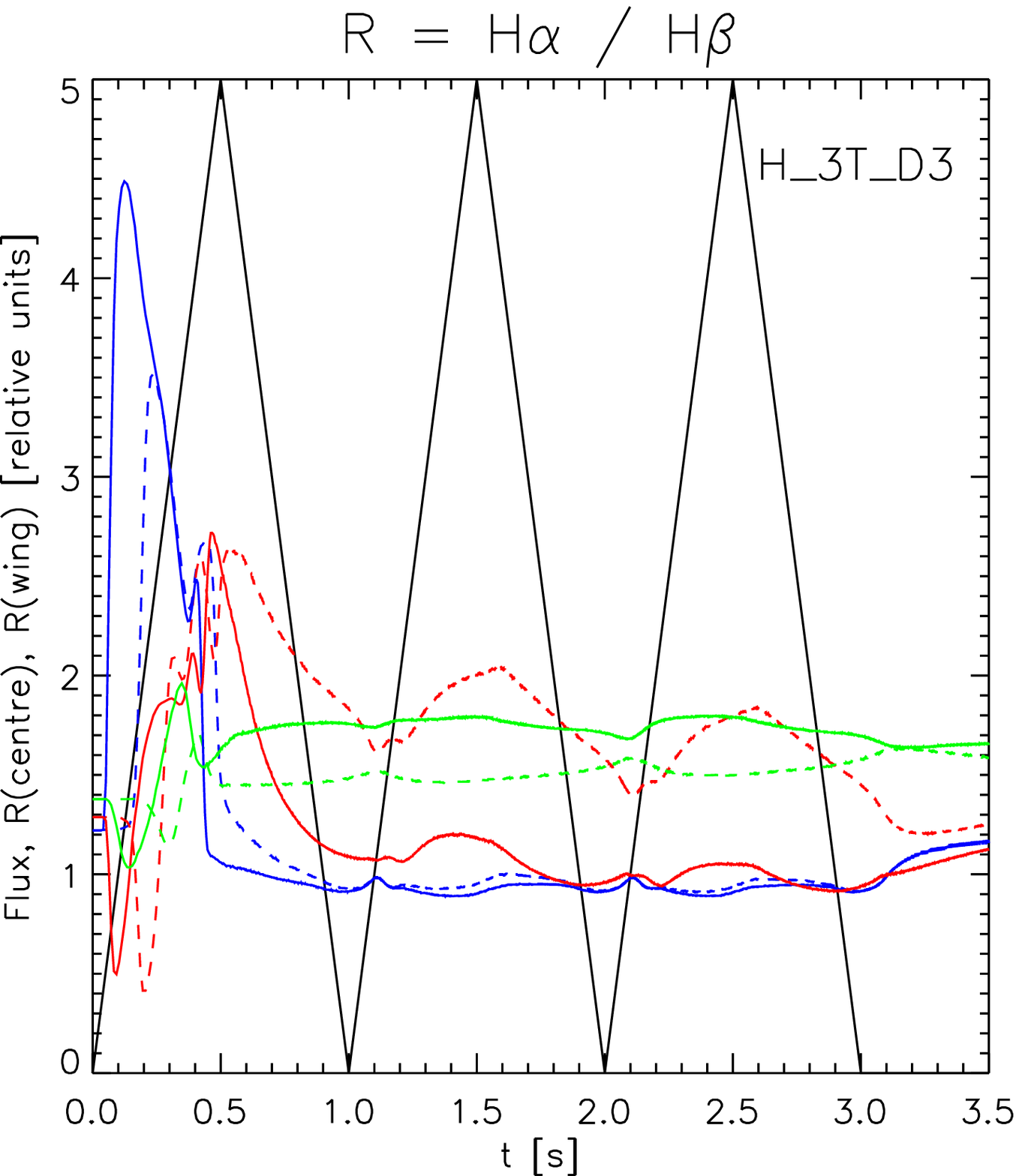}
\includegraphics[width=4.cm]{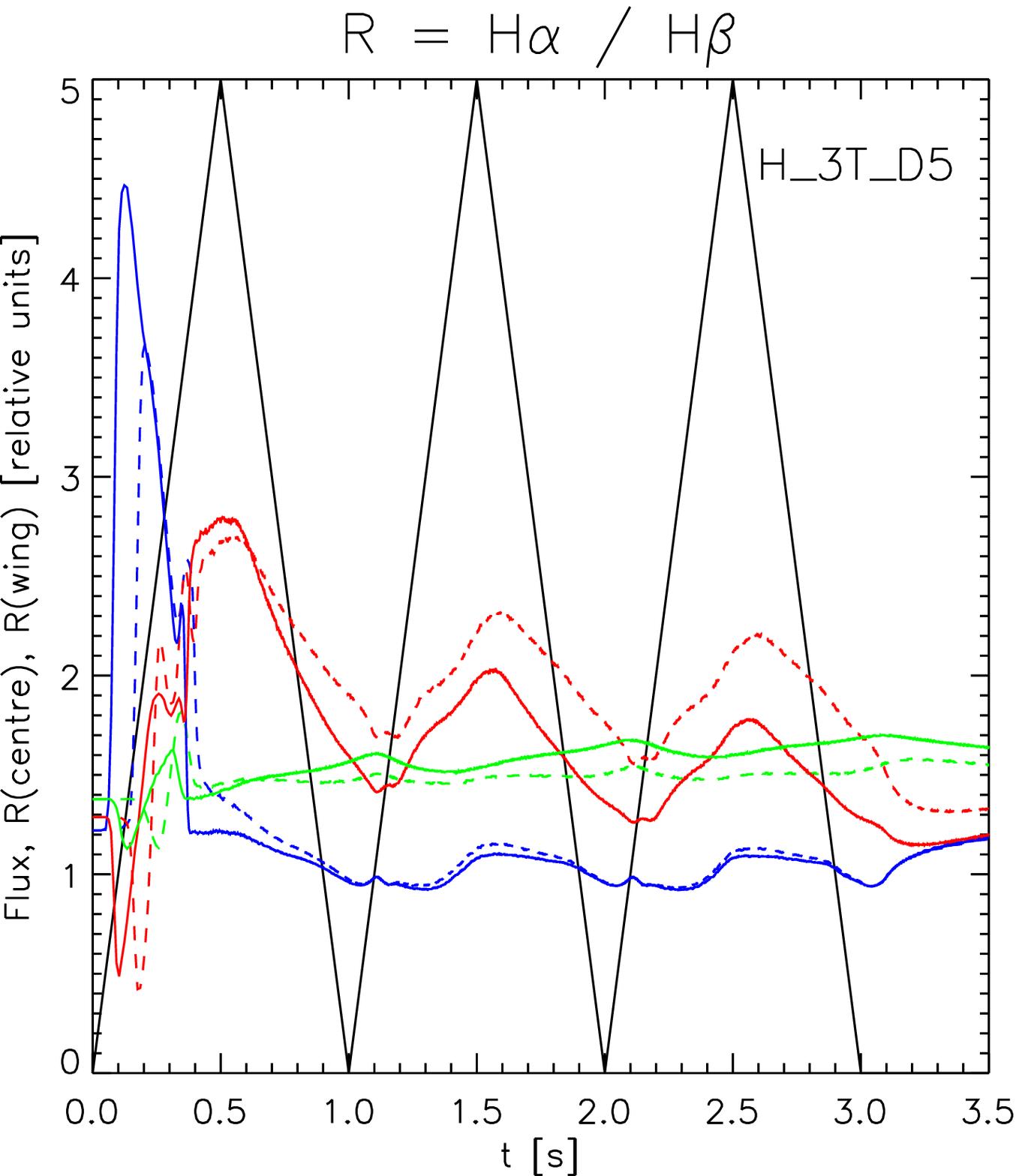}
\end{center}
\caption{The intensity ratios 
$R_{\alpha\beta}(\Delta\lambda,t)$ for three selected wavelengths $\Delta\lambda$. Solid lines denote models with $\cnt$
and dashed without $\cnt$. The black solid line shows the time modulation of the beam flux. 
{\em Top}: trapezoidal time modulation. 
{\em Bottom}: modulation by triangular peaks. 
}
\label{fig:lineratios}
\end{figure*}

Figure~\ref{fig:lineratios} shows the time evolution of H$\alpha$/H$\beta$ at several $\Delta\lambda$,
$R_{\alpha\beta}(\Delta\lambda,t)$ (indicated also in Fig.~\ref{fig:3Dratio}),
for all considered models of the beam heating (see Table~\ref{modelparam}).
In these plots, a sidelobe
at a certain $\Delta\lambda$ would appear as an increased line above the others, varying according to the beam time modulation.

To consider a sidelobe being caused by the electron beam, the sidelobe should correspond to a simulation with $\cnt$ included. 
For some beam parameters $R_{\alpha\beta}(0.5\ \mbox{\AA},t)$ 
exhibits the
reported sidelobe behaviour with the exception of the high-flux models with $\delta=3$ - see 
Fig.~\ref{fig:3Dratio} (right) or the corresponding
third panel in the top and bottom row in Fig.~\ref{fig:lineratios} -- where $R_{\alpha\beta}(0.5\ \mbox{\AA},t)$  
rapidly drops below 
$R_{\alpha\beta}(\Delta\lambda>0.5\ \mbox{\AA},t)$ and no sidelobe exists in $R_{\alpha\beta}$ on a beam timescale. 
H$\alpha$/H$\gamma$ ratio shows a similar behaviour to $R_{\alpha\beta}$
whereas H$\beta$/H$\gamma$ ratio is much weakly sensitive to beam parameters.

On the other hand, neglecting the effect of $\cnt$, i.e. assuming other agents than electron beams for the flare energy transport, 
sidelobes at $\Delta\lambda=0.5$~\AA\ are present at all models. Thus, the observed sidelobes cannot be considered as
a unique signature of beam presence in the atmosphere but they are probably related to an impulsive heating.

Note that the maximal sidelobes from our simulations may appear, for some models, at wavelength positions slightly
different from $\Delta\lambda\sim 0.5$~\AA.

Furthermore, other kinds of intensity ratios, e.g. a relative line intensity with respect to the line centre value such as 
$R_\alpha(\Delta\lambda,t) = I(\Delta\lambda,t) / I(0\ \mbox{\AA},t)$, do not show any systematic difference between the models
with and without $\cnt$ either. Therefore, the intensity ratios do not provide 
a reliable diagnostic tool suitable for analysing
the presence of the electron beams in the Balmer line formation regions.
\subsubsection{Wavelength-integrated intensity}
\begin{figure}
\centerline{
\includegraphics[height=5cm]{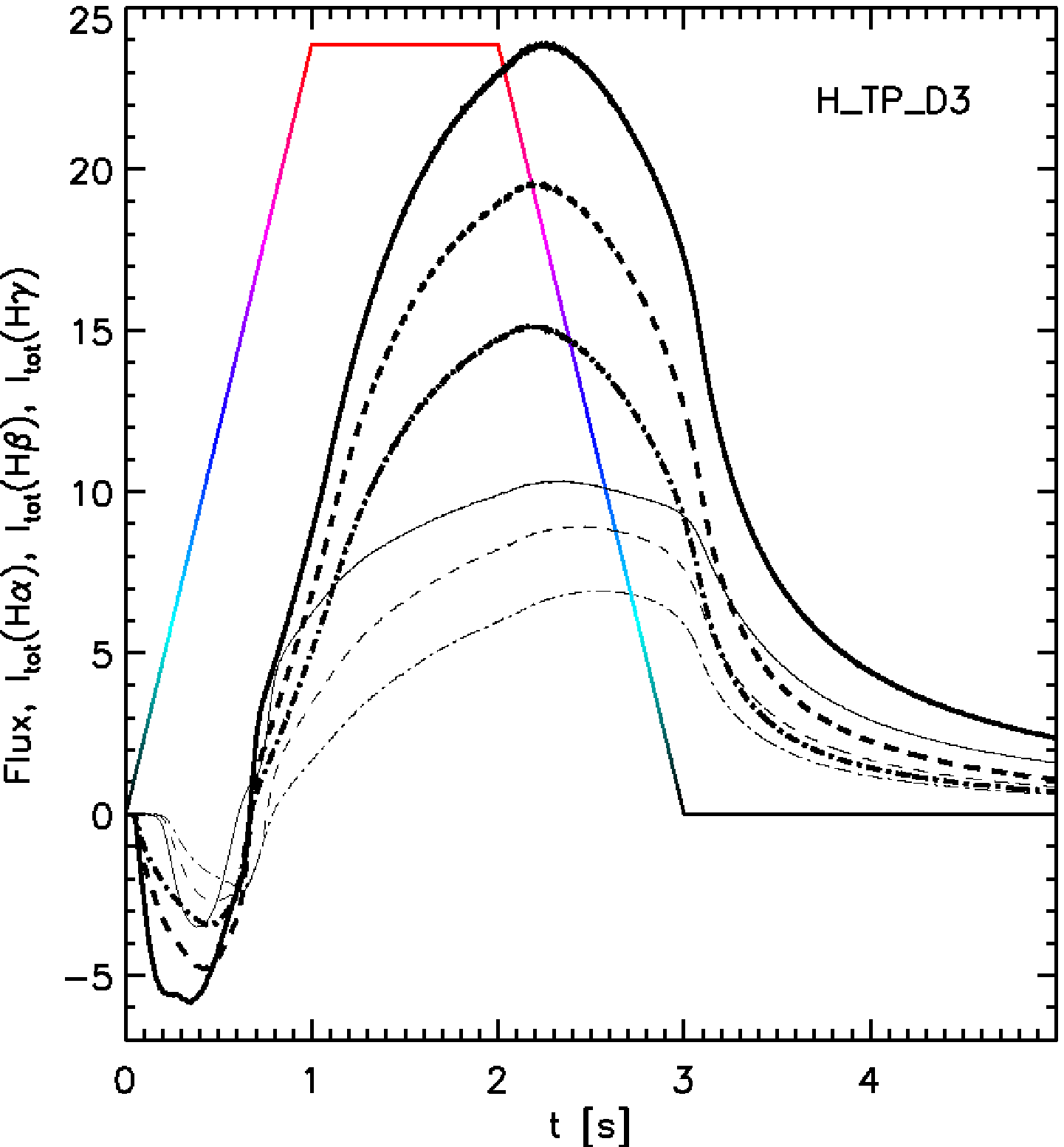}
\includegraphics[height=5cm]{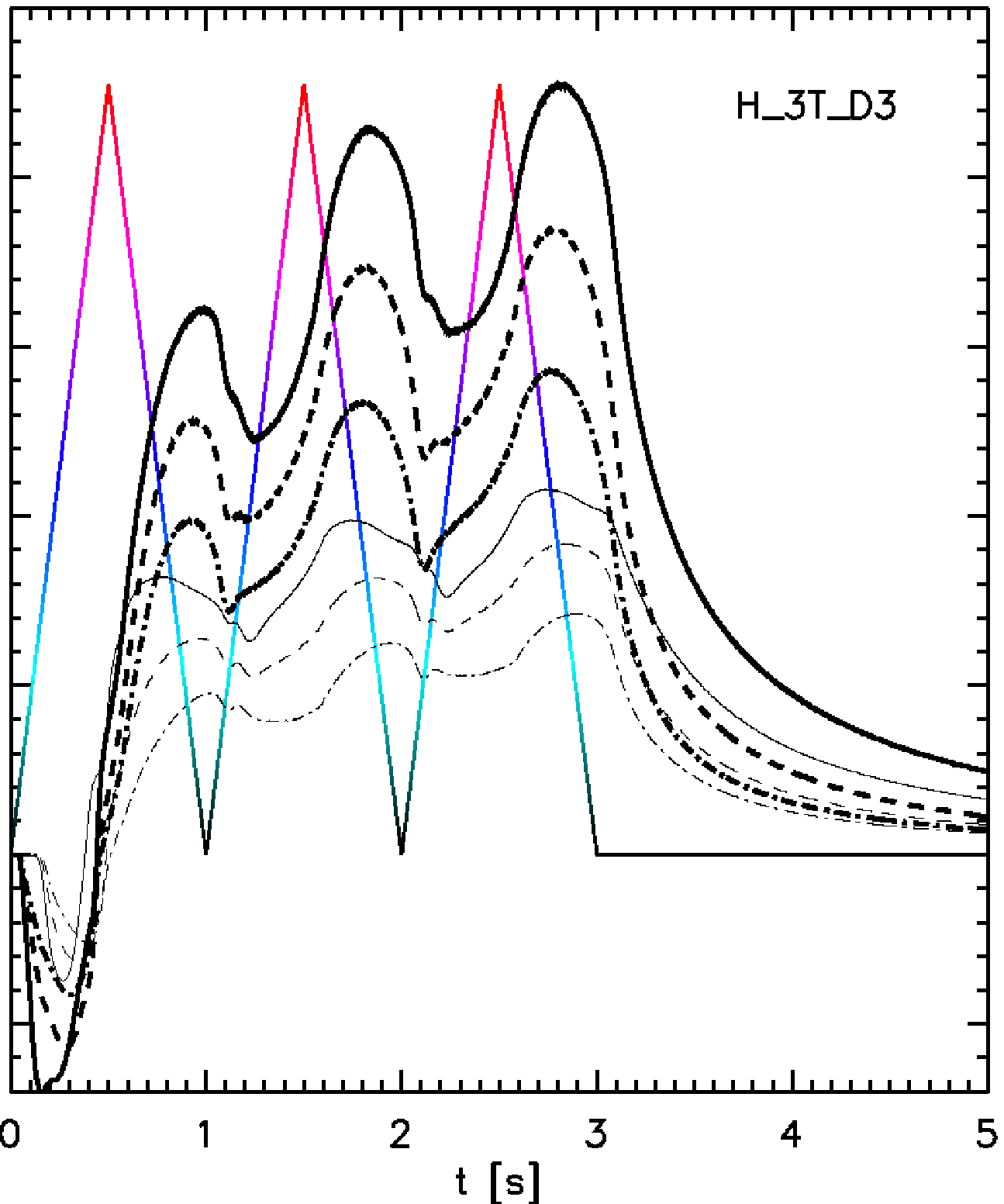}}
\caption{
Time evolution of H$\alpha$ (solid), H$\beta$ (dashed),
and H$\gamma$ (dot-dashed) wavelength-integrated intensities.
The modulation of the beam flux is shown as colour-changing broken line,
the colour indicates the flux value. Thick lines denote models with $\cnt$ and thin without $\cnt$.
Total deposited energy $E_\mathrm{tot}$ is the same for both models.
}
\label{fig:ew}
\end{figure}
Wavelength-integrated intensity $I_\mathrm{tot}$  (proportional to equivalent width)
was recently proposed by \citet{cheng06} as a tool for diagnostics of the non-thermal effects in the solar flares.
On the basis of static semiempirical models, they propose to judge the relative importance of thermal and non-thermal heating
in flares by analysis of $I_\mathrm{tot}$(H$\alpha$) and $I_\mathrm{tot}$(CaII 8542\AA) which show significantly different sensitivity to $\cnt$.
Using this idea, we analyse in detail the wavelength-integrated intensity
\begin{equation}\label{eq:ew}
I_\mathrm{tot} = k \int\limits_{-\Delta\lambda_\mathrm{max}}^{+\Delta\lambda_\mathrm{max}} [I(\lambda,t) - I(\lambda,t=0)]\ \dd\lambda
\end{equation}
of the studied Balmer lines for more general radiative hydrodynamic models.
$k$ is a normalisation to scale intensities, here we use line centre intensity at $t=0$~s.

\begin{figure}
\centerline{\includegraphics[width=8.5cm]{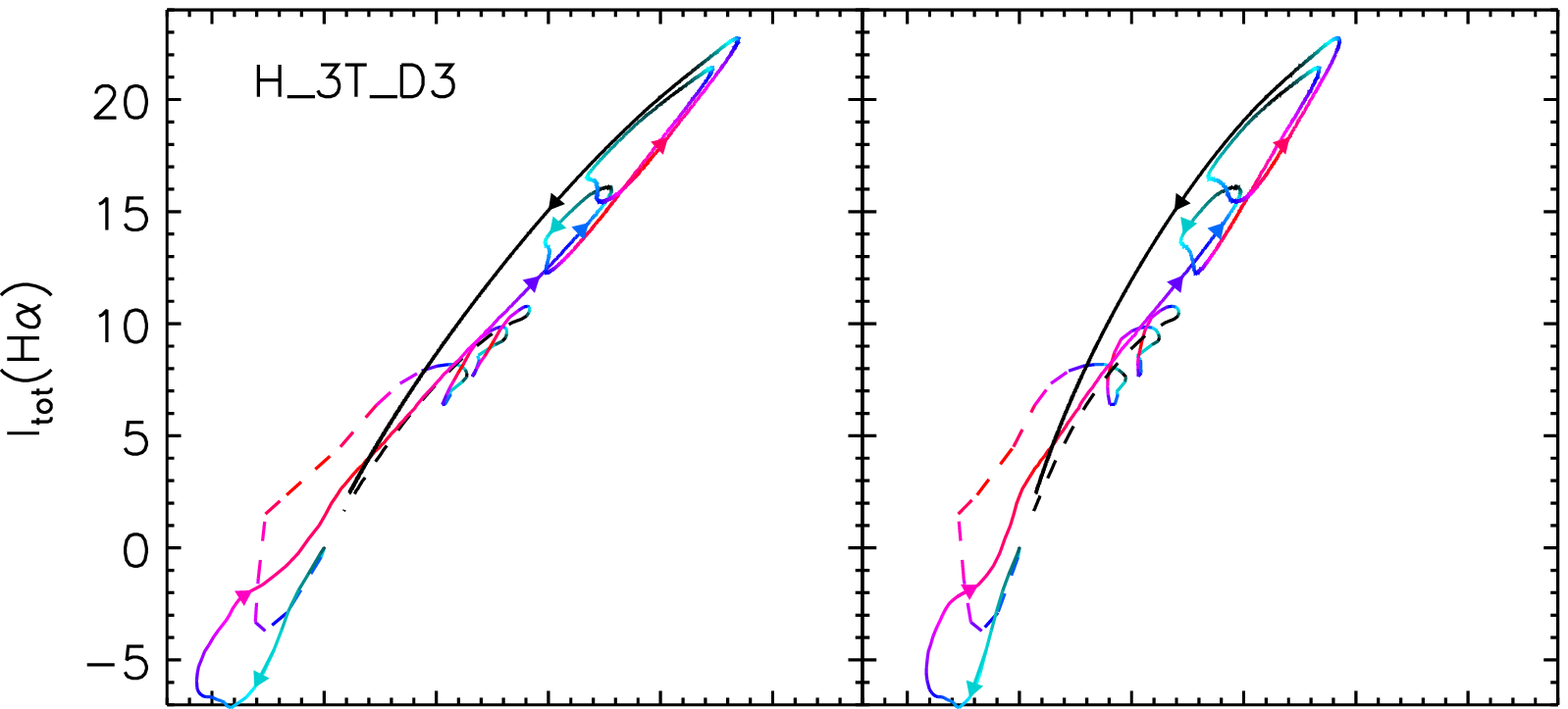}}
\centerline{\includegraphics[width=8.5cm]{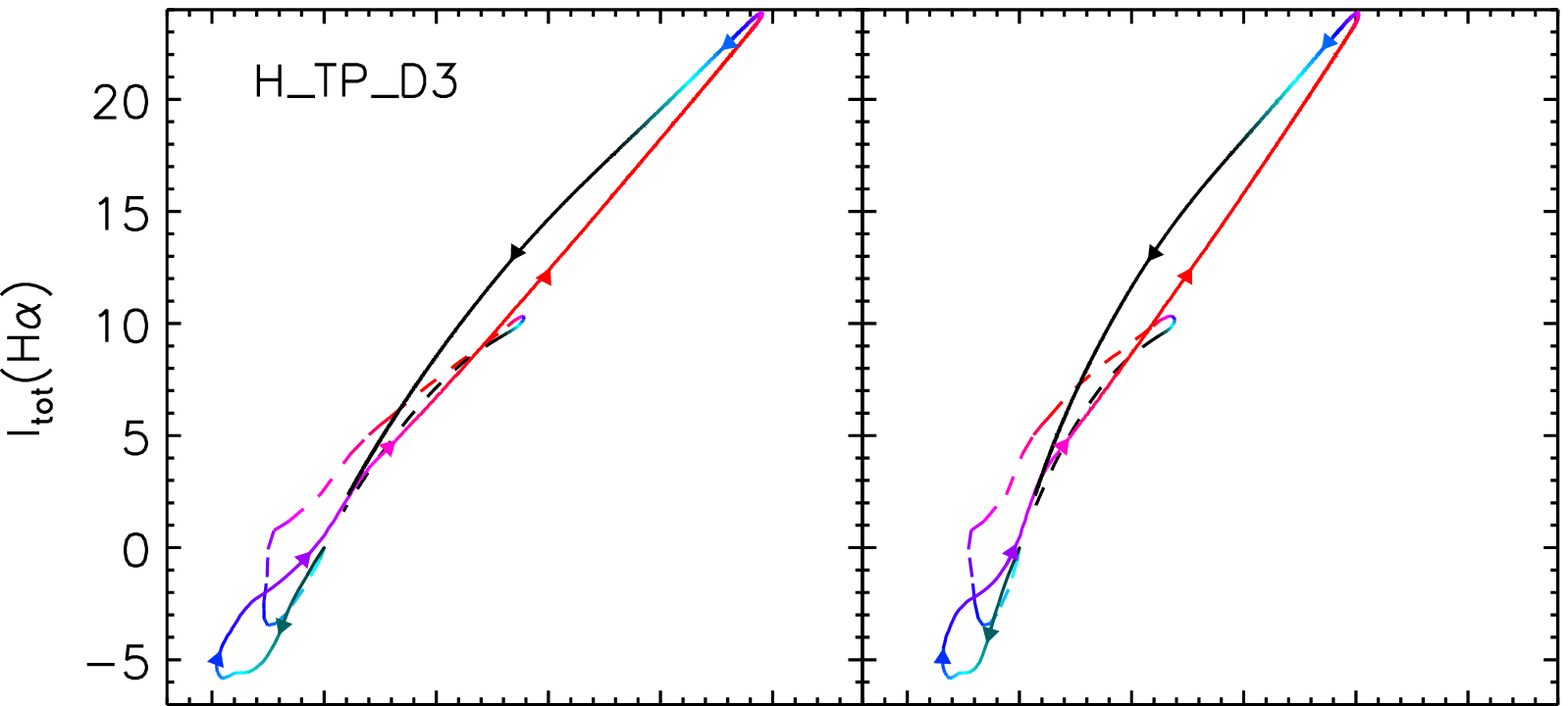}}
\centerline{\includegraphics[width=8.5cm]{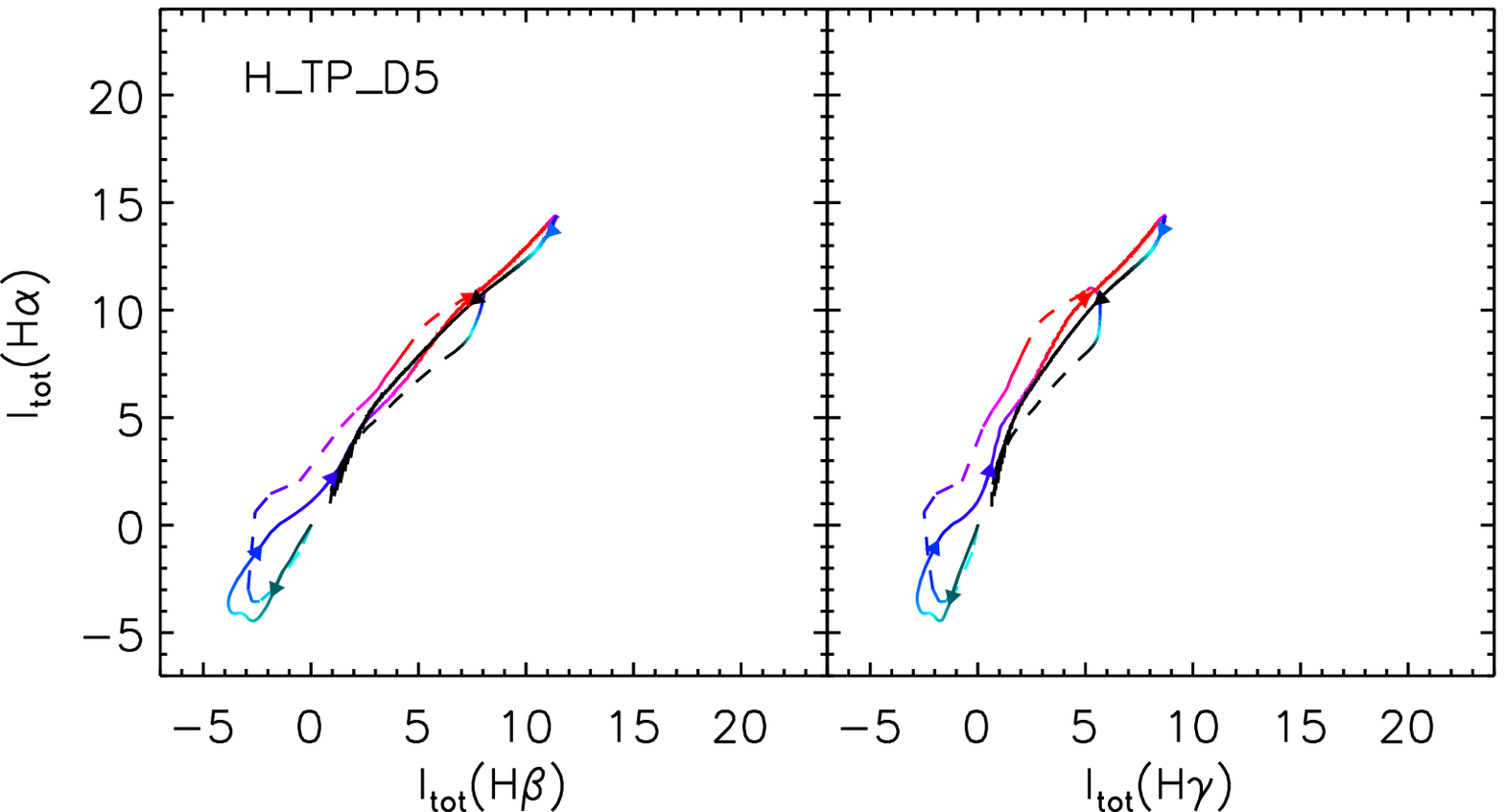}}
\caption{Time evolution of $I_\mathrm{tot}$(H$\alpha$), $I_\mathrm{tot}$(H$\beta$), and $I_\mathrm{tot}$(H$\gamma$).
The colour corresponds to the beam flux value  -- see also colouring in Fig.~\ref{fig:ew}. 
Arrows indicate the direction of time evolution. Solid lines denote models with $\cnt$,
dashed lines the models without $\cnt$.
{\em Top:} H\_3T\_D3 model. {\em Centre:} H\_TP\_D3 model. {\em Bottom:} H\_TP\_D5 model.
Total deposited energy $E_\mathrm{tot}$ is the same for all panels.
}
\label{fig:ew_composed}
\end{figure}

From our simulations it follows that maximum reached values of $I_\mathrm{tot}$ are predominantly given by the total deposited energy
$E_\mathrm{tot}$ (see Eq.~\ref{eq:etot} and Fig.~\ref{fig:ew}). 
Detailed time evolution of the beam flux is reflected as ``loopy structures'' in $I_\mathrm{tot}$ - $I_\mathrm{tot}$
plots -- see the top row of Fig.~\ref{fig:ew_composed} -- and the time variation of $I_\mathrm{tot}$ correlates with the beam
flux variation, see Fig.~\ref{fig:ew}.
However, there is no unique dependence of Balmer line $I_\mathrm{tot}$ on $F_\mathrm{max}$, see 
the right panel of Fig.~\ref{fig:ew} which shows gradual increase of $I_\mathrm{tot}$ with local minima and maxima corresponding
to the time modulation of the beam flux.
Moreover, in the case of the high-flux models, $I_\mathrm{tot}$ depend also on $\delta$. As a consequence of
increased wing emission for lower $\delta$, see Fig.~\ref{fig:lineint_time}, $I_\mathrm{tot}$ of all studied Balmer lines 
reach large values for flatter electron spectra (compare centre and bottom panel in Fig.~\ref{fig:ew_composed}).

Taking $\cnt$ into account leads to a significant increase of $I_\mathrm{tot}$ - see Figs.~\ref{fig:ew} and \ref{fig:ew_composed}, 
but similar increase can be caused by stronger heating by other mechanisms than electron beams. Due to this reasons,
neither wavelength-integrated intensities of Balmer lines are good indicators of electron beam presence in the Balmer line
formation regions.
\section{Conclusions}\label{sec:discussion}
Presented radiative hydrodynamic simulations revealed the complexity of the 
response of hydrogen Balmer lines to the electron beam heating. At the same time, they proved to be 
a very useful tool to obtain answers to questions raised in the Introduction.

\begin{enumerate}
\item We showed that the Balmer line intensities do vary on beam flux variation time scales, i.e. 
on a subsecond time scale.
The time variations are caused by time evolution of the temperature structure, electron density and influence
of the non-thermal collisional rates. Depending on the amount of the beam flux, 
time evolution of line intensities may also exhibit both fast (pulse like), and gradual (e.g. an increase
of intensity on a time scale larger than the beam flux time variation) components,
see e.g. the case of model L\_3T\_D3 in Fig.~\ref{fig:lineint_time}. 
Such behaviour is known for the H$\alpha$ line from observations \citep{trott00}.
Therefore, we conclude that the fast pulse-like variations seem to be a good indicator 
of the particle beams, namely when correlated with HXR or radio pulsations.
\item Influence of the non-thermal rates on the Balmer lines depends on the beam parameters, both
the energy flux and power-law index. $\cnt$ significantly alter the ionisation structure, leading
to a modification of the line formation regions which are not ionised due to the heating.
Depending on the beam parameters, $\cnt$ can affect line centres, wings or both, but generally
$\cnt$ result in an increased emission from a secondary formation region in the chromosphere.
\item Concerning the diagnostic tools based on Balmer lines, except for the close correlation 
of the time variation of the beam flux and the line intensities,
we did not found any systematic behaviour that would uniquely indicate the presence of the non-thermal 
electrons in the atmosphere solely from observations of Balmer lines. 
Complementary information such
as hard X-ray emission or spectral lines having different sensitivity to $\cnt$, e. g. Ca II (8542~\AA),
are needed to assess the presence of the non-thermal particles. 
\end{enumerate}
In this model study, we analysed the influence of the beam heating and the non-thermal collisional rates on
hydrogen Balmer lines in the case of prescribed fast beam flux modulation. The next step is
to compare the observed line emission with the simulated one using the non-LTE RHD models for beam parameters 
inferred from hard X-ray or radio emission.
In this way, the role of different flare energy transport mechanisms
e.g. such as alternative heating of the chromosphere by Alfv\'{e}n waves recently proposed by \citet{fle08}
can be adequately addressed. 
We plan to apply our code to fast time variations of H$\alpha$ and hard X-ray emissions observed during solar flares
\citep[e.g.][]{2007A&A...461..303R}.
\begin{acknowledgements}
We thank the referee, S.~L.~Hawley, for many valuable comments.
This work was supported by grants 205/04/0358, 205/06/P135, 205/07/1100 of
the Grant Agency of the Czech Republic and the research project AV0Z10030501 
(Astronomick\'{y} \'{u}stav). Computations were performed on 
OCAS  (Ond\v{r}ejov Cluster for Astrophysical Simulations) 
and Enputron, a computer cluster for extensive computations 
(Universita J. E. Purkyn\v e).
\end{acknowledgements}

\bibliographystyle{aa}
\bibliography{1559}

\begin{thebibliography}{40}
\expandafter\ifx\csname natexlab\endcsname\relax\def\natexlab#1{#1}\fi

\bibitem[{{Abbett} \& {Hawley}(1999)}]{abb99}
{Abbett}, W.~P. \& {Hawley}, S.~L. 1999, \apj, 521, 906

\bibitem[{{Allred} {et~al.}(2005){Allred}, {Hawley}, {Abbett}, \&
  {Carlsson}}]{all05}
{Allred}, J.~C., {Hawley}, S.~L., {Abbett}, W.~P., \& {Carlsson}, M. 2005,
  \apj, 630, 573

\bibitem[{{Bai}(1982)}]{bai82}
{Bai}, T. 1982, \apj, 259, 341

\bibitem[{{Canfield} {et~al.}(1984){Canfield}, {Gunkler}, \&
  {Ricchiazzi}}]{can84}
{Canfield}, R.~C., {Gunkler}, T.~A., \& {Ricchiazzi}, P.~J. 1984, \apj, 282,
  296

\bibitem[{{Cheng} {et~al.}(2006){Cheng}, {Ding}, \& {Li}}]{cheng06}
{Cheng}, J.~X., {Ding}, M.~D., \& {Li}, J.~P. 2006, \apj, 653, 733

\bibitem[{{Czaykowska} {et~al.}(1999){Czaykowska}, {de Pontieu}, {Alexander},
  \& {Rank}}]{czay99}
{Czaykowska}, A., {de Pontieu}, B., {Alexander}, D., \& {Rank}, G. 1999, \apjl,
  521, L75

\bibitem[{{Ding} {et~al.}(2001){Ding}, {Qiu}, {Wang}, \& {Goode}}]{ding01}
{Ding}, M.~D., {Qiu}, J., {Wang}, H., \& {Goode}, P.~R. 2001, \apj, 552, 340

\bibitem[{{Doschek} {et~al.}(1996){Doschek}, {Mariska}, \& {Sakao}}]{dos96}
{Doschek}, G.~A., {Mariska}, J.~T., \& {Sakao}, T. 1996, \apj, 459, 823

\bibitem[{{Emslie}(1978)}]{em78}
{Emslie}, A.~G. 1978, \apj, 224, 241

\bibitem[{{Fang} {et~al.}(1993){Fang}, {Henoux}, \& {Gan}}]{fang93}
{Fang}, C., {Henoux}, J.~C., \& {Gan}, W.~Q. 1993, \aap, 274, 917

\bibitem[{{Fisher} {et~al.}(1985){Fisher}, {Canfield}, \&
  {McClymont}}]{ficacly85}
{Fisher}, G.~H., {Canfield}, R.~C., \& {McClymont}, A.~N. 1985, \apj, 289, 414

\bibitem[{Fletcher \& Hudson(2008)}]{fle08}
Fletcher, L. \& Hudson, H.~S. 2008, A\&A, 675, 1645

\bibitem[{{Hawley} \& {Fisher}(1994)}]{1994ApJ...426..387H}
{Hawley}, S.~L. \& {Fisher}, G.~H. 1994, \apj, 426, 387

\bibitem[{{Heinzel}(1991)}]{hein91}
{Heinzel}, P. 1991, \solphys, 135, 65

\bibitem[{{Heinzel}(1995)}]{hein95}
{Heinzel}, P. 1995, \aap, 299, 563

\bibitem[{{Heinzel} \& {Karlick\'{y}}(1992)}]{heinkar92}
{Heinzel}, P. \& {Karlick\'{y}}, M. 1992, in Lecture Notes in Physics, Berlin
  Springer Verlag, Vol. 399, IAU Colloq. 133: Eruptive Solar Flares, ed.
  Z.~{Svestka}, B.~V. {Jackson}, \& M.~E. {Machado}, 359--+

\bibitem[{{Hoyng} {et~al.}(1981){Hoyng}, {Duijveman}, {Machado}, {Rust},
  {Svestka}, {Boelee}, {de Jager}, {Frost}, {Lafleur}, {Simnett}, {van Beek},
  \& {Woodgate}}]{hoy81}
{Hoyng}, P., {Duijveman}, A., {Machado}, M.~E., {et~al.} 1981, \apjl, 246, L155

\bibitem[{{Hudson} \& {F{\'a}rn{\'{\i}}k}(2002)}]{hufa02}
{Hudson}, H.~S. \& {F{\'a}rn{\'{\i}}k}, F. 2002, in ESA Special Publication,
  Vol. 506, Solar Variability: From Core to Outer Frontiers, ed. J.~{Kuijpers},
  261--264

\bibitem[{{Karlick\'{y}}(1990)}]{ka90}
{Karlick\'{y}}, M. 1990, \solphys, 130, 347

\bibitem[{{Karlick\'{y}} \& {H\'{e}noux}(1992)}]{ka92}
{Karlick\'{y}}, M. \& {H\'{e}noux}, J.-C. 1992, \aap, 264, 679

\bibitem[{Karlick\'{y} {et~al.}(2004)Karlick\'{y}, Ka\v{s}parov\'{a}, \&
  Heinzel}]{ka04a}
Karlick\'{y}, M., Ka\v{s}parov\'{a}, J., \& Heinzel, P. 2004, A\&AL, 416, L13

\bibitem[{{Kashapova} {et~al.}(2008){Kashapova}, {Kotr{\v c}}, \&
  {Kupryakov}}]{kash08}
{Kashapova}, L.~K., {Kotr{\v c}}, P., \& {Kupryakov}, Y.~A. 2008, Annales
  Geophysicae, 26, 2975

\bibitem[{{Ka{\v s}parov{\'a}} {et~al.}(2003){Ka{\v s}parov{\'a}}, {Heinzel},
  {Varady}, \& {Karlick{\'y}}}]{kas03}
{Ka{\v s}parov{\'a}}, J., {Heinzel}, P., {Varady}, M., \& {Karlick{\'y}}, M.
  2003, in Astronomical Society of the Pacific Conference Series, Vol. 288,
  Stellar Atmosphere Modeling, ed. I.~{Hubeny}, D.~{Mihalas}, \& K.~{Werner},
  544

\bibitem[{{Kopp} \& {Pneuman}(1976)}]{kopne76}
{Kopp}, R.~A. \& {Pneuman}, G.~W. 1976, \solphys, 50, 85

\bibitem[{{Kotr\v{c}} {et~al.}(2008){Kotr\v{c}}, {Kashapova}, \&
  {Kuprjakov}}]{2008ESPM...12.2.61K}
{Kotr\v{c}}, P., {Kashapova}, L.~K., \& {Kuprjakov}, Y.~A. 2008, 12th European
  Solar Physics Meeting, Freiburg, Germany, held September, 8-12, 2008.~Online
  at http://espm.kis.uni-freiburg.de/, p.2.61, 12, 2

\bibitem[{{Nagai} \& {Emslie}(1984)}]{naem84}
{Nagai}, F. \& {Emslie}, A.~G. 1984, \apj, 279, 896

\bibitem[{{Nejezchleba}(1998)}]{tomnej98}
{Nejezchleba}, T. 1998, \aaps, 127, 607

\bibitem[{{Oran} \& {Boris}(1987)}]{orbo87}
{Oran}, E.~S. \& {Boris}, J.~P. 1987, NASA STI/Recon Technical Report A, 88,
  44860

\bibitem[{{Oran} \& {Boris}(2000)}]{orbo87book}
{Oran}, E.~S. \& {Boris}, J.~P. 2000, {Numerical Simulation of Reactive Flow}
  (Numerical Simulation of Reactive Flow, by Elaine S.~Oran and Jay P.~Boris,
  pp.~550.~ISBN 0521581753.~Cambridge, UK: Cambridge University Press, November
  2000.)

\bibitem[{{Peres} {et~al.}(1982){Peres}, {Serio}, {Vaiana}, \&
  {Rosner}}]{pes82}
{Peres}, G., {Serio}, S., {Vaiana}, G.~S., \& {Rosner}, R. 1982, \apj, 252, 791

\bibitem[{{Radziszewski} {et~al.}(2007){Radziszewski}, {Rudawy}, \&
  {Phillips}}]{2007A&A...461..303R}
{Radziszewski}, K., {Rudawy}, P., \& {Phillips}, K.~J.~H. 2007, \aap, 461, 303

\bibitem[{{Rosner} {et~al.}(1978){Rosner}, {Tucker}, \& {Vaiana}}]{rtv78}
{Rosner}, R., {Tucker}, W.~H., \& {Vaiana}, G.~S. 1978, \apj, 220, 643

\bibitem[{{Rybicki} \& {Hummer}(1991)}]{ryhu91}
{Rybicki}, G.~B. \& {Hummer}, D.~G. 1991, \aap, 245, 171

\bibitem[{{Shibata}(1996)}]{shi96}
{Shibata}, K. 1996, Advances in Space Research, 17, 9

\bibitem[{{Sturrock}(1968)}]{stu68}
{Sturrock}, P.~A. 1968, in IAU Symposium, Vol.~35, Structure and Development of
  Solar Active Regions, ed. K.~O. {Kiepenheuer}, 471

\bibitem[{Tandberg-Hanssen \& Emslie(1988)}]{ta88}
Tandberg-Hanssen, E. \& Emslie, G.~A. 1988, The physics of solar flares
  (Cambridge University Press)

\bibitem[{{Trottet} {et~al.}(2000){Trottet}, {Rolli}, {Magun}, {Barat},
  {Kuznetsov}, {Sunyaev}, \& {Terekhov}}]{trott00}
{Trottet}, G., {Rolli}, E., {Magun}, A., {et~al.} 2000, \aap, 356, 1067

\bibitem[{{Turkmani} {et~al.}(2005){Turkmani}, {Vlahos}, {Galsgaard},
  {Cargill}, \& {Isliker}}]{turk05}
{Turkmani}, R., {Vlahos}, L., {Galsgaard}, K., {Cargill}, P.~J., \& {Isliker},
  H. 2005, \apjl, 620, L59

\bibitem[{{{\v S}t{\v e}p{\'a}n} {et~al.}(2007){{\v S}t{\v e}p{\'a}n}, {Ka{\v
  s}parov{\'a}}, {Karlick{\'y}}, \& {Heinzel}}]{st07}
{{\v S}t{\v e}p{\'a}n}, J., {Ka{\v s}parov{\'a}}, J., {Karlick{\'y}}, M., \&
  {Heinzel}, P. 2007, \aap, 472, L55

\bibitem[{{Vernazza} {et~al.}(1981){Vernazza}, {Avrett}, \& {Loeser}}]{val81}
{Vernazza}, J.~E., {Avrett}, E.~H., \& {Loeser}, R. 1981, \apjs, 45, 635

\end{thebibliography}
\end{document}